\documentclass{article}

\usepackage[margin=1in]{geometry}
\usepackage[utf8]{inputenc} 
\usepackage[T1]{fontenc}    
\usepackage{hyperref}       
\usepackage{url}            
\usepackage{booktabs}       
\usepackage{amsfonts}       
\usepackage{nicefrac}       
\usepackage{microtype}      
\usepackage{xcolor}         
\usepackage{graphicx}
\usepackage{wrapfig}
\usepackage{capt-of}
\usepackage{subcaption}

\usepackage{multirow}           
\usepackage{duckuments}         

\usepackage{amsmath}
\usepackage{mathtools}
\usepackage{enumitem}
\usepackage{amsthm}
\usepackage{pifont}
\usepackage[linesnumbered,vlined,ruled,commentsnumbered]{algorithm2e}
\usepackage{tcolorbox}
\usepackage{tikz}
\usetikzlibrary{automata, positioning}
\usepackage[export]{adjustbox}
\usepackage[square, numbers]{natbib}
\newcommand\xtw[1]{\textcolor{blue}{[xw: #1]}}
\newcommand\unsure[1]{\textcolor{olive}{[check: #1]}}

\newtheorem{theorem}{Theorem}[section]

\newtheorem{definition}[theorem]{Definition}

\usepackage{float}

\newtheorem{example}{Example}[section]

\newcommand{\yuhong}[1]{{\textcolor{red}{Y: #1}}}

\author{Yuhong Luo \thanks{Rutgers University, \texttt{y.luo@rutgers.edu}} \and Daniel Schoepflin \thanks{DIMACS, Rutgers University, \texttt{daniel.r.schoepflin@gmail.com}} \and Xintong Wang \thanks{Rutgers University, \texttt{xintong.wang@rutgers.edu}}}

\title{Algorithmic Collusion at Test Time: A Meta-game Design and Evaluation}

\begin{document}

\maketitle
\begin{abstract}
The threat of algorithmic collusion, and whether it merits regulatory intervention, remains debated, as existing evaluations of its emergence often rely on long learning horizons, assumptions about counterparty rationality in adopting collusive strategies, and symmetry in hyperparameters and economic settings among players.
To study collusion risk, we introduce a meta-game design for analyzing algorithmic behavior under test-time constraints.
We model agents as possessing pretrained policies with distinct strategic characteristics (e.g., competitive, naively cooperative, or robustly collusive), and formulate the problem as selecting a meta-strategy that combines a pretrained, initial policy with an in-game adaptation rule.
We seek to examine whether collusion can emerge under rational choices and how agents co-adapt toward cooperation or competition. 
To this end, we sample normal-form empirical games over meta-strategy profiles, 
compute relevant game statistics (e.g., payoffs against individuals and regret against an equilibrium mixture of opponents), and construct empirical best-response graphs to uncover strategic relationships.
We evaluate reinforcement-learning, UCB, and LLM-based strategies in repeated pricing games under symmetric and asymmetric cost settings, and present findings on the feasibility of algorithmic collusion and the effectiveness of pricing strategies in practical ``test-time'' environments. 
 The source code is available at: \url{https://github.com/chailab-rutgers/CollusionMetagame}.
\end{abstract}

\section{Introduction}
The use of algorithms to automate economic decisions---such as pricing, bidding, and bargaining---has become increasingly prevalent.
Rather than following static rules, modern systems leverage optimization techniques (e.g., reinforcement learning) or AI models with reasoning capabilities (e.g., large language models) to learn from their environment, anticipate the behavior of other participants, and adapt their strategies to improve outcomes.

This growing autonomy introduces new, critical risks including the undesired and uncommunicated cooperation among algorithms, commonly known as \textit{algorithmic collusion}.
Recognized as a major concern in the era of advanced AI~\citep{hammond2025multiagentrisksadvancedai}, algorithmic collusion poses unique challenges beyond human collusion~\citep{cramton2000collusive, agranov2015collusion}, as algorithms may learn to coordinate without explicit communication or intent~\citep{oecd}.
%
%
%
Prior work demonstrates that collusive behavior can arise in simulated pricing and auction environments using common algorithms such as Q-learning~\citep{Calvano2020Artificial} and large language models (LLMs)~\citep{Fish2024Algorithmic}.
However, whether such behavior persists under realistic conditions and deployments remains unclear.
For example, \citet{Calvano2020Artificial} document convergence to collusion between two Q-learning agents after around 1.5 million interaction rounds, reflecting a prohibitively long learning horizon with significant early-stage exploration costs.

A more practical evaluation considers test-time interactions, where pretrained policies are randomly rematched and allowed to interact for a limited number of rounds~\citep{Calvano2020Artificial, Eschenbaum2022Robust}.
These studies show that while collusion may fade at first, it can re-emerge after roughly 40,000 rounds of play.
Importantly, prior analyses largely assume symmetric algorithmic configurations (e.g., identical hyperparameters) and leave open a central question: whether collusion constitutes a rational, stable outcome when agents strategically behave under test-time constraints.

In this work, we seek to evaluate the viability and effectiveness of algorithmic collusion in ``test-time'' settings, where an agent must adapt within a limited number of interactions to opponents whose policies and economic settings (e.g., cost or quality) may differ from those encountered during training.
%
To this end, we propose a meta-game framework.
Different learning algorithms (e.g., Q-learning, no-regret methods, and LLMs) are first used at training time to generate candidate \emph{initial policies} for deployment.
These policies are then grouped into strategic categories based on performance metrics, including their propensity to collude with a training-time partner and their robustness against a best-response opponent.
At test time, a \emph{strategy} is formed by pairing an initial policy sampled from a selected category with an adaptation rule (e.g., a learning rate) drawn from a set of feasible update procedures for repeated play.
A \emph{meta-strategy} governs both the choice of initial policy category and the manner in which the policy adapts during interaction.

We are interested in analyzing these competing meta-strategies by modeling their interactions as a game, referred to as the \textit{meta-game}.
By sampling strategies for each player together with initial states, we generate meta-game instances from which evaluation metrics can be computed and game-theoretic analysis can be further conducted.
Through this framework, we provide a statistical characterization of meta-strategy performance and address the central question: Can algorithmic collusion emerge within a limited time horizon under rational strategic choices?

\paragraph{Our contribution.}
We propose metrics to characterize and evaluate a (pretrained) policy
and introduce 
a meta-game framework to assess algorithmic collusion at test time through interactions among meta-strategies.
We conduct extensive experiments using a diverse set of algorithms (including Q-learning, UCB, and LLMs) to generate initial policies for repeated pricing games.
Our results provide empirical evidence on meta-strategy performance, and offer insights into the conditions under which algorithmic collusion may emerge and persist.
%
%
We highlight several key findings below: 
\begin{itemize}[leftmargin=*]
    \item 
    Q-learning can produce pretrained policies that \textit{robustly collude} with their best-response counterparts, whereas UCB-based policies, despite colluding with their training-time partners, are often exploitable by best-response opponents. 
    \item Under symmetric cost settings, each algorithm we evaluate---Q-learning, UCB, LLM---admits at least one pure or mixed Nash equilibrium among meta-strategies that leads to a collusive outcome, indicating that collusion can arise from rational strategic choices.\looseness=-1
    \item For Q-learning meta-strategies, collusion diminishes under shorter interaction horizons and pessimistic Q-value initialization, which can be interpreted as reflecting an agent's prior belief that its counterparty is less likely to collude.
    Notably, in contrast to prior studies based on symmetric algorithmic setups that observe sustained collusion in asymmetric cost settings~\cite{Calvano2020Artificial}, our findings reveal that rational strategy selection substantially suppresses collusion under asymmetry.
    
       \item While UCB meta-strategies exhibit stronger collusion than Q-learning overall, Q-learning with random initialization can best respond to most UCB meta-strategies, calling into question UCB's competitiveness under test-time conditions.
       \item LLM-based agents can demonstrate adaptive behavior guided by pre-game history: policies that exhibit greater collusion during pre-game interactions tend to sustain or re-establish collusion even after episodes of exploitation at test time.
       
\end{itemize}

\subsection{Related Work}

\paragraph{Algorithmic collusion.}

Collusion can arise as an equilibrium among rational players in infinitely repeated games.
Building on the Folk Theorem, prior work shows that cooperation can be sustained when algorithms embed credible punishment for deviations~\citep{Usui2021Symmetric, Hansen2021Frontiers, lamba2022pricingalgorithms, ABADA2024927}.
More recent studies find that collusive equilibria may also emerge without explicit threat mechanisms, as shown in a Stackelberg framework~\citep{Arunachaleswaran2024Algorithmic}, and examine how algorithmic (non-)collusion can be detected and regulated through data audits~\citep{Hartline2024Regulation, hartline_regulation_2025}.

Collusive behavior has been documented in e-commerce and dynamic pricing settings~\citep{ezrachi_algorithmic_2015, Byrne2019Learning, musolff_algorithmic_2022, Assad2024Algorithmic,gianluca2022}.
Following~\citet{Calvano2020Artificial}, a growing body of literature investigates algorithmic collusion in the laboratory across settings ranging from stylized games, such as the repeated Prisoner’s Dilemma~\citep{Meylahn2022Limiting, Wolfram2023Intrinsic}, to more realistic auctions~\citep{Banchio2022Artificial}. 
These studies also consider different algorithms, including reinforcement learning~\citep{klein_autonomous_2021, hettich_algorithmic_2021}, bounded-regret methods~\citep{Hansen2021Frontiers}, and LLMs~\citep{Fish2024Algorithmic, agrawal2025evaluatingllmagentcollusion}.
Several follow-up studies suggest that prior findings may overstate the threat of algorithmic collusion, as they rely on restrictive modeling assumptions~\citep{abada_algorithmic_2024}.
Collusion typically emerges after long training horizons---on the order of millions of repeated-game rounds---and tends to depend on symmetric algorithmic configurations, suggesting implicit coordination or prior communication~\citep{lambin_less_2023}. 
Moreover, the algorithmic configurations that give rise to collusion may be irrational in practice, being outperformed by more robust or commonly used alternatives~\citep{Abada2023Artificial, bichler2024onlineoptimizationalgorithmsrepeated}. Thus, the emergence of algorithmic \textit{tacit} collusion among rational agents under realistic conditions and finite interaction horizons remains an open question.

\paragraph{Strategy selection, adaptation, and evaluation.}
Our work focuses on strategy performance at deployment, resembling a test-time or tournament environment.
Prior research has examined strategy selection and evaluation through meta-game frameworks~\citep{kiekintveld_selecting_nodate, Li2024Meta} and empirical game-theoretic analysis (EGTA)~\citep{wellman2025empiricalgametheoreticanalysissurvey}, applied to domains such as trading agent competitions~\citep{Reeves2005Generating, Wellman2005Strategic, McBurney2006Selecting}, financial markets~\citep{Wah2016,Wang2017,Wang2018,Wang2020b}, supply chain management~\citep{Jordan2007Empirical}, negotiation~\citep{Li2024Meta}, and auctions~\citep{Vorobeychik2006Empirical}.
Related to our setting is~\citet{carissimo2025algorithmiccollusionalgorithmorchestration}, who model collusion as coordination in Q-learning hyperparameters (e.g., discount and exploration factors) during training.  
Our study differs by focusing on test-time adaptation among pretrained agents with heterogeneous initializations, distinguishing deployment behavior from training dynamics.

Related online learning literature also examines \textit{adaptability} and \textit{non-exploitability} in repeated games~\citep{Crandall2010Learning, Crandall2014Towards, DiGiovanni2022Balancing}, and develops approximate best-response methods for equilibrium analysis in complex games such as poker and Go~\citep{lisyeqilibrium2017, Timbers2022Approximate, martin2024approxedapproximateexploitabilitydescent}.
These insights inform our design of metrics for characterizing pretrained policies and their adaptation strategies.

\section{Preliminaries}
\label{sec:background}


We consider an $n$-player simultaneous-move repeated game progressing from time $0$ to an end time unknown to the players. 
At each round $t$, a player $j$ commits to an \emph{action} (e.g., price) $p_{j,t} \in \mathcal{P}$, where $\mathcal{P}$ denotes a discrete action space.

%
%
%
The game admits a \emph{Stage-Game Nash Equilibrium} (SGNE) in which no player can profitably deviate in a single round, given that stage games are independent across periods.
Let $r^N_j$ denote player $j$'s \emph{competitive payoff}, $\Bar{r}^N$ the average competitive payoff, and $p^N_j$ the \emph{competitive action} of player $j$ under the SGNE. 
In pricing settings with homogeneous goods and symmetric costs, this corresponds to the \textit{Bertrand Nash equilibrium}.

Players may nonetheless achieve higher payoffs through coordinated behavior that avoids the SGNE.
 The optimal joint outcome for all players in a stage game can be derived by maximizing the joint payoffs; we refer to this outcome as \emph{the monopoly} which characterizes full collusion. 
In pricing contexts, such outcomes are detrimental to consumers. 
 Let $r^M_j$ denote player $j$'s \emph{monopoly payoff}, $\Bar{r}^M$ the average monopoly payoff, and  $p^M_j$ the associated \emph{monopoly action} of player $j$. 

The repeated pricing game can be modeled as an extensive-form game whose central solution concept is the \textit{subgame perfect equilibrium} (SPE)---a strategy profile that constitutes a Nash equilibrium in every subgame.
According to the Folk Theorem~\citep{friedman1971non}, many SPEs exist, with each supported by \textit{credible threats}, i.e., responses that are rational for the player making them~\citep{schelling1956essay}.
By contrast, non-credible threats cannot sustain an SPE, and monopoly outcomes often depend on such threats.
Indeed, in our experiments, we find no evidence of credible threats among pretrained policy pairs, suggesting that realistic collusion may not rely on them.

Following~\citet{Calvano2020Artificial}, we define the Collusion Index (CoI) as a measure of collusion: when CoI $= 0\%$, players are fully competitive by playing SGNE,
whereas CoI $= 100\%$ corresponds to full collusion at the monopoly outcome.

\begin{definition}[Collusion Index (CoI)]
    \label{def:coi}
   \[CoI \coloneqq \frac{\Bar{r} - \Bar{r}^N}{\Bar{r}^M-\Bar{r}^N}\text{, where $\Bar{r}$ is the average per-player payoff.}\]
\end{definition}


We define a state of the game $s \in \mathcal{S}$ as actions (i.e., prices) of all players in one round, ordered by the player IDs. In this work, we focus on two-player repeated games. We let $S_t$ be the random variable of the state of the game at timestamp $t$.
A policy $\pi_{j,t}\in\Pi$ for player $j$ at time $t$ maps the state $S_t = s$ to a probability distribution over actions.
%
%
Given a discount factor $\gamma$ and a fixed policy profile $(\pi_j,\pi_{-j})$, let $V^{\pi_j|\pi_{-j}}(s)$ denote the expected utility of player $j$ starting from state $s$.


\begin{definition} [State-Value Function]  Suppose  player $j$ interacts with an opponent following a fixed policy $\pi_{-j}$. The state-value function under $\pi_j$ is defined as 
    \begin{align*}
    V^{\pi_j | \pi_{-j}}(s) &= \mathbb{E}\bigg[\sum_{k=0}^\infty \gamma^k r_{j}(S_{t+k}) \bigg| S_t=s, \pi_j, \pi_{-j} \bigg] \\
    &= r_j(s) + \gamma \sum_{p_j\in \mathcal{P}}\sum_{p_{-j}\in \mathcal{P}}\pi_j(s, p_j) \pi_{-j}(s, p_{-j})V^{\pi_j| \pi_{-j}}((p_j, p_{-j})).
\end{align*}
\end{definition}

If state $(p_j, p_{-j})$ is an absorbing state, i.e.,  $\pi_j((p_j,p_{-j}), p_j) = 1$ and $\pi_{-j}((p_j,p_{-j}), p_{-j}) = 1$, then $V^{\pi_j | \pi_{-j}}((p_j,p_{-j})) = \frac{r_j((p_j,p_{-j}))}{1-\gamma}$.

\section{Meta-game Design for Algorithmic Collusion}
\label{sec:metagame}
This section introduces our meta-game design to reason about how players select and adapt their policies in ``test-time'' environments with limited interactions,  
and assess whether rational choices can lead to collusive outcomes.

\subsection{Initial Policy, Strategy, and Meta-strategy}\label{sec:meta_strategy}
\subsubsection{Initial Policy}
As in many multi-agent domains, agents begin play with a pretrained policy derived from prior experience, we consider agents that start with an initial policy generated via a pretraining phase. 
We model this by applying a learning algorithm $\mathcal{A}_{\theta}$ and a set of random seeds $\mathcal{K}$ to a game of interest $\mathcal{G}$ to generate a set of initial policy profiles $\Pi$.
We assume each pretraining process, i.e., a stochastic procedure that produces a policy profile $\pi_\kappa = \{\pi^\kappa_1, ..., \pi^\kappa_N\} = \mathcal{A}_{\theta}(\mathcal{G}, \kappa)$, is subject to no time constraint or cost of learning.\looseness=-1

\subsubsection{Strategy}
Without adaptation, however, pretrained policies can fail against unfamiliar opponents, resulting in unexpected state transitions and suboptimal outcomes at test time.%
\footnote{Indeed, in experiments, we find that two independently pretrained Q-learning agents, when paired against each other without adaptation, can yield payoffs lower than competitive pricing (see Fig.~\ref{fig:pre_adapt} in the appendix).} 
Achieving strong performance thus requires an update procedure to adapt the initial policy to the specific opponent during play.
We define an agent's strategy for the game at test time $\mathcal{G^*}$ as the combination of a (pretrained) initial policy and an update procedure. 
This strategy is evaluated over a shorter interaction horizon, where learning incurs costs and performance at each round counts.

Rather than directly modifying the policy, we consider strategies that operate on an underlying \textit{internal representation} that encodes the policy.
We denote this representation by $Z_{j,t}$ for player $j$ at time $t$, and it is updated after each round. 
For example, in Q-learning, the internal representation consists of Q-values, from which the agent selects the action with the highest value.

Formally, a strategy is specified by two functions, typically determined by the learning algorithm and its hyperparameters:
\begin{itemize}
    \item A \textit{decoding function} $\phi$ which maps the internal representation to a policy, $\phi(Z_{j, t})=\pi_{j, t}$, and  
    \item An \textit{update function} $\omega$ which updates the representation based on the latest experience (i.e., the state), $\omega(Z_{j, t}, S_{j, t})=Z_{j, t+1}$.
\end{itemize}
At any time $t$, a player’s strategy is fully specified by its current representation $Z_{j,t}$ and its decoding and update functions $(\phi_j, \omega_j)$. 
In Section~\ref{sec:exp}, we will illustrate how different algorithms, including Q-learning, UCB, and LLMs, instantiate these components.

\subsubsection{Meta-strategy}
Pretraining can generate a large set of initial policies, each of which may be paired with infinitely many update procedures (e.g., varying learning rates). 
As a result, the induced strategy space becomes prohibitively large, rendering direct and systematic analysis intractable.
To manage this complexity, we group initial policies according to their performance along two key strategic dimensions: their ability to achieve cooperation (i.e., how effectively a policy learns to cooperate with a specific partner) and their ability to avoid exploitation (i.e., how robustly the policy performs against a best-response opponent). 
We begin by introducing metrics to evaluate these dimensions.

\begin{definition}[Paired Cooperativeness (PC)]
\label{def:PC} 
Let $\pi_j$ and $\pi_{-j}$ denote a pair of policies. 
The paired cooperativeness between $\pi_j$ and $\pi_{-j}$ is defined as the respective mean state values when interacting with one another, i.e., 
\[\text{PC}(\pi_j, \pi_{-j})=\Big(\overline{V}^{\pi_j|\pi_{-j}}, \overline{V}^{\pi_{-j}|\pi_{j}}\Big), \text{ where }
\overline{V}^{\pi_x|\pi_y}=\frac{\sum_{s\in\mathcal{S}}V^{\pi_x|\pi_{y}}(s)}{|\mathcal{S}|}.\]
\end{definition}

\textit{Paired Cooperativeness} (PC) measures the expected utility of each policy in a pair when the initial state is drawn uniformly at random. 
Intuitively, jointly pretrained policy pairs tend to exhibit high PC values, with positively correlated mean state values, as their trajectories are trained to converge to collusive outcomes from any initial state. 
In contrast, independently pretrained policies paired at test time are less likely to reach collusive absorbing states, leading to lower PC values and increasing the likelihood of one policy exploiting the other.

With PC capturing how well two policies sustain cooperation, we further introduce \textit{cooperative robustness} (CR), which evaluates a policy’s performance against its best response \textit{and} its tendency to cooperate with that best response.

\begin{definition}[Cooperative Robustness (CR)]
    \label{def:CR}  
    Let $\pi_b$ denote the worst-case best-response policy to $\pi_j$.
    \[\pi_b = \arg\min_{\pi \in BR(\pi_j)} \overline{V}^{\pi_j|\pi}, \text{ where } BR(\pi_j) \coloneqq \arg\max_{\pi  \in \Pi} V^{\pi|\pi_j}(s), \forall s \in \mathcal{S}.\]
    The best responses can be obtained via value iteration~\citep{sutton2018reinforcement}.
    The cooperative robustness between $\pi_j$ and $\pi_b$ is defined as their respective mean state values when interacting with each other: 
    \[\text{CR}(\pi_j, \pi_b)=\Big(\overline{V}^{\pi_j|\pi_b}, \overline{V}^{\pi_b|\pi_{j}}\Big).\]
    
\end{definition}

Intuitively, the relative magnitudes of $\overline{V}^{\pi_j|\pi_b}$ and $\overline{V}^{\pi_b|\pi_j}$ characterize the strategic nature of the interaction between the two policies. 
When these values are highly asymmetric, it indicates an exploitative relationship, typically with $\pi_b$ exploiting $\pi_j$, which reflects $\pi_j$'s lack of robustness. 
When both values are low, the interaction tends to settle in a competitive absorbing state, indicating robustness but limited cooperative gains. 
In contrast, when both values are high and comparable, the two policies exhibit robust cooperation, achieving mutually beneficial outcomes while remaining stable against unilateral deviations.%
\footnote{See Fig.~\ref{fig:cat_Q} in the appendix for examples of Q-learning policies in different strategic categories.}

We characterize pretrained initial policies along these two strategic dimensions and group them into distinct categories. 
Policies within each category are then paired with a discrete set of update procedures. 
We define a \textit{meta-strategy} as a family of strategies formed by combining an initial policy category---defined by specific strategic attributes (e.g., in PC and CR)---with a corresponding update rule.
Practically, each meta-strategy specifies how an agent \textit{selects and adapts} an initial policy from the pool of candidates generated during pretraining.
%
%

\medskip

\begin{figure*}
\centering
\includegraphics[trim={0.8cm 1.1cm 0.15cm 1.2cm}, clip,width=0.85\textwidth]{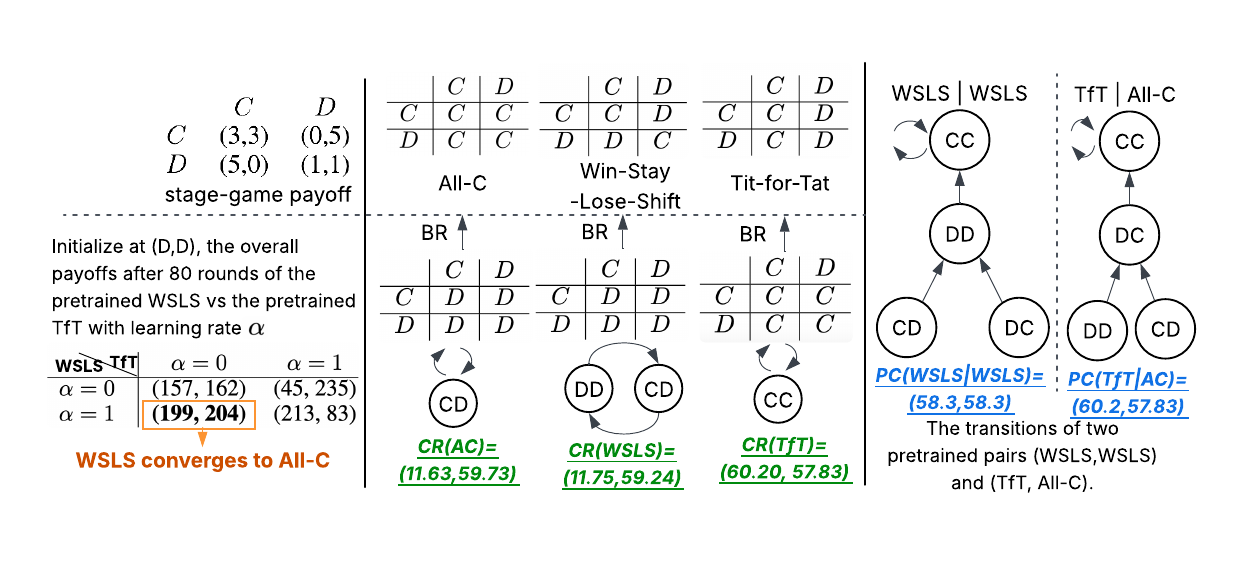}
\caption{
A toy meta-game for repeated Prisoner's Dilemma on canonical strategies. 
The NE of this meta-game leads to cooperation. 
See Example~\ref{exp:toy} for more discussion.}\label{fig:toy}
\end{figure*}

Below we provide a motivating example based on canonical strategies in the repeated Prisoner's Dilemma, coupled with simple update rules, as illustrated in Fig.~\ref{fig:toy}.

%
\begin{example}[A toy meta-game for repeated Prisoner's Dilemma]\label{exp:toy}
Consider the following initial policies for repeated Prisoner's Dilemma: Win-Stay, Lose-Shift (\texttt{WSLS}), Tit-for-Tat (\texttt{TfT}), and Always-Cooperate (\texttt{AC}). 
Suppose they are obtained from two pretrained policy pairs that successfully learned to cooperate, i.e., (\texttt{WSLS}, \texttt{WSLS}) and (\texttt{TfT}, \texttt{AC}).

We categorize these policies. 
All policies achieve high PC with their respective pretrained partners due to cooperative behavior.
However, \texttt{WSLS} and \texttt{AC} both yield low mean state values against their best responses, resulting in highly asymmetric CR, as they can be exploited by a best-response policy (e.g., Always-Defect).
By contrast, \texttt{TfT} attains high CR, as its best response is to always cooperate.

Consider four meta-strategies constructed by pairing high-PC–low-CR and high-PC–high-CR initial policies with either slow or fast in-game update rules.
Suppose \texttt{WSLS} and \texttt{TfT} are sampled as representative initial policies from the two categories
and are paired with low and high learning rates.
Fig.~\ref{fig:toy} presents one part of the empirical meta-game defined over this restricted, sampled strategy space.
Simulated profile payoffs indicate a meta-game equilibrium in which \texttt{WSLS} updates its policy and \texttt{TfT} remains unchanged.%
\footnote{
Q-learning is adopted for both pretraining and test-time adaptation with $\gamma=0.95$. 
We normalize initial Q-values following the procedure described in Appendix~\ref{sec:Q_Normal} and set exploration $\varepsilon=0$. 
We evaluate strategies at $t=80$, when both strategies converged.}


We are interested in sampling and aggregating many such payoffs to construct empirical meta-games and analyze the relative performance of meta-strategies.
For instance, in a different meta-game, \texttt{AC} could be sampled as an alternative initial policy from the high-PC–low-CR category.
\end{example}

%

\subsection{Empirical Game-theoretic Analysis}
\label{sec:egta}
Our meta-game evaluation leverages empirical game-theoretic analysis (EGTA), a methodology for analyzing complex game situations and strategic interactions through agent-based simulation~\cite{Wellman2006Methods, wellman2025empiricalgametheoreticanalysissurvey}.
EGTA constructs an empirical normal-form game over a selected set of strategies using payoff data generated from simulations of the underlying environment.

We apply EGTA to a repeated two-player, general-sum game (e.g., repeated pricing or auction) to investigate algorithmic collusion. 
Given a set of $M$ meta-strategies, our goal is to evaluate their relative performance in the test-time, base game $\mathcal{G}^*$. 
Let $\hat{\psi}^m$ denote a strategy sampled from meta-strategy $\mathcal{M}^m$
and let $\hat{\Psi}=\{\hat{\psi}^1, \dots, \hat{\psi}^M\}$ represent the set of all sampled strategies. 
We then construct an empirical meta-game $\mathcal{MG}(\hat{\Psi})$ over this sampled strategy space.

When the base game $\mathcal{G^*}$ is symmetric (e.g., agents have the same cost and quality parameters), we estimate the meta-game payoff function by simulating two-player strategy profiles over $\hat{\Psi}$ to play $\mathcal{G^*}$ (due to symmetry, the assignment of strategies to players does not matter).%
\footnote{To prevent collusion arising from both players using an identical or jointly pretrained policy, if the same policy or its pretrained pair has been drawn, we redraw an instance. 
}
The empirical meta-game payoff matrix is estimated by repeatedly simulating $\mathcal{G}^*$ for each two-player strategy profile over different initial states sampled from $\mathcal{S}$ across runs.
When the base game $\mathcal{G}^*$ is asymmetric, agents are pretrained on different parameters and may possess distinct strategies. 
We therefore sample $\hat{\Psi}_1=\{\hat{\psi}_1^1, \dots, \hat{\psi}_1^M\}$ and $\hat{\Psi}_2=\{\hat{\psi}_2^1, \dots, \hat{\psi}_2^M\}$ from their respective meta-strategy sets to construct the empirical game.


\subsection{A Meta-game Evaluation Framework}\label{sec:metric}

Given a two-player, general-sum repeated game $\mathcal{G}^*$ and meta-strategies $\{\mathcal{M}^1, ..., \mathcal{M}^M\}$, we construct empirical meta-games via the following sampling and evaluation procedure in Algorithm~\ref{alg:metagame}.


\begin{algorithm}[h]
\SetKwInOut{Input}{Input}
\SetKwInOut{Output}{Output}
\Input{Meta-strategy set $\Psi$, base game $\mathcal{G}^*$, number of meta iterations $N_\text{meta}$, number of base game runs $N_\text{base}$}
\Output{Empirical meta-games and game-analysis stats $X$}

\Repeat{$N_\text{meta}$}{
    Construct a strategy set $\hat{\Psi} = \{\hat{\psi}^1, \dots, \hat{\psi}^M\}$ by uniformly sampling one strategy from each meta-strategy in $\Psi$.

    \Repeat{$N_\text{base}$}{
        Sample an initial state $s_0 \sim \mathcal{S}$ for the base game $\mathcal{G}^*$.

        Estimate the empirical $\mathcal{MG}(\hat{\Psi})$ by simulating two-player profiles as described in Section~\ref{sec:egta}, and record the resulting profile payoffs.

        Compute the desired statistics $X$ from $\mathcal{MG}(\hat{\Psi})$.
    }
}
\caption{Meta-game evaluation procedure}
\label{alg:metagame}
\end{algorithm}

\paragraph{Game-analysis statistics.}

We construct \emph{weighted best-response graphs} using the payoff matrices of meta-games generated by Algorithm~\ref{alg:metagame}.
%
For each meta-game, the \emph{best-response score} from strategy $u$ to $v$ is defined as the ratio of $u$'s average payoff against $v$ to the highest average payoff achievable by any strategy against $v$.
The best-response scores across all meta-games are aggregated to determine the edge weights. 
If $\mathcal{G^*}$ is asymmetric, we maintain two directed best-response graphs, one for each player role.
%
%

We evaluate the \emph{uniform score} of a meta-strategy, i.e., the expected payoff of a meta-strategy when the opponent is drawn from a uniform distribution over all meta-strategies. 
%
%
%
As uniform scores represent a naive belief of the opponent distribution,
we also adopt an alternative metric, NE-regret, as proposed by~\citet{Jordan2007Empirical}.

\begin{definition} [NE-Regret]
Suppose $\sigma^*$ is a symmetric mixed-strategy NE of $\mathcal{MG}(\Psi)$. The NE-regret of a pure meta-strategy $\mathcal{M}_j$ is defined as
the difference between the expected payoff of player $j$ of the equilibrium profile ($\sigma^*$, $\sigma^*$) and that of the mixed-strategy profile ($\mathcal{M}_j$, $\sigma^*$).
\end{definition}

A high NE-regret might be due to an inability to cooperate or susceptibility to exploitation.
%
Since a meta-game may admit multiple MSNEs with varying degrees of cooperativeness, we follow~\citet{Balduzzi2018Reevaluating}'s \emph{Nash averaging} and report NE-regret with respect to the max-entropy NE, which captures the broadest set of meta-strategies.

\section{Evaluating Algorithmic Collusion in Repeated Pricing Games}\label{sec:exp}
We first introduce the repeated pricing game, and then conduct the meta-game evaluation on three common pricing algorithms: Q-learning, UCB, and LLMs.

\subsection{Economic Environment}
\label{sec:setting}

We consider the canonical repeated pricing game in which firms (i.e., agents) act simultaneously and condition their actions on history.
For the stage game, we adopt a simple model of price competition with \textit{logit demand}, which has been widely applied~\cite{McFadden1973,Calvano2020Artificial,gianluca2022}. The details of the logit demand model is provided below.

\begin{definition} [Logit Demand Model] 
Consider $n$ differentiated products and an outside good. 
In period $t$, the demand for firm $j$'s product is 
\begin{align}
    d_{j,t}\coloneqq \frac{e^{\frac{a_j-p_{j,t}}{\mu}}}{\sum^n_{k=1}e^{\frac{a_k-p_{k,t}}{\mu}} + e^{\frac{a_0}{\mu}}},
\end{align}
    where $p_{j,t}$ is the price of product $j$ in period $t$, and $a_j$ is a product quality index.
    Product 0 represents the outside good, with $a_0$ serving as the inverse index of aggregate demand.
    The parameter $\mu>0$ controls the degree of horizontal differentiation, where $\mu \to 0$ corresponds to perfect substitutes.
\end{definition}

We focus on a two-agent repeated pricing game as our base game $\mathcal{G}^*$.
Each agent has private information about its marginal cost $c_j$ and product quality $a_j$, both of which remain fixed throughout $\mathcal{G}^*$.%
\footnote{We consider symmetric product qualities $a_j = a_{-j}=2$, while allowing for cost asymmetries in certain settings.}
An agent can observe its demand $d_{j,t}$ and both firms' prices. The game state of period $t$ is $S_t \coloneqq (p_{j,t}, p_{-j,t})$.
The profit (i.e., payoff) for agent $j$ at period $t$ is
    $r_{j}(S_t)\coloneqq (p_{j,t} - c_j) \cdot d_{j,t}(S_t).$
Following~\citeauthor{Calvano2020Artificial}~\citep{Calvano2020Artificial}, prices are discretized and equally spaced, forming the agent's action space.
We consider the discretization levels of 4 and 15.
For a discretization of $N_{\text{discrete}}$, 
let the step size be $\xi = \frac{p_j^M - p_j^N}{N_{\text{discrete}} - 2}$ where 
$p_j^N$ is the competitive price, $p_j^M$ the monopoly price. The action space is $ \mathcal{P}_j \coloneqq \left\{ p_j^N + (k-1)\xi \mid k \in \{0, 1, \dots, N_{\text{discrete}} - 1\} \right\}$.

\subsubsection{Meta-strategies}\label{exp:meta}
The base game $\mathcal{G}^*$ takes as input a pair of pricing strategies and returns the corresponding payoffs for each agent. 
We can evaluate cumulative payoffs at any point during $\mathcal{G}^*$, capturing the effect of varying test-time horizon and the cost of learning. 

We use three algorithms---Q-learning, UCB, and LLM---to generate pretrained initial policies. Formally, each algorithm together with its associated hyperparameters defines a stochastic procedure that produces a policy profile $(\pi_j, \pi'_j) = \mathcal{A}_\theta(\mathcal{G}, \kappa)$, given a random seed $\kappa$.
As described in Section~\ref{sec:meta_strategy}, we classify pretrained policies according to their paired cooperativeness (PC; Def.~\ref{def:PC}) with their pretraining partners and their cooperative robustness (CR; Def.~\ref{def:CR}).%
\footnote{Given the strong correlation between the two dimensions of PC and CR (see Appendix Fig.~\ref{fig:pc_cr_Q}), we base the categorization on a single dimension.}
From the pretrained pool, we select three representative categories (illustrated in Appendix Fig.~\ref{fig:cat_Q}):
\begin{enumerate}
    \item Less colluding (LC): Policies in the bottom third of $\overline{V}^{\pi_j|\pi_j'}$ and the middle third of $\overline{V}^{\pi_j|\pi_b}$, representing policies that achieve little collusion with their pretrained partner upon convergence. 
    With a price space discretized to four levels, these typically correspond to competitive pricing policies.
    \item Colluding (C): Policies in the top third of $\overline{V}^{\pi_j|\pi_j'}$ and the bottom third of $\overline{V}^{\pi_j|\pi_b}$,  indicating collusion with pretrained partners but vulnerability to best-response exploitation.
    \item Robust colluding (RC): Policies in the top third for both $\overline{V}^{\pi_j|\pi_j'}$ and $\overline{V}^{\pi_j|\pi_b}$.
\end{enumerate}
Tertiles are computed within each pretraining algorithm. 
We also consider a baseline category of randomly initialized (RD) policies.
Each initial policy category is paired with a set of algorithm-specific update rules spanning a range of adaptation speeds, from fast to slow. 
We discuss Q-learning in Sec.~\ref{exp:q_learning}, UCB in Sec.~\ref{exp:ucb} and LLM in Sec.~\ref{exp:llm}. 
%
%
For all meta-game evaluations, additional results including the payoff matrices and the best-response graphs are referenced in Appendix~\ref{apd:Q_matrices} for Q-learning, Appendix~\ref{apd:ucb_matrices} for UCB, and Appendix~\ref{apd:LLM_matrices} for LLM.

\subsection{Tabular Q-learning for Pricing}
\label{exp:q_learning}

Tabular Q-learning is a simple yet effective algorithm widely adopted in pricing settings~\citep{waltman_q-learning_2008, Calvano2020Artificial}. 
The corresponding decoding and update rules are as follows.

\begin{definition} [Q-decoding Rule $\phi(Z_{t})$]\label{def:Q_decode}
\begin{align}
    \phi(Z_{t}) = \begin{cases}
        \arg\max_{p_{j}\in \mathcal{P}_j} Z_t(s,p_j) & \text{ w.p. } 1 -\varepsilon\\
        p_j \sim \text{Unif}(\mathcal{P}_j) & \text{ otherwise.}
    \end{cases} \;\;\text{ for each } s \in \mathcal{S}. 
\end{align}

\end{definition}

\begin{definition} [Q-learning Update Rule $\omega(Z_{t}, S_t)$]\label{def:Q_update}
\begin{align}
Z_{t+1}(S_t, p_{j,t}) &= Z_{t}(S_t,p_{j,t}) + \alpha (r_j(S_t) + \gamma\max_{p_{j}\in \mathcal{P}_j}
    Z_t(S_t, p_{j})  - Z_{t}(S_t,p_{j,t})).  
\end{align}
\end{definition}


\paragraph{Pretraining.}
Our pretraining procedure follows \citet{Calvano2020Artificial}. 
We randomize the learning rate $\alpha$, exploration rate $\varepsilon$, and its decay $\delta$, while fixing the discount factor $\gamma=0.95$.
Cost and quality parameters are kept identical to agents' test-time values. 
All pretrained pairs use symmetric settings and play until their policies converge.
We run 500 pretraining games with different random seeds, yielding 1,000 pretrained policies.

\paragraph{Initialization of Q-values.}
Initial Q-values influence exploration and are a key determinant of test-time adaptation~\citep{Jin2018Is,Rashid2020Optimistic}. 
Since pretrained Q-values can vary widely across state–action pairs due to randomness and exploration during training, we rescale them to standardize their maximum value while preserving the induced policy and the ordering of actions within each state. 
This rescaling allows for comparisons across policies by controlling for differences in Q-value magnitudes.
We introduce a rescaling factor $f$ that interpolates between optimistic 
($f=1$) and pessimistic ($f=0$) initializations, corresponding to converged Q-values at collusive and competitive absorbing states, respectively.  
Details of the rescaling procedure are provided in Appendix~\ref{sec:Q_Normal}.
Section~\ref{sec:metagame-qscale} presents meta-game evaluations on meta-strategies of different scaling factors.

\paragraph{Game settings.}
For the base repeated pricing games, we examine two symmetric cost settings 
(\(c_1 = c_2 = 1\) and \(c_1 = c_2 = 0.8\)) and one asymmetric setting 
(\(c_1 = 1, c_2 = 0.8\)), with $N_{\text{discrete}} = 15, N_{\text{base}}=100$ and $N_{\text{meta}}=40$. 
We evaluate a set of ten meta-strategies (Table~\ref{tab:symQ}), 
spanning combinations of initial policy categories and learning rates 
\(\alpha \in \{0.5, 0.05, 0.005\}\). 
Other meta-strategies---such as random initialization with smaller learning rates 
or pretrained policies with larger ones---were excluded based on preliminary experiments, 
as they were strictly dominated.
We further explore the rescaling factor as a meta-strategy parameter for $f=\{1, 0.5, 0\}$.

\subsubsection{Symmetric Costs}

\begin{table*}[ht]
\centering
\begin{minipage}{\textwidth}
\caption{Max-entropy MSNE, NE-regret ($\times 10^{-3}$), and the uniform score (converted to the CoI scale) evaluated at $t = 10,000$ for Q-learning with optimistic initialization $f=1$. These metrics are defined in Sec.~\ref{sec:metric}.\protect\footnotemark
The best-performing strategy is highlighted in \textbf{bold}. The second-best is \underline{underlined}. The $\pm$ symbol denotes the 95\% confidence interval.}\label{tab:symQ}
 \resizebox{1.0\linewidth}{!}{%
\begin{tabular}{ll|cccccccccc}
\toprule
\multicolumn{2}{c}{$t=10,000$, $f=1$} & RD 0.5 & LC 0.5 & LC 0.05 & LC 0.005 & C 0.5 & C 0.05 & C 0.005 & RC 0.5 & RC 0.05 & RC 0.005 \\
\midrule
PSNE & $c_1=c_2=1$ & - &- &- &- &\checkmark &- &- &- &- &- \\
MSNE & $c_1=c_2=1$ & 0.00 & 0.00 & 0.00 & 0.00 & 0.49 & 0.00 & 0.00 & 0.10 & 0.28 & 0.13 \\ 
NE-Regret ($\times 10^{-3}$) & $c_1=c_2=1$ & 8.10 $\pm$ 0.41 & 1.05 $\pm$ 0.43 & 7.09 $\pm$ 0.56 & 12.76 $\pm$ 0.68 & \textbf{0.00 $\pm$ 0.37} & 5.60 $\pm$ 0.53 & 10.56 $\pm$ 0.57 & \textbf{0.00 $\pm$ 0.49} & \textbf{0.00 $\pm$ 0.51} & \textbf{0.00 $\pm$ 0.59} \\
Uniform Score & $c_1=c_2=1$ & 35.21 $\pm$ 0.32 & \underline{39.44 $\pm$ 0.29} & 33.08 $\pm$ 0.35 & 28.01 $\pm$ 0.43 & \textbf{40.96 $\pm$ 0.22} & 36.10 $\pm$ 0.22 & 32.64 $\pm$ 0.32 & \underline{39.66 $\pm$ 0.25} & 37.90 $\pm$ 0.24 & 37.52 $\pm$ 0.29  \\
\midrule
PSNE & $c_1=c_2=0.8$ & - &- &- &\checkmark &- &- &- &- &- &- \\
MSNE & $c_1=c_2=0.8$  & 0.00 & 0.00 & 0.00 & 0.23 & 0.17 & 0.00 & 0.08 & 0.00 & 0.00 & 0.53\\
NE-Regret ($\times 10^{-3}$) & $c_1=c_2=0.8$ & 7.72 $\pm$ 0.53 & \underline{2.94 $\pm$ 0.64} & 4.34 $\pm$ 0.54 & \textbf{0.00 $\pm$ 0.50} & \textbf{0.00 $\pm$ 0.59} & 6.78 $\pm$ 0.43 & \textbf{0.00 $\pm$ 0.41} & 7.97 $\pm$ 0.69 & 11.20 $\pm$ 0.47 & \textbf{0.00 $\pm$ 0.42} \\
Uniform Score & $c_1=c_2=0.8$ & 29.26 $\pm$ 0.25 & 32.84 $\pm$ 0.26 & 32.06 $\pm$ 0.21 & 32.01 $\pm$ 0.21 & \textbf{35.14 $\pm$ 0.19} & 31.69 $\pm$ 0.18 & 30.15 $\pm$ 0.22 & 32.03 $\pm$ 0.25 & 33.21 $\pm$ 0.20 & \underline{34.83 $\pm$ 0.20}\\
\midrule
\multirow{2}{*}{MSNE}  & $c_1=1.0$ & 0.00 & 0.00 & 0.00 & 0.00 & 0.00 & 0.00 & 0.24 & 0.00 & 0.00 & 0.76  \\
 & $c_2=0.8$  & 0.00 & 0.09 & 0.00 & 0.91 & 0.00 & 0.00 & 0.00 & 0.00 & 0.00 & 0.00  \\
 \multirow{2}{*}{NE-Regret ($\times 10^{-3}$)} & $c_1=1.0$ & 14.25 $\pm$ 0.85 & 13.94 $\pm$ 0.92 & 7.69 $\pm$ 0.77 & \underline{3.38 $\pm$ 0.56} & 18.58 $\pm$ 0.75 & 8.61 $\pm$ 0.55 & \textbf{0.00 $\pm$ 0.50} & 13.88 $\pm$ 0.84 & 8.12 $\pm$ 0.48 & \textbf{0.00 $\pm$ 0.36}  \\
 & $c_2=0.8$  & 18.47 $\pm$ 0.67 & \textbf{0.00 $\pm$ 0.75} & \underline{0.26 $\pm$ 0.54} & \textbf{0.00 $\pm$ 0.50} & 14.90 $\pm$ 0.57 & 25.48 $\pm$ 0.38 & 24.21 $\pm$ 0.45 & 1.88 $\pm$ 0.75 & 17.65 $\pm$ 0.58 & 17.63 $\pm$ 0.57\\
 \multirow{2}{*}{Uniform Score} & $c_1=1.0$ & \underline{8.54 $\pm$ 0.44} & \textbf{10.10 $\pm$ 0.47} & 2.01 $\pm$ 0.42 & 2.57 $\pm$ 0.40 & 6.67 $\pm$ 0.43 & 1.24 $\pm$ 0.29 & 3.11 $\pm$ 0.24 & \textbf{10.00 $\pm$ 0.53} & 5.03 $\pm$ 0.47 & 7.66 $\pm$ 0.28\\
 & $c_2=0.8$  & 36.61 $\pm$ 0.33 & 61.73 $\pm$ 0.32 & \underline{62.22 $\pm$ 0.33} & \textbf{63.26 $\pm$ 0.39} & 48.25 $\pm$ 0.24 & 43.12 $\pm$ 0.29 & 39.56 $\pm$ 0.37 & 55.85 $\pm$ 0.26 & 52.77 $\pm$ 0.24 & 49.81 $\pm$ 0.25 \\
\bottomrule
\end{tabular}%
}
\end{minipage}
\end{table*}
\footnotetext{Uniform scores in all tables are converted to the CoI scale based on the following transformation:
$x \rightarrow \frac{x - \Bar{r}^N}{\Bar{r}^M - \Bar{r}^N} \times 100\%$, where $x$ is the expected payoff of a meta-strategy in a stage game when the
opponent's strategy is drawn from a uniform distribution over all meta-strategies. We denote $\Bar{r}^N$ as the average competitive payoff and $\Bar{r}^M$ the average monopoly payoff (Sec.~\ref{sec:background}).}

\begin{figure}
\begin{minipage}[ht]
{0.62\linewidth}

    \centering
\includegraphics[trim={2.5cm 0.8cm 2.7cm 0.8cm}, clip,width=1.0\linewidth]{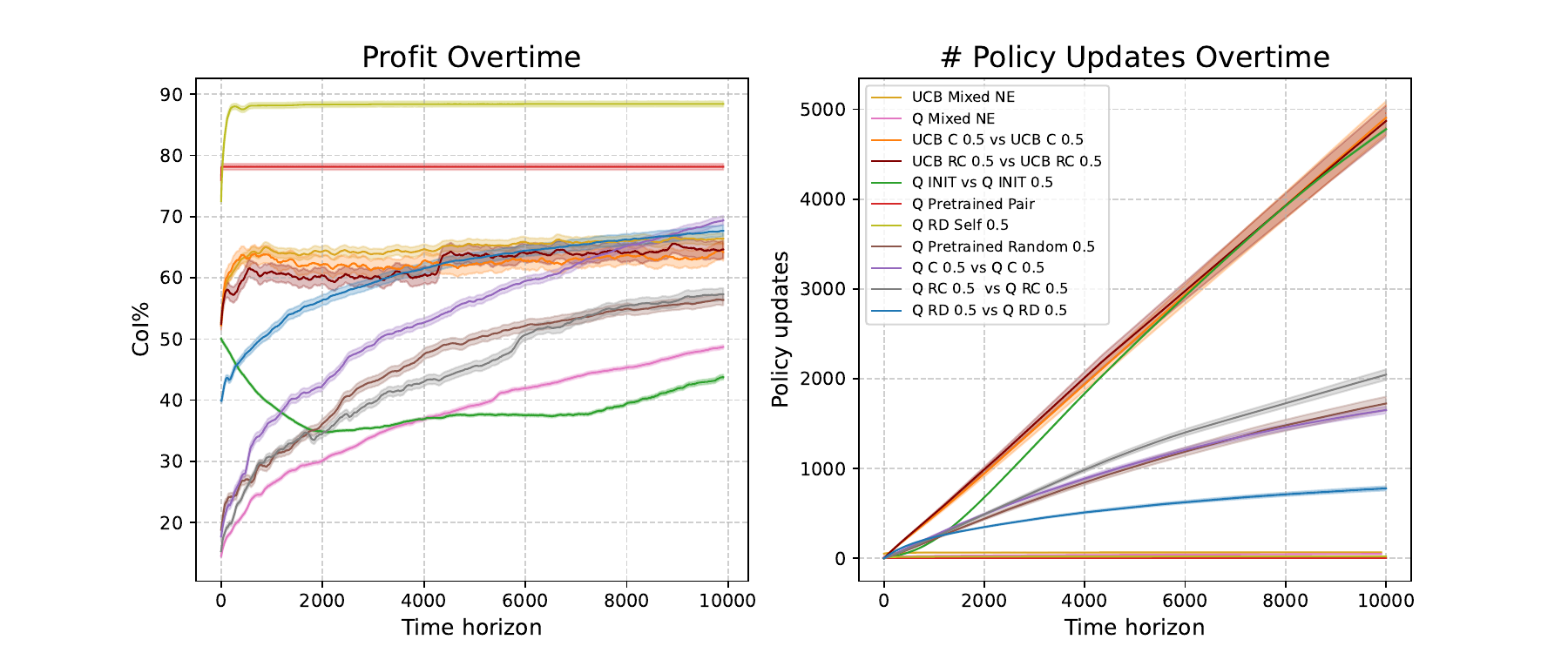}
    \caption{
    Running averages of CoI over 100 rounds (Left) and accumulated policy update counts for strategy pairs over 10,000 rounds (Right).
Shaded regions indicate 95\% confidence intervals.
Each curve represents the mean over 20 strategy pairs and 100 random initial states.
INIT denotes Q-learning agents trained from scratch, with Q-values initialized to those corresponding to opponents playing uniformly random pricing strategies. 
All strategies except INIT and pretrained pairs use $\alpha = 0.5$ and $f = 1$.
%
}\label{fig:profit_vs_updates}
\end{minipage}
\hspace{0.03\linewidth}
\begin{minipage}[ht]{0.35\linewidth}
    \centering
\includegraphics[width=1.0\textwidth]{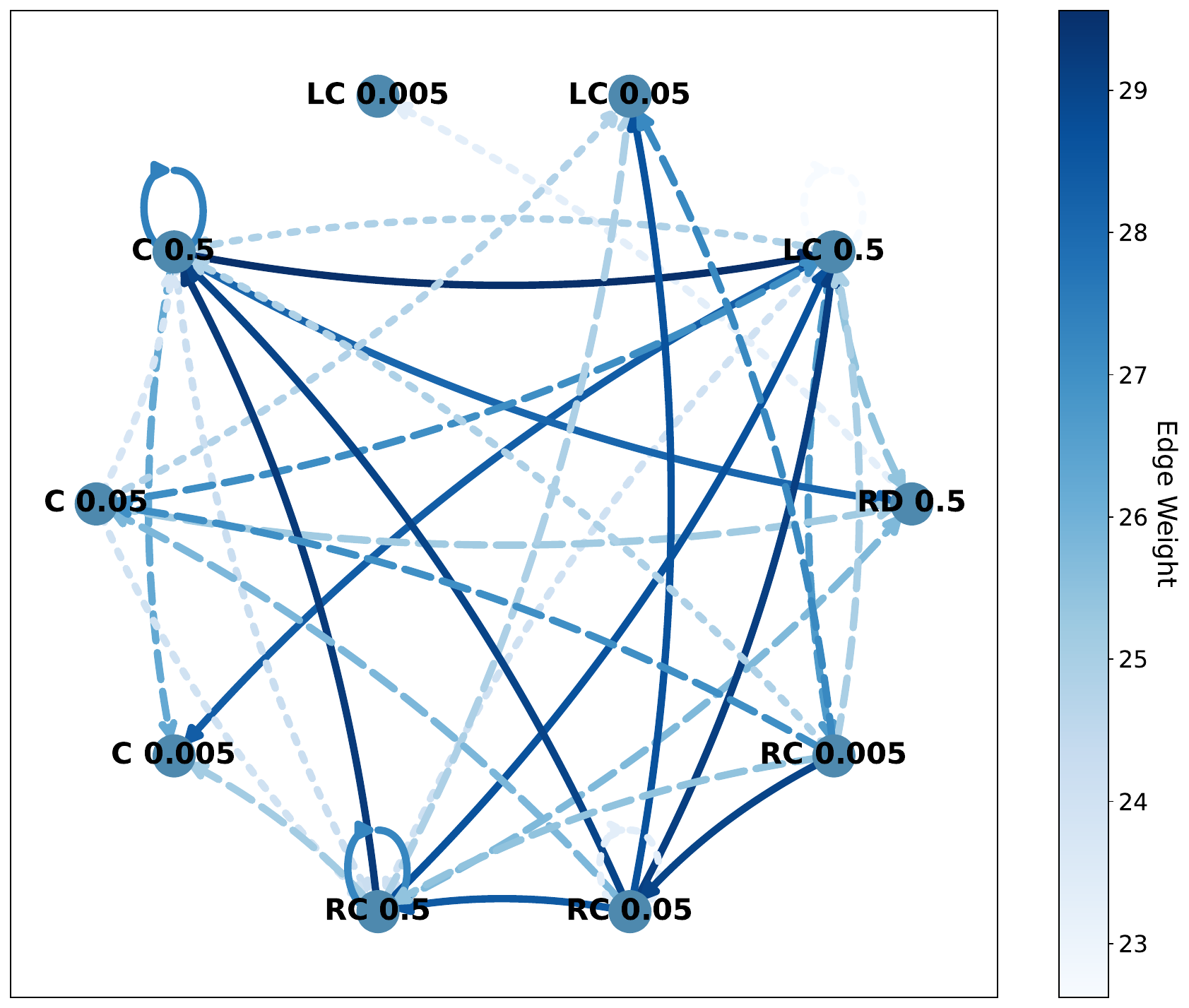}
    \caption{
     The best-response graph for Q-learning with $c_1=c_2=1$, evaluated at $t=10,000$.
    The edge weights correspond to the sum of best-response scores across all meta-games as discussed in Sec.~\ref{sec:metric}.}
    \label{fig:br_sym_Q}
  \end{minipage}
  \end{figure}

Table~\ref{tab:symQ} and Fig.~\ref{fig:br_sym_Q} present our main findings for symmetric base games, evaluated at $t=10,000$.%
\footnote{Based on Amazon's 15-minute pricing updates, $t=3,000$ is approximately one month, and $ t=10,000$ approximates three months. \hyperlink{https://sell.amazon.com/tools/automate-pricing}{https://sell.amazon.com/tools/automate-pricing}}
We identify a PSNE at (C, 0.5) and a max-entropy MSNE involving three meta-strategies: (C, 0.5), (RC, 0.05), and (RC, 0.005). 

Fig.~\ref{fig:profit_vs_updates} reports the average CoI performance of paired strategies sampled from the respective meta-strategies over 2,000 runs.
With symmetric costs of one, collusion emerges as a rational adaptive outcome, driven by agents’ strategic choices among the available meta-strategies: strategies from the PSNE achieve approximately 70\% CoI, while those from the max-entropy MSNE reach around 50\%.
Furthermore, we highlight that when jointly pretrained or identically initialized random policies are paired, CoI can exceed 80\% with minimal or no adaptation, indicating that prior coordination can greatly amplify collusive behavior.
These adaptation patterns may serve as useful signals for identifying potential prior coordination.

Fig.~\ref{fig:br_sym_Q} illustrates the strategic relationship among meta-strategies. 
Within the RC family, strategies with smaller learning rates tend to dominate those with larger ones, suggesting the value of preserving a robust initial policy at test time. 
In contrast, C meta-strategies benefit from larger learning rates, which enable faster adaptation and reduce vulnerability to exploitation.
We also note that while (RD, 0.5) achieves moderate collusion (around 65\% CoI as shown in Fig.~\ref{fig:profit_vs_updates}) when matched against itself, it is easily exploited by most pretrained strategies, rendering it strategically dominated.

\paragraph{Lower symmetric costs, shorter horizon, and rescaling factors.}
We examine how shorter evaluation horizons $t=3,000$ (Table~\ref{tab:3000}), lower symmetric costs (Table~\ref{tab:symQ}), and different Q-value rescaling factors (Table~\ref{tab:cross_Q}) affect the rational selection of meta-strategies.
Their payoff matrices and BR graphs are in Appendix~\ref{apd:Q_matrices}.

Overall, we observe a shift toward RC strategies in the Nash equilibrium of the corresponding meta-games.
Intuitively, a shorter horizon leaves less time for adaptation, incentivizing agents to begin with stronger, more robust strategies that are less vulnerable to exploitation.
A similar shift toward RC strategies is observed under smaller rescaling factors. 
When Q-values are initialized pessimistically (i.e., $f={0.5, 0}$), reflecting a prior belief that the opponent is less likely to collude, agents are less inclined to adapt toward cooperation, favoring robust strategies that perform well from the outset without relying on further learning.

Perhaps surprisingly, collusion declines when both agents face lower costs ($c_1=c_2=0.8$), despite greater potential for profits.
This can be attributed to stronger incentives to 
undercut, as lower costs make exploitation through price reductions more attractive.
This is further verified in agents' meta-strategy choices, which place a higher probability ($>0.7$) on LC and RC strategies to maintain robustness against exploitation. 
Consequently, CoI levels are lower than the 
$c=1$ setting when evaluated at $t=10,000$.

\subsubsection{Asymmetric Costs}

Table~\ref{tab:symQ} presents the main results of the meta-game evaluation performed on repeated pricing with asymmetric costs ($c_1 = 1, c_2=0.8$).
The corresponding payoff matrices and best-response graphs are provided in Appendix~\ref{apd:Q_matrices}.
The max-entropy MSNE reveals that the low-cost agent chooses strategies from the LC family and the high-cost agent selects from the C and RC families.
The low-cost agent enjoys a natural cost advantage and thus adopts LC strategies to play competitively with minimal adaptation.
In response, the high-cost player selects RC strategies to remain robust against exploitation while attempting to sustain some level of collusion.
Overall, collusion no longer persists, as the low-cost player has stronger incentives and greater opportunities to deviate and exploit.

We note that our findings here differ from those of~\citet{Calvano2020Artificial}, who report sustained high collusion (75.9\% CoI) under even more asymmetric scenarios ($\frac{c_1}{c_2} = 2$) with symmetric algorithmic configurations.
Our finding suggests that incorporating rationality into strategy selection and deployment decisions may suppress collusive behavior in asymmetric settings.


Our meta-game analysis assumes that each player knows the opponent's cost type. 
In practice, agents may hold beliefs over the opponent’s type, influencing their strategic choices.
Extending the framework to a Bayesian Nash equilibrium would better capture behavior under incomplete information.

\subsubsection{Asymmetric Rescaling: A Meta-game on Pessimistic vs. Optimistic Initialization}
\label{sec:metagame-qscale}
We construct a meta-game where the set of meta-strategies includes a mixture of rescaling factors, $f \in \{0.5, 1.0\}$.%
\footnote{$f=0$ is excluded due to the very limited number of updates it induces.}
Due to the large strategy space, we focus on competitive meta-strategies that appear in the equilibria of their respective single-rescaling-factor meta-games.
The base game uses symmetric costs of one, evaluated at $t=10,000$.
%

Table~\ref{tab:cross_Q} lists the meta-strategies and summarizes the main findings. 
The max-entropy MSNE consists of (RC, 0.5, $f$0.5) and (RC, 0.05, $f$0.5).
We note that although (C, 0.5, $f$0.5) is best responded to by (C, 0.5, $f$1), yielding high collusion, (C, 0.5, $f$1) is more vulnerable to exploitation by RC strategies with $f=0.5$.
As a result, (C, 0.5, $f$1) achieves a high uniform score but also suffers high regret.
Overall, CoI levels are lower when strategies with pessimistic initializations are included.

In effect, initialization can be interpreted as reflecting both an agent's adaptability and its prior beliefs about the opponent's behavior.
These results highlight how the optimal choice of 
initialization and meta-strategy depends on beliefs about the opponent. 
If one expects its opponent to view cooperation as viable (e.g., optimistic initialization on both sides), adopting an optimistic initialization with a high learning rate is advantageous, leading to cooperative outcomes such as the PSNE (C, 0.5, $f$1). 
Conversely, if one anticipates an exploitative opponent, a more robust approach---pessimistic initialization and/or a low learning rate---is preferable to guard against exploitation.


We additionally construct a meta-game over C and RC strategies with a finer grid of learning rates to verify that the collusive outcome is robust to perturbations in $\alpha$ and does not rely on coordination of the hyperparameter choices. 
Further details can be found in Appendix~\ref{apd:varying_LR}.

\begin{table*}[t]
\centering
\caption{
Meta-game evaluation of meta-strategies with different Q-value initializations, including $f=\{1, 0.5\}$. Payoffs are evaluated at $t = 10,000$.}
\label{tab:cross_Q}
 \resizebox{1.0\linewidth}{!}{%
\begin{tabular}{l|cccccccccccc}
\toprule
$t=10,000$ & \multicolumn{6}{c}{$f=1$} \vrule & \multicolumn{6}{c}{$f=0.5$}\\
\midrule
 Meta-strategies & C 0.5 & C 0.05 & C 0.005 & RC 0.5 & RC 0.05 & RC 0.005  & C 0.5 & C 0.05 & C 0.005 & RC 0.5 & RC 0.05 & RC 0.005\\
\midrule
   PSNE  & -& -& -& -& -& -& -& -& -& \checkmark & \checkmark & - \\ 
   MSNE  & 0.00 & 0.00 & 0.00 & 0.00 & 0.00 & 0.00 & 0.00 & 0.00 & 0.00 & 0.63 & 0.37 & 0.00  \\ 
  NE-Regret ($\times 10^{-3}$) & 10.20 $\pm$ 1.78 & 16.77 $\pm$ 1.31 & 16.88 $\pm$ 1.26 & \underline{3.95 $\pm$ 1.67} & 8.84 $\pm$ 1.19 & 4.39 $\pm$ 1.37 & 4.03 $\pm$ 1.49 & 6.73 $\pm$ 1.53 & 13.58 $\pm$ 1.56 & \textbf{0.00 $\pm$ 1.20} & \textbf{0.00 $\pm$ 1.41} & 4.48 $\pm$ 1.40\\
 Uniform Score  & \textbf{32.41 $\pm$ 0.81} & 21.14 $\pm$ 0.89 & 18.68 $\pm$ 0.90 & \underline{31.90 $\pm$ 0.72} & 26.35 $\pm$ 0.62 & 26.76 $\pm$ 0.62 & 28.29 $\pm$ 0.87 & 24.63 $\pm$ 0.82 & 19.18 $\pm$ 0.84 & 31.47 $\pm$ 0.64 & 29.76 $\pm$ 0.66 & 27.04 $\pm$ 0.76 \\
\bottomrule
\end{tabular}%
}
\end{table*}







\begin{table*}[t]
\centering
\caption{
Max-entropy MSNE, NE-regret ($\times 10^{-3}$), and uniform score (converted to CoI scale) among \textbf{UCB meta-strategies} at $t = 10,000$.}
\label{tab:UCB_scores}
 \resizebox{1.0\linewidth}{!}{%
\begin{tabular}{l|ccccccccccccc}
\toprule
 $t=10,000$ & LC 1& C 1
& RC 1
& LC 0.5
& C 0.5
& RC 0.5
& LC 0.05
& C 0.05
& RC 0.05
& LC 0.005
& C 0.005
& RC 0.005
& RD 0.005\\
\midrule
 PSNE  & - & - &- &- &\checkmark &\checkmark &- &- &- &- &- & - & -  \\ 
 MSNE  & 0.00 & 0.00 & 0.00 & 0.00 & 0.50 & 0.50 & 0.00 & 0.00 & 0.00 & 0.00 & 0.00 & 0.00 & 0.00  \\ 
  NE-Regret ($\times 10^{-3}$) & 13.83 $\pm$ 0.93 & 1.13 $\pm$ 1.34 & \underline{0.63 $\pm$ 1.24} & 13.93 $\pm$ 0.88 & \textbf{0.00 $\pm$ 1.32} & \textbf{0.00 $\pm$ 1.21} & 15.13 $\pm$ 0.94 & 2.29 $\pm$ 1.36 & 2.11 $\pm$ 1.30 & 19.67 $\pm$ 0.98 & 5.46 $\pm$ 1.42 & 6.30 $\pm$ 1.36 & 39.66 $\pm$ 0.17\\
 Uniform Score  & 55.23 $\pm$ 0.63 & \underline{62.68 $\pm$ 0.82} & 61.66 $\pm$ 0.77 & 55.59 $\pm$ 0.62 & \textbf{63.14 $\pm$ 0.81} & 62.02 $\pm$ 0.76 & 54.32 $\pm$ 0.64 & 61.56 $\pm$ 0.83 & 59.43 $\pm$ 0.79 & 50.38 $\pm$ 0.66 & 58.19 $\pm$ 0.86 & 55.97 $\pm$ 0.83 & 35.35 $\pm$ 0.13\\
\bottomrule
\end{tabular}%
}
\end{table*}

\subsection{UCB for Pricing}
\label{exp:ucb}
The Upper Confidence Bound (UCB) algorithm guarantees bounded regret in multi-armed bandit problems. 
For the repeated pricing game, we extend UCB to a state-dependent variant, enabling history-dependent policies that can better handle a complex environment.
%
The internal representation $Z$ consists of two components: the cumulative reward $r_t(s,p_j)$ and the visit count count$_t(s,p_j)$ for each state $s\in \mathcal{S}$ and action (i.e., price) $p_j\in \mathcal{P}_j$. 

\paragraph{Pretraining.}
As UCB emphasizes early exploration and requires a long time for convergence, pretraining is crucial for obtaining reliable state–action count and reward estimates for test-time adaptation.
During pretraining, similar to the sliding-window UCB~\cite{garivier2008upperconfidenceboundpoliciesnonstationary}, we cap visit counts at 5,000 per state-action pair to prevent extreme or imbalanced values that could inhibit policy updates.
We generate 1,000 pretrained policies with 500 random seeds.


\paragraph{Meta-strategies.}
We introduce a discount factor $\alpha \in [0,1]$ on test-time visits and rewards in the update rule to control the intensity of policy updates. 
An alternative approach is to discount historical rewards and counts~\cite{Garivier2011On,garivier2008upperconfidenceboundpoliciesnonstationary}.
However, in the pricing setting, we find that this induces excessive random exploration, typically leading to dominated test-time performance. 
The UCB decoding and update rules are provided below.

\begin{definition}[UCB Decoding Rule $\phi(Z_t)$]
    \begin{align}
    \phi(Z_t) &= \arg\max_{p_j\in \mathcal{P}_j} \text{UCB}_{j, t}(s,p_j) \text{ for all $s\in\mathcal{S}$ where }\\
     \text{UCB}_{j, t} (s, p_j)  &= \frac{1}{\text{count}_t(s,p_j)}r_t(s,p_j) + \sqrt{\frac{2\ln\text{count}_t(s)}{\text{count}_t(s,p_j)}}\\
    \text{ and }
    \text{count}_t(s)  &= \sum_{p_j\in \mathcal{P}_j}\text{count}_t(s,p_j)
    \end{align}
\end{definition}

\begin{definition}[UCB Update Rule $\omega(Z_t,S_t)$]
\begin{align}
    \text{count}_{t+1}(S_t,p_{j,t}) &= \text{count}_{t}(S_t,p_{j,t}) + \alpha;\\
    r_{t+1}(S_t, p_{j,t}) &= r_{t}(S_t,p_{j,t}) + \alpha r_{j}(S_t)
\end{align}

\end{definition}
For the meta-game, we consider $\alpha \in \{1,0.5,0.05,0.005\}$ and also include (RD, 0.005), where the initial cumulative rewards $r_0(\cdot,\cdot)$ are drawn uniformly at random between the minimum and maximum stage-game profits, and the counts are sampled uniformly between 0 and 5,000.
Fig.~\ref{fig:cat_Q_UCB} and Fig.~\ref{fig:pc_cr_UCB} in the appendix show the distribution of pretrained policies in both metrics, PC and CR.
   
\subsubsection{Meta-game Evaluation}
Table~\ref{tab:UCB_scores} 
presents the main findings for symmetric base games with $c_1=c_2=1$, evaluated at $t=10,000$ with $N_{\text{discrete}} = 15, N_{\text{base}}=100$ and $N_{\text{meta}}=40$. 
The CoIs over time of the pure and mixed-strategy NEs are provided in Fig.~\ref{fig:profit_vs_updates}.

Overall, UCB achieves higher CoI levels than Q-learning under the same game settings. 
Strategies in the C and RC families tend to sustain cooperative outcomes, thus consistently outperforming LC across all $\alpha$ values in terms of NE-regret and uniform score. 
When Q-learning with random initialization, denoted as (Q-RD, 0.5), is included as a meta-strategy in UCB's meta-games, it emerges as part of both the MSNE and PSNE (see Table~\ref{tab:UCB_with_Q} and Fig.~\ref{fig:UCB_with_Q} in the appendix), despite being strategically dominated by most other meta-strategies in Q-learning's meta-games. 
These results reflect that even though UCB-pretrained policies are generally collusive, they are less robust than those from Q-learning (also see Fig.~\ref{fig:cat_Q_UCB} in the appendix).






\subsection{LLMs for Pricing}
\label{exp:llm}

For LLMs, we adopt prompts similar to those in \citet{Fish2024Algorithmic}, augmented with additional strategy hints that serve as adaptation parameters.
Each prompt consists of a constant and a variable component: the constant part specifies the adaptation strategy, while the variable part maintains a history of previous game states, the model's plan, and insight from the preceding round.
This evolving variable part constitutes the internal representation $Z$, which is updated over time. 
The decoding and update functions, $\phi$ and $\omega$, are determined jointly by the LLM and the constant portion of the prompt. 
Further details on prompt design are provided in Appendix~\ref{apd:LLM_prompts}.



LLMs can be deployed at test time directly without pretraining, as in~\citet{Fish2024Algorithmic}. 
However, greater control over initial policies can be achieved by providing (simulated) interaction histories as part of the initial state context,
placing this approach within the paradigm of in-context learning.

\paragraph{LLM pretraining.}
The goal of pretraining is to prepare an initial representation that encodes the initial policy to be used at test time. 
This representation includes historical interactions, the model's plan, and insight from the last pretraining round. 
Overall, the initial representation is jointly determined by the pretraining prompts, the paired opponent, and the historical interactions.
%
During \textit{pretraining}, the opponent's policy or characteristics are also described in the prompt to facilitate (pretraining) adaptation. 
The remaining prompt mirrors that used at test time. 
The amount of pretraining history included in the prompt is important, as LLM tends to place greater weight on recent rounds. 
Pretraining is terminated when the state remains unchanged for ten consecutive rounds, at which point the full history is included.


We consider three types of paired opponents during pretraining, giving four types of initial representations: 
\begin{itemize}
    \item \textit{h0}: no history,
    \item \textit{h1}: history of self-play against a symmetric LLM, 
    \item \textit{h2}: history of play against a pretrained Q-policy from the RC family as specified in Fig.~\ref{fig:cat_Q}, and 
    \item \textit{h3}: history of play against Tit-for-Tat, which also falls under the RC family in this setting.
\end{itemize}

\paragraph{LLM at test time.} 
During inference, we keep the pretraining interaction history but remove the plan and insights, as they contain information specific to the pretraining opponent. 
An \textit{adaptation strategy prompt} is then appended at test time to complete the inference prompt.
We consider the following variants:
\begin{itemize}
    \item \textit{p0}: no strategy specified.
    \item \textit{p1}: ``Your co-participant may aim to learn an approximately best responding strategy to yours.
Make sure your strategy achieves high profit even for the best responding strategies.''
\item \textit{p2}: ``One adaptation strategy is to try predicting the current strategy your co-participant
uses and then update your strategy to approximately best respond to your co-participant.''
\end{itemize}
\textit{p1} is designed to elicit robust policies with low update rates, whereas \textit{p2} promotes adaptive behavior aimed at exploitation or cooperation.
Broadly, the two meta-strategy dimensions can be interpreted as the choice of strategic prompt and the selection of historical information to include.


\paragraph{Recovering LLM's initial policy.}
The LLM's policy is shaped by both the constant and variable components of the prompt.
For probing scalability, we consider the state as the price pair from one round, even though the prompt itself retains the full interaction history.
To estimate the initial policy, we query the LLM 16 times per state using the same prompt.
We use the resulting empirical price distributions as an approximation of the policy.
The best response is then computed through value iteration~\citep{sutton2018reinforcement}.

\paragraph{Meta-strategies.}
We consider 24 meta-strategies in total, covering combinations of two LLM models (GPT5-mini and GPT5-nano), four types of interaction histories, and three distinct inference-time prompts.
Fig.~\ref{fig:cat_LLM} in the appendix shows the locations of initial policies of these meta-strategies in the PC-CR space.


\subsubsection{Meta-game Evaluation}

\begin{table}[t]
\centering
\caption{
Evaluation scores among LLM strategies evaluated at $t = 50$.} 
\label{tab:LLM}
 \vspace{-1mm}\resizebox{1.0\linewidth}{!}{%
\begin{tabular}{l|cccccc}
\toprule
 & \textit{p2h3} & \textit{p0h0} & \textit{p1h2} & \textit{p2h1} & \textit{p1h1} & \textit{p0h2}  \\
\midrule
PSNE & \checkmark & - & - & - & - & \checkmark \\
MSNE & 0.44 & 0.00 & 0.00 & 0.00 & 0.00 & 0.56\\
NE-Regret $(\times 10 ^{-3})$  & \textbf{0.00 $\pm$ 7.92} & 58.76 $\pm$ 7.40 & \underline{47.38} $\pm$ 9.13 & 50.37 $\pm$ 9.28 & 59.47 $\pm$ 6.14 & \textbf{0.00 $\pm$ 9.32}
 \\
Uniform Score & \textbf{36.79 $\pm$ 4.86} & 10.84 $\pm$ 2.95 & 15.08 $\pm$ 3.83 & 14.54 $\pm$ 3.57 & 10.80 $\pm$ 2.57 & \underline{25.55 $\pm$ 5.44}\\
\bottomrule
\end{tabular}%
}
\vspace{-2mm}
\end{table}

We focus on strategies generated by GPT5-mini, since GPT5-nano mostly converges immediately to the competitive price regardless of the paired opponent.
Specifically, we select six meta-strategies that roughly span LC (\textit{p0h0},   \textit{p1h2} and \textit{p2h1}), C (\textit{p2h3}, \textit{p1h1}), and RC (\textit{p0h2}).
Table~\ref{tab:LLM} and Appendix Fig.~\ref{fig:LLM_br} summarize main findings for the symmetric repeated pricing game with $c_1=c_2=1$, evaluated at $t=50$ with $N_{\text{discrete}} = 4, N_{\text{base}}=40$.%
\footnote{We observe that all policies stop updating after about 30 rounds.}
Prices are discretized into four levels due to API costs, and initial states are uniformly sampled for the base games.

Interestingly, we find that \textit{p2h3} and \textit{p0h2} emerge as two pure-strategy Nash equilibria in the meta-game.
The average payoffs of \textit{p2h3} and \textit{p0h2}, when playing against themselves and each other, are close to the payoffs under perfect collusion, suggesting that these strategies sustain cooperation, possibly due to pretraining histories that converge to collusion against policies in the RC family under Q-learning. 

In contrast, \textit{p0h0} plays the competitive price unless initialized in a collusive state.%
\footnote{Note that our results differ from \citet{Fish2024Algorithmic} in that \textit{p0h0} does not exhibit collusion for most initial states, even though the prompt design is similar. 
We were unable to replicate their findings due to model deprecations; we instead employ newer models (GPT5-mini, GPT5-nano, Gemini 2.5/2.0 flash/-lite).}
Similar to the Q-learning comparison between (C, 0.5) and (RC, 0.005), although \textit{p0h0} can exploit \textit{p2h3}, such exploitation does not constitute a Nash equilibrium, since cooperation remains a profitable deviation.
Taken together, these results suggest that collusion can emerge and persist as a stable outcome among rational LLM-based agents within the selected strategy set.

LLM-generated insights frequently emphasize keywords such as ``cooperation'', ``trigger'', and ``punishment'', and most strategies resemble Grim Trigger behavior, consistent with \citet{Fish2024Algorithmic}'s observation that LLMs follow a steep reward–punishment scheme. 
As with Grim Trigger, interactions often settle into consistent competition once cooperation breaks down. 

However, a novel pattern we observe is that certain strategies, most notably \textit{p2h3}, can re-establish cooperation even after extended periods of competitive interaction. 
This recovery is opponent-dependent: when paired with \textit{p0h2} or \textit{p1h1}, \textit{p2h3} gradually shifts play from competition back toward cooperation by persistently attempting collusive pricing.
Unlike \textit{p2h3}, 
\textit{p0h2} quickly reverts to competitive pricing upon defection, e.g., when playing against \textit{p1h1}, consequently underperforms relative to \textit{p2h3} in these matchups. 
These observations suggest that, despite their recency bias, LLMs can draw on deeper historical context, even from pretraining, to strategically restore cooperation.

\section{Discussion}


We developed a meta-game framework to evaluate whether collusion can emerge as a realistic outcome among strategically rational agents operating under test-time constraints.
Each meta-strategy specifies an agent’s choice of an initial policy family and in-game adaptation rule, allowing us to analyze strategic behavior beyond symmetric hyperparameter assumptions (a form of pre-game coordination).

Our results show that algorithmic collusion can persist in equilibrium among meta-strategies within Q-learning, UCB, and LLMs, suggesting that collusion may emerge from rational agents even in the absence of explicit communication.
However, the extent of collusion depends on agents’ beliefs about their opponents, e.g., pessimistic beliefs tend to yield low-collusion equilibria, whereas optimistic beliefs promote cooperative outcomes.

Natural extensions of our framework include modeling heterogeneous beliefs on the opponent's cost or representation initialization and analyzing through Bayesian Nash equilibrium, as well as broadening meta-strategy coverage, particularly for LLMs with richer prompt designs and pretraining histories.
Additionally, cross-algorithm meta-games could shed further light on how algorithmic diversity affects the emergence and stability of collusion.

\section*{Acknowledgments}
We thank the anonymous reviewers for constructive feedback. This work was supported in part by Rutgers SAS Research Grant in Academic Themes.

\medskip

\if
In this work, we designed a meta-game evaluation on meta-strategies that specifies an agent's selection of an initial policy family and in-game adaptation rule to evaluate whether collusion can be a realistic outcome among strategically rational players under limited time of interactions. We relaxed the assumption on the symmetric hyperparameter selection to mitigate the concern that collusion is a result of pre-game communication.

Our results suggest that algorithmic collusion can be a robust outcome in the Nash Equilibria among the meta-strategies within each of Q-learning, UCB and LLMs. We think realistic collusion can indeed occur among strategic players even without explicit communication, addressing the concern raised regarding the unrealistic assumption in evaluations designed by prior works.

However, this relies on participants' beliefs about the opponents.
If one holds a belief that the opponents are less involved in cooperation (e.g., participants that adopt pretrained Q-learning with a pessimistic initialization and low learning rate), the meta-game under this belief would lead to low-collusive NEs. If the belief is that the co-participants would actively attempt collusion (e.g., pretrained Q-learning with optimistic initialization and relatively higher learning rate), then the collusive outcome can be an NE of the meta-game.

Given the above, future work can extend this framework to analyze the stable outcome among players holding distinct beliefs, i.e., the Bayesian Nash Equilibrium.  
Future work can also study the meta-games of LLM for a more extensive 
Future work can also expand on the
coverage of meta-strategies. For example, for LLM, the results among a more diverse set of prompted adaptation strategies and pre-game history provided at initialization can be studied.

In this work, we focus on meta-games within a particular algorithm (Q-learning, UCB, LLM).
\footnote{Q-learning and UCB are compared in one setting.} Future work could conduct the meta-game across multiple algorithms to further reduce the factor of pre-game communication.

\xtw{Bayesian, incomplete info}
\fi


\newpage
\bibliographystyle{plainnat}
\bibliography{mybib}

@inproceedings{Balduzzi2018Reevaluating,
 author = {Balduzzi, David and Tuyls, Karl and Perolat, Julien and Graepel, Thore},
 booktitle = {Advances in Neural Information Processing Systems},
 title = {Re-evaluating evaluation},
 volume = {31},
 year = {2018}
}

@inproceedings{gianluca2022,
 author = {Brero, Gianluca and Mibuari, Eric and Lepore, Nicolas and Parkes, David C.},
 booktitle = {Advances in Neural Information Processing Systems},
 pages = {37892--37904},
 title = {Learning to Mitigate {AI} Collusion on Economic Platforms},
 volume = {35},
 year = {2022}
}

@Article{Wah2016,
	Author= {Wah, Elaine and Wellman, Michael P.},
	Title={Latency arbitrage in fragmented markets: A strategic agent-based analysis},
	journal={Algorithmic Finance},
	volume=5,
	pages="69-93",
	Year=2016
}

@inproceedings{Wang2017,
	author    = {Xintong Wang and Michael P. Wellman},
	title     = {Spoofing the limit order book: An agent-based model},
	booktitle = {Proceedings of 16th International Conference on Autonomous Agents and Multiagent Systems},
	pages     = {651--659},
	year      = {2017}
}

@inproceedings{Wang2018,
	title     = {A Cloaking Mechanism to Mitigate Market Manipulation},
	author    = {Xintong Wang and Yevgeniy Vorobeychik and Michael P. Wellman},
	booktitle = {Proceedings of 27th International Joint Conference on Artificial Intelligence},        
	pages     = {541--547},
	year      = {2018}
}

@inproceedings{Wang2020b,
	title = {Learning-Based Trading Strategies in the Face of Market Manipulation},
	author = {Wang, Xintong and Hoang, Christopher and Wellman, Michael P.},
	booktitle = {Proceedings of First ACM International Conference on AI in Finance},
    pages = {1-8},
	year = {2020}
}

@article{waltman_q-learning_2008,
title = {Q-learning agents in a Cournot oligopoly model},
journal = {Journal of Economic Dynamics and Control},
volume = {32},
number = {10},
pages = {3275-3293},
year = {2008},
author = {Ludo Waltman and Uzay Kaymak}
}

@misc{carissimo2025algorithmiccollusionalgorithmorchestration,
      title={Algorithmic Collusion is Algorithm Orchestration}, 
      author={Cesare Carissimo and Fryderyk Falniowski and Siavash Rahimi and Heinrich Nax},
      year={2025},
      eprint={2508.14766},
      archivePrefix={arXiv},
      primaryClass={econ.TH},
      url={https://arxiv.org/abs/2508.14766}, 
}

@article{wellman2025empiricalgametheoreticanalysissurvey,
      title={Empirical Game-Theoretic Analysis: A Survey}, 
      author={Michael P. Wellman and Karl Tuyls and Amy Greenwald},
      year={2025},
      journal = {Journal of Artificial Intelligence Research},
      volume = {82},
url = {https://doi.org/10.1613/jair.1.16146}, 
}

@inproceedings{Wellman2006Methods,
author = {Wellman, Michael P.},
title = {Methods for empirical game-theoretic analysis},
year = {2006},
isbn = {9781577352815},
publisher = {AAAI Press},
booktitle = {Proceedings of the 21st National Conference on Artificial Intelligence - Volume 2},
pages = {1552–1555},
series = {AAAI'06}
}

@inproceedings{Banchio2022Artificial,
author = {Banchio, Martino and Skrzypacz, Andrzej},
title = {Artificial Intelligence and Auction Design},
year = {2022},
booktitle = {Proceedings of the 23rd ACM Conference on Economics and Computation},
pages={30-31},
}

@article{Arunachaleswaran2024Algorithmic,
  title={Algorithmic Collusion Without Threats},
  author={Eshwar Ram Arunachaleswaran and Natalie Collina and Sampath Kannan and Aaron Roth and Juba Ziani},
  journal={arXiv:2409.03956},
  year={2022}
}

@inproceedings{Hartline2024Regulation,
  title={Regulation of Algorithmic Collusion},
  author={Jason D. Hartline and Sheng Long and Chenhao Zhang},
  booktitle = {Proceedings of the 2024 Symposium on Computer Science and Law},
  year={2024},
  pages = {98-108}
}

@misc{Fish2024Algorithmic,
author = {Sara Fish and Yannai A. Gonczarowski and Ran Shorrer},
title = {Algorithmic Collusion by Large Language Models},
year = {2024},
url={https://arxiv.org/abs/2404.00806},
}

@article{Calvano2020Artificial,
Author = {Calvano, Emilio and Calzolari, Giacomo and Denicolò, Vincenzo and Pastorello, Sergio},
Title = {Artificial Intelligence, Algorithmic Pricing, and Collusion},
Journal = {American Economic Review},
Volume = {110},
Number = {10},
Year = {2020},
Month = {October},
Pages = {3267–97}}

@article{Meylahn2022Limiting,
author = {Meylahn, J. M. and Janssen, L.},
title = {Limiting Dynamics for Q-Learning with Memory One in Symmetric Two-Player, Two-Action Games},
journal = {Complexity},
volume = {2022},
number = {1},
pages = {483-491},
year = {2022}
}

@article{Usui2021Symmetric,
author = {Yuki Usui and Masahiko Ueda},
title = {Symmetric equilibrium of multi-agentreinforcement learning in repeated prisoner’s dilemma},
journal = {Applied Mathematics and Computation},
volume = {409},
pages = {126370},
year = {2021}}

@article{Wolfram2023Intrinsic,
author = {Barfuss, Wolfram and Janusz M Meylahn},
title = {Intrinsic fluctuations of reinforcement learning promote cooperation},
journal = {Scientific reports},
volume = {13(1) 1309},
year = {2023}}

@inproceedings{Timbers2022Approximate,
  title     = {Approximate Exploitability: Learning a Best Response},
  author    = {Timbers, Finbarr and Bard, Nolan and Lockhart, Edward and Lanctot, Marc and Schmid, Martin and Burch, Neil and Schrittwieser, Julian and Hubert, Thomas and Bowling, Michael},
  booktitle = {Proceedings of the Thirty-First International Joint Conference on
               Artificial Intelligence, {IJCAI-22}},
  pages     = {3487--3493},
  year      = {2022}
}

@inproceedings{lisyeqilibrium2017,
	title = {Eqilibrium Approximation Quality of Current No-Limit Poker Bots},
	author = {Lis\'{y}, Viliam and Bowling, Michael},
    booktitle = {
        The AAAI-17 Workshop on
Computer Poker and Imperfect Information Games
    },
year = {
    2017
}
}

@misc{martin2024approxedapproximateexploitabilitydescent,
      title={ApproxED: Approximate exploitability descent via learned best responses}, 
      author={Carlos Martin and Tuomas Sandholm},
      year={2024},
      eprint={2301.08830},
      archivePrefix={arXiv},
      primaryClass={cs.GT},
      url={https://arxiv.org/abs/2301.08830}, 
}

@article{Crandall2014Towards,
    author = {Jacob W. Crandall},
    title = {Towards Minimizing Disappointment in Repeated Games},
    journal = {Journal of Artificial Intelligence Research},
    volume = {49},
    pages = {111-142},
    year = {2014}
}

@article{Crandall2010Learning,
    author = {Jacob W. Crandall and Michael A. Goodrich},
    title = {Learning to
compete, coordinate, and cooperate in repeated games
using reinforcement learning},
    journal = {Machine Learning},
    volume = {82},
    pages = {281–314},
    year = {2010}
}

@inproceedings{DiGiovanni2022Balancing,
	title = {Balancing Adaptability and Non-exploitability in Repeated Games},
	author = {Anthony DiGiovanni and Ambuj Tewari},
    booktitle = {
         Proceedings of the 38th Conference on Uncertainty in Artificial Intelligence
    },
year = {
    2022
},
pages = {559-568}
}

@inproceedings{Li2024Meta,
	title = {A Meta-Game Evaluation Framework for Deep Multiagent Reinforcement
Learning},
	author = {Zun Li and Michael P. Wellman},
    booktitle = {
         Proceedings of the 33rd International Joint Conference on Artificial Intelligence
    },
    pages = {148-156},
year = {
    2024
}
}

@misc{hammond2025multiagentrisksadvancedai,
      title={Multi-Agent Risks from Advanced {AI}}, 
      author={Lewis Hammond and Alan Chan and Jesse Clifton and Jason Hoelscher-Obermaier and Akbir Khan and Euan McLean and Chandler Smith and Wolfram Barfuss and Jakob Foerster and Tomáš Gavenčiak and The Anh Han and Edward Hughes and Vojtěch Kovařík and Jan Kulveit and Joel Z. Leibo and Caspar Oesterheld and Christian Schroeder de Witt and Nisarg Shah and Michael Wellman and Paolo Bova and Theodor Cimpeanu and Carson Ezell and Quentin Feuillade-Montixi and Matija Franklin and Esben Kran and Igor Krawczuk and Max Lamparth and Niklas Lauffer and Alexander Meinke and Sumeet Motwani and Anka Reuel and Vincent Conitzer and Michael Dennis and Iason Gabriel and Adam Gleave and Gillian Hadfield and Nika Haghtalab and Atoosa Kasirzadeh and Sébastien Krier and Kate Larson and Joel Lehman and David C. Parkes and Georgios Piliouras and Iyad Rahwan},
      year={2025},
      eprint={2502.14143},
      archivePrefix={arXiv},
      primaryClass={cs.MA},
      url={https://arxiv.org/abs/2502.14143}, 
}

@article{Byrne2019Learning,
Author = {Byrne, David P. and de Roos, Nicolas},
Title = {Learning to Coordinate: A Study in Retail Gasoline},
Journal = {American Economic Review},
Volume = {109},
Number = {2},
Year = {2019},
Month = {February},
Pages = {591–619}
}

@Article{Assad2024Algorithmic,
journal={Journal of Political Economy},
author={Stephanie Assad and Robert Clark and Daniel Ershov and Lei Xu},
title={Algorithmic Pricing and Competition: Empirical Evidence from the German Retail Gasoline Market},
year={2024},
pages={723-771},
volume={132},
number={3}
}

@misc{Eschenbaum2022Robust,
author={Nicolas Eschenbaum and Filip Mellgren and Philipp Zahn},
title={Robust Algorithmic Collusion},
year={2022},
url={https://arxiv.org/abs/2201.00345},
}

@misc{abada_algorithmic_2024,
	title = {Algorithmic Collusion: Where Are We and Where Should We Be Going?},
	author = {Abada, Ibrahim and Harrington Jr, Joseph E. and Lambin, Xavier and Meylahn, Janusz M},
	year = {2024},
    url = { http://dx.doi.org/10.2139/ssrn.4891033},
}

@article{ABADA2024927,
title = {Collusion by mistake: Does algorithmic sophistication drive supra-competitive profits?},
journal = {European Journal of Operational Research},
volume = {318},
number = {3},
pages = {927-953},
year = {2024},
author = {Ibrahim Abada and Xavier Lambin and Nikolay Tchakarov}
}

@misc{lamba2022pricingalgorithms,
      title={Pricing with algorithms}, 
      author={Rohit Lamba and Sergey Zhuk},
      year={2022},
      eprint={2205.04661},
      archivePrefix={arXiv},
      primaryClass={econ.TH},
      url={https://arxiv.org/abs/2205.04661}, 
}

@article{Hansen2021Frontiers,
author = {Hansen, Karsten T. and Misra, Kanishka and Pai, Mallesh M.},
title = {Frontiers: Algorithmic Collusion: Supra-competitive Prices via Independent Algorithms},
journal = {Marketing Science},
volume = {40},
number = {1},
pages = {1-12},
year = {2021}}

@inproceedings{musolff_algorithmic_2022,
	title = {Algorithmic Pricing Facilitates Tacit Collusion: Evidence from E-Commerce},
	pages = {32--33},
	booktitle = {Proceedings of the 23rd {ACM} Conference on Economics and Computation},
	publisher = {{ACM}},
	author = {Musolff, Leon},
year = {2022}
}

@article{klein_autonomous_2021,
 author = {Timo Klein},
 journal = {The {RAND} Journal of Economics},
 number = {3},
 pages = {538--558},
 title = {Autonomous algorithmic collusion: Q-learning under sequential pricing},
 volume = {52},
 year = {2021}
}

@article{hettich_algorithmic_2021,
	title = {Algorithmic Collusion: Insights from Deep Learning},
	author = {Hettich, Matthias},
	year = {2021},
    url={https://ssrn.com/abstract=3785966}
}

@inproceedings{hartline_regulation_2025,
	title = {Regulation of Algorithmic Collusion, Refined: Testing Pessimistic Calibrated Regret},
	pages = {108--120},
	booktitle = {Proceedings of the 2025 Symposium on Computer Science and Law},
	author = {Hartline, Jason D. and Wang, Chang and Zhang, Chenhao},
	year = {2025}
}

@article{lambin_less_2023,
author = {Lambin, Xavier},
year = {2023},
month = {01},
pages = {},
title = {Less Than Meets the Eye: Simultaneous Experiments as a Source of Algorithmic Seeming Collusion},
journal = {{SSRN} Electronic Journal},
doi = {10.2139/ssrn.4498926}
}

@article{Abada2023Artificial,
author = {Abada, Ibrahim and Lambin, Xavier},
title = {Artificial Intelligence: Can Seemingly Collusive Outcomes Be Avoided?},
journal = {Management Science},
volume = {69},
number = {9},
pages = {5042-5065},
year = {2023}
}

@misc{garivier2008upperconfidenceboundpoliciesnonstationary,
      title={On Upper-Confidence Bound Policies for Non-Stationary Bandit Problems}, 
      author={Aurélien Garivier and Eric Moulines},
      year={2008},
      eprint={0805.3415},
      archivePrefix={arXiv},
      primaryClass={math.ST},
      url={https://arxiv.org/abs/0805.3415}, 
}

@InProceedings{Garivier2011On,
author="Garivier, Aur{\'e}lien
and Moulines, Eric",
title="On Upper-Confidence Bound Policies for Switching Bandit Problems",
booktitle="Algorithmic Learning Theory",
year="2011",
publisher="Springer Berlin Heidelberg",
pages="174--188"
}

@book{sutton2018reinforcement,
  title={Reinforcement learning: An introduction},
  author={Sutton, Richard S. and Barto, Andrew G.},
  year={2018},
  publisher={MIT Press}
}

@inproceedings{
Rashid2020Optimistic,
title={Optimistic Exploration even with a Pessimistic Initialisation},
author={Tabish Rashid and Bei Peng and Wendelin Boehmer and Shimon Whiteson},
booktitle={International Conference on Learning Representations},
year={2020}}

@inproceedings{Jin2018Is,
 author = {Jin, Chi and Allen-Zhu, Zeyuan and Bubeck, Sebastien and Jordan, Michael I},
 booktitle = {Advances in Neural Information Processing Systems},
 title = {Is Q-Learning Provably Efficient?},
 volume = {31},
 year = {2018}
}

@incollection{McFadden1973,
	title = {Conditional logit analysis of qualitative choice behavior},
	booktitle = {Fontiers in {Econometrics}},
	publisher = {Academic press},
	author = {McFadden, Daniel},
	year = {1974},
	pages = {105--142}
}

@misc{agrawal2025evaluatingllmagentcollusion,
      title={Evaluating {LLM} Agent Collusion in Double Auctions}, 
      author={Kushal Agrawal and Verona Teo and Juan J. Vazquez and Sudarsh Kunnavakkam and Vishak Srikanth and Andy Liu},
      year={2025},
      eprint={2507.01413},
      archivePrefix={arXiv},
      primaryClass={cs.GT},
      url={https://arxiv.org/abs/2507.01413}, 
}

@misc{bichler2024onlineoptimizationalgorithmsrepeated,
      title={Online Optimization Algorithms in Repeated Price Competition: Equilibrium Learning and Algorithmic Collusion}, 
      author={Martin Bichler and Julius Durmann and Matthias Oberlechner},
      year={2024},
      eprint={2412.15707},
      archivePrefix={arXiv},
      primaryClass={cs.GT},
      url={https://arxiv.org/abs/2412.15707}, 
}

@inproceedings{Jordan2007Empirical,
author = {Jordan, Patrick R. and Kiekintveld, Christopher and Wellman, Michael P.},
title = {Empirical game-theoretic analysis of the TAC Supply Chain game},
year = {2007},
booktitle = {Proceedings of the 6th International Joint Conference on Autonomous Agents and Multiagent Systems}
}

@inproceedings{kiekintveld_selecting_nodate,
author = {Kiekintveld, Christopher and Wellman, Michael P.},
title = {Selecting strategies using empirical game models: an experimental analysis of meta-strategies},
year = {2008},
booktitle = {Proceedings of the 7th International Joint Conference on Autonomous Agents and Multiagent Systems - Volume 2},
pages = {1095–1101}
}

@inproceedings{McBurney2006Selecting,
author = {Phelps, S. and Marcinkiewicz, M. and Parsons, S.},
title = {A novel method for automatic strategy acquisition in N-player non-zero-sum games},
year = {2006},
booktitle = {Proceedings of the Fifth International Joint Conference on Autonomous Agents and Multiagent Systems},
pages = {705–712}
}

@inproceedings{Vorobeychik2006Empirical,
author = {Vorobeychik, Yevgeniy and Kiekintveld, Christopher and Wellman, Michael P.},
title = {Empirical mechanism design: methods, with application to a supply-chain scenario},
year = {2006},
booktitle = {Proceedings of the 7th ACM Conference on Electronic Commerce},
pages = {306–315}
}

@inproceedings{Wellman2005Strategic,
author = {Estelle, Joshua and Vorobeychik, Yevgeniy and Wellman, Michael P. and Singh, Satinder and Kiekintveld, Christopher and Soni, Vishal},
title = {Strategic interactions in the TAC 2003 supply chain tournament},
year = {2004},
booktitle = {Proceedings of the 4th International Conference on Computers and Games},
pages = {316–331}
}

@article{Reeves2005Generating,
    author = {Daniel M. Reeves},
    title = {Generating Trading Agent Strategies:
Analytic and Empirical Methods for Infinite and Large
Games},
    journal = {PhD thesis, University of Michigan},
    year = {2005}
}

@article{ezrachi_algorithmic_2015,
	title = {Algorithmic Collusion: Problems and Counter-Measures},
	author = {Ezrachi, Ariel and Stucke, Maurice E.},
	year = {2015},
}

@article{oecd,
title = {Algorithmic Competition, OECD Competition Policy Roundtable Background Note},
url={www.oecd.org/daf/competition/algorithmic-competition-2023.pdf},
author={OECD},
year={2023}
}

@article{cramton2000collusive,
  title={Collusive bidding: Lessons from the {FCC} spectrum auctions},
  author={Cramton, Peter and Schwartz, Jesse A},
  journal={Journal of regulatory Economics},
  volume={17},
  number={3},
  pages={229--252},
  year={2000},
  publisher={Springer}
}

@article{agranov2015collusion,
author = {Agranov, Marina and Yariv, Leeat},
year = {2015},
pages = {93-108},
title = {Collusion through Communication in Auctions},
volume = {107},
journal = {Games and Economic Behavior}
}

@article{friedman1971non,
  title={A non-cooperative equilibrium for supergames},
  author={Friedman, James W},
  journal={The Review of Economic Studies},
  volume={38},
  number={1},
  pages={1--12},
  year={1971},
  publisher={Wiley-Blackwell}
}

@article{schelling1956essay,
 author = {Thomas C. Schelling},
 journal = {The American Economic Review},
 number = {3},
 pages = {281--306},
 publisher = {American Economic Association},
 title = {An Essay on Bargaining},
 volume = {46},
 year = {1956}
}
\newpage
\appendix

\section{Deferred Content and Results from Section~\ref{exp:q_learning} on Q-learning} 

\subsection{Test-time Payoff Comparison: Joint vs.~Independent Pretraining}
\begin{figure*}[ht]
    \centering
    \includegraphics[trim={0.5cm 0.5cm 0.9cm 1.1cm}, clip, width=0.6\linewidth]{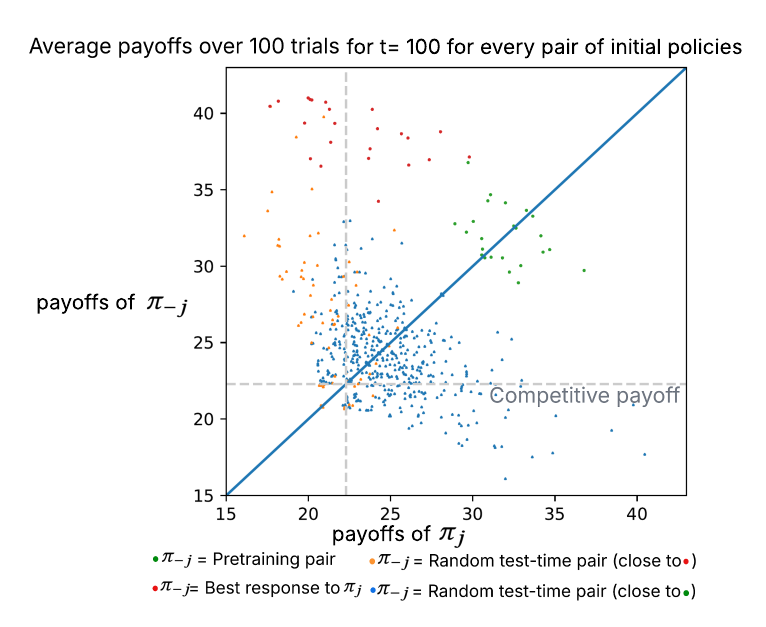}
    \caption{We collect 12 pairs of pretrained Q-learning policies with 15 discretized actions and use them as initial policies for test-time evaluation. We measure total payoffs from $t=0$ to $t=100$, averaged over 100 random seeds. In \textit{green}, each policy is paired with its original pretraining pair. In \textit{red}, each pretrained policy ($\pi_j$) is paired with its best response ($\pi_{-j}$). 
    In \textit{orange} and \textit{blue}, we show payoffs for randomly sampled, independently pretrained pairs without adaptation: \textit{orange} points lie closer (in L2 distance) to the \textit{red} cluster, while \textit{blue} points are closer to the \textit{green}.
    }
\label{fig:pre_adapt}
\end{figure*}

The majority of randomly paired policies achieve payoffs below the competitive benchmark (i.e., when both policies play the competitive price). 
This suggests that while pretrained policies can achieve high-level collusion, pairing them with an independently pretrained policy disrupts their ability to collude instantly.

\subsection{Normalization of Q-values}\label{sec:Q_Normal}

For Q-learning, the initial policy is encoded in the initial Q-values. However, Q-values accumulated during pretraining can be irregular due to the variable number of visits and the number of iterations required for convergence. To allow a reliable comparison, we apply the following normalization. Let $\Tilde{Z}_{j}$ denote the Q-values accumulated during pretraining for player $j$, and denote the converged policies pair as ($\pi_j, \pi'_j$).
\begin{align}
    Z_{j}(s,p) &\coloneqq Q^N + f\cdot (Q^M - Q^N) - \max_{s^*} V^{\pi_j|\pi'_j}(s^*) + \min_{p'}Q^{\pi_j| \pi'_{j}}(s,p') \\
    & + \bigg(\Tilde{Z}_{j}(s,p) - \min_{p'}\Tilde{Z}_{j}(s,p') \bigg)
     \times \frac{V^{\pi_j| \pi'_{j}}(s)
    - \min_{p'}Q^{\pi_j| \pi'_{j}}(s,p')}{\max_{p'}\Tilde{Z}_{j}(s,p') - \min_{p'}\Tilde{Z}_{j}(s,p')}
\end{align}
where $Q^M = \bar{r}^M/(1-\gamma)$ and $Q^N = \bar{r}^N/(1-\gamma)$ represent the theoretical discounted expected returns associated with absorbing into the perfectly colluding and competitive states, respectively.
 
 After normalization, the original argmax of $\Tilde{Z}_j$ is preserved: $Z_j(s,\pi_j(s)) = V^{\pi_j | \pi'_j}(s) + Q^N + f \cdot (Q^M - Q^N) - \max_{s^*} V^{\pi_j|\pi'_j}(s^*)$, and the minimum satisfies $\min_{p'} Z_j(s,p) = \min_{p'}Q^{\pi_j| \pi'_j}(s,p') + Q^N + f\cdot (Q^M - Q^N) - \max_{s^*} V^{\pi_j|\pi'_j}(s^*)$.
 The maximum Q-value (at the argmax action) maps to the true value function $V^{\pi_j | \pi'_j}(s)$ and the minimum Q-value maps to $\min_{p'}Q^{\pi_j| \pi'_j}(s,p')$ with an offset $Q^N + f\cdot (Q^M - Q^N) - \max_{s^*}V^{\pi_j|\pi'_j}(s^*)$.
 
 The scaling factor $f$ is set to 1 in main experiments, and we vary $f$ for further analysis in Sec.~\ref{sec:exp}. This normalization ensures that Q-values across all meta-strategies share a common scale, so that a given learning rate induces comparable adaptation speeds.
%
 When $f$ is large, the Q-values are optimistically initialized,  which is a well-known technique for promoting efficient exploration~\citep{Jin2018Is, Rashid2020Optimistic}.





\subsection{Categorizing Q-learning Pretrained Policies via PC and CR}

\begin{figure}[ht]
      \centering
    \includegraphics[trim={0.75cm 0.75cm 0.6cm 1cm},clip,width=0.6\linewidth]{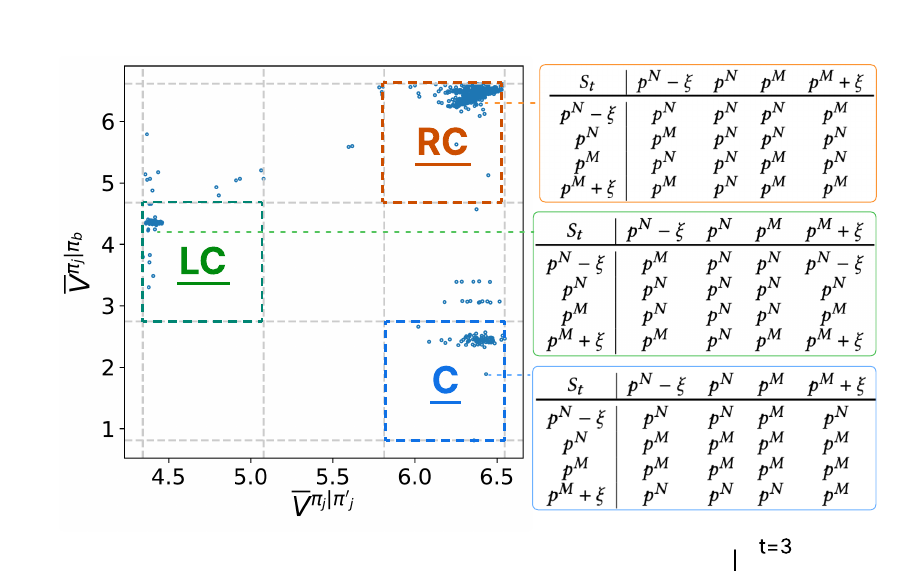}
    \captionof{figure}{With 4 discretized actions, the 500 Q-learning pretrained policies form three distinct clusters when laid out along PC (Def.~\ref{def:PC}) and CR (Def.~\ref{def:CR}), corresponding to the competitive (LC), naively collusive (C), and robustly collusive (RC) categories introduced in Sec.~\ref{exp:meta}.}
    \label{fig:cat_Q}    
\end{figure}

 Fig.~\ref{fig:cat_Q} Left shows $\overline{V}^{\pi_j, \pi_j'}$ v.s.~$\overline{V}^{\pi_j, \pi_b}$ along with the LC, C and RC categorization for $N_{\text{discrete}}=4$. 
 Fig.~\ref{fig:cat_Q} Right illustrates three representative policies in each category.
 The top policy (RC) sustains collusion with its pretraining partner and remains collusive even when faced with a best-responding opponent, the middle policy (LC) is resistant to exploitation but defaults to competitive pricing rather than colluding, and the bottom policy colludes with its pretraining partner but is vulnerable to exploitation (C).

\begin{figure}[ht]
    \centering
    \includegraphics[trim={1cm 1cm 1cm 1cm},clip, width=0.6\linewidth]{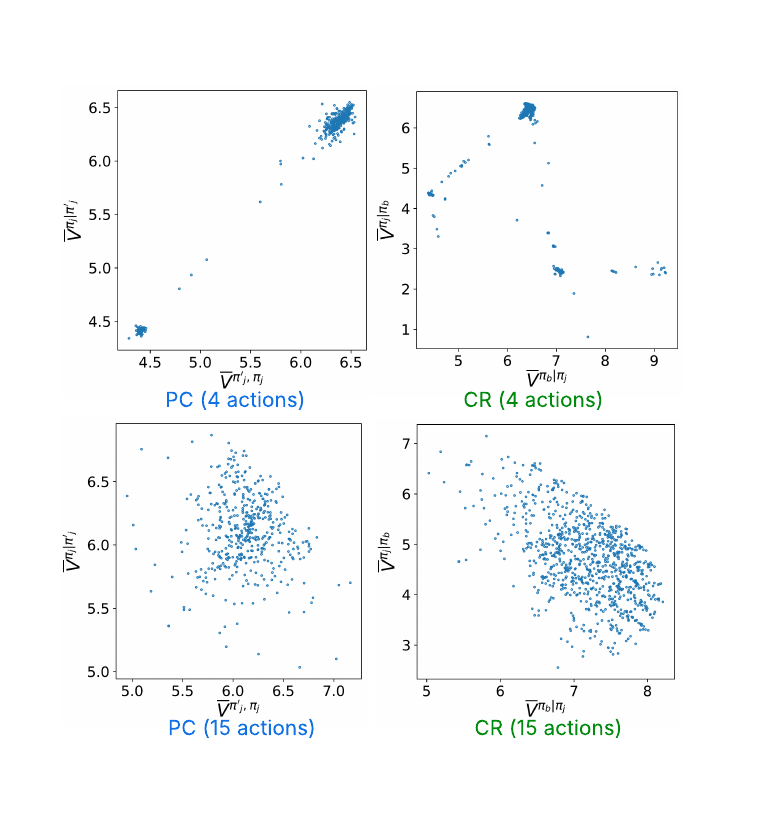}
    \caption{The PC (Def.~\ref{def:PC}) and CR (Def.~\ref{def:CR}) of 1,000 pretrained Q-learning policies with 4 and 15 discretized actions, respectively, under the experiment setting described in Sec.~\ref{exp:q_learning}.}
    \label{fig:pc_cr_Q}
\end{figure}

In Fig.~\ref{fig:pc_cr_Q}, we further show the PC and CR of pretrained policies under $N_{\text{discrete}}=4$ and $N_{\text{discrete}}=15$.

\begin{figure}[ht]
    \centering
    \includegraphics[trim={1cm 1.cm 1cm 1.25cm}, clip, width=0.6\linewidth]{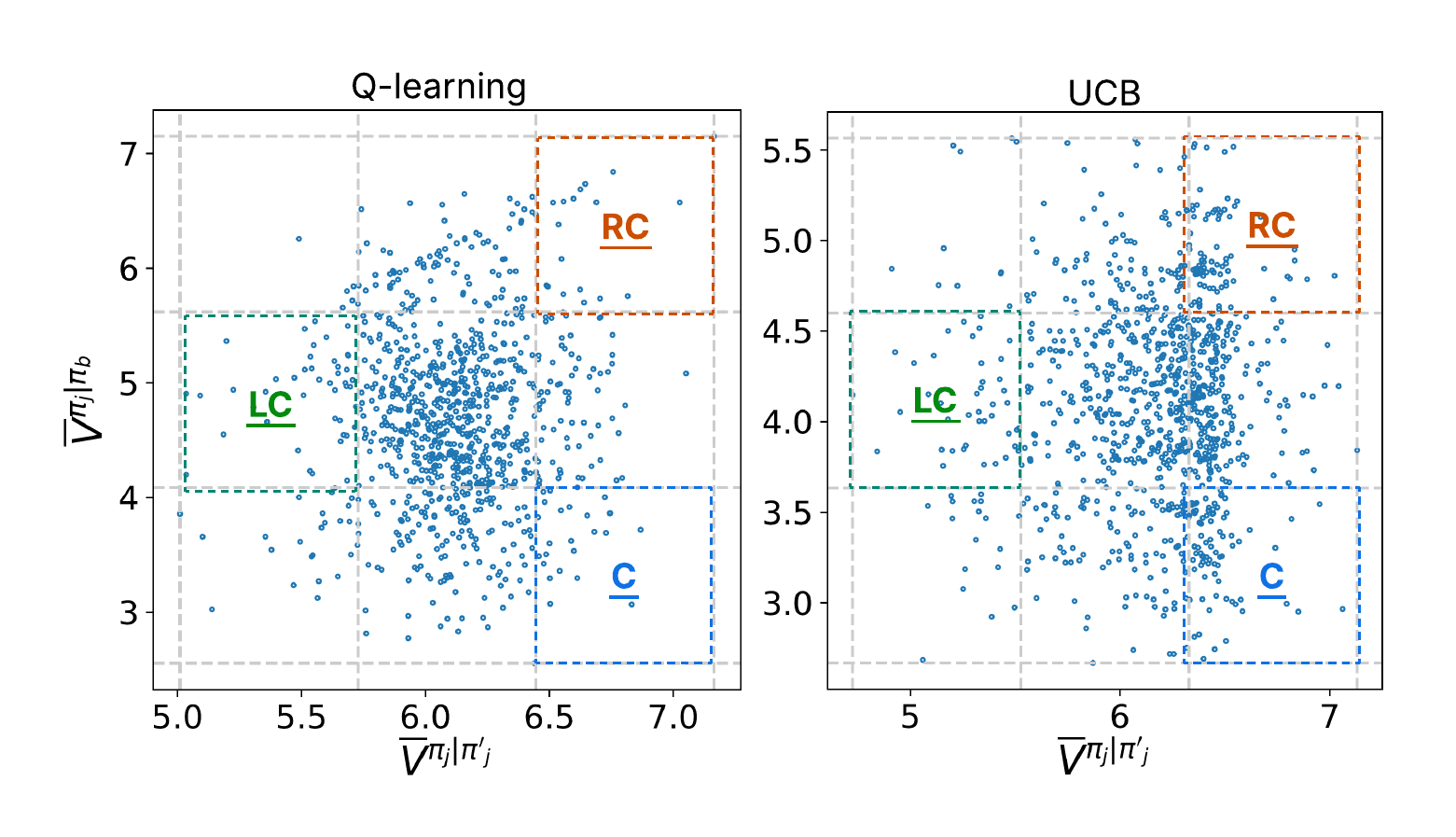}
    \caption{Categorization of pretrained policies with 15 discretized actions obtained via Q-learning (Left) and UCB (Right), under the experiment settings described in Sec.~\ref{exp:q_learning} and \ref{exp:ucb}.}
    \label{fig:cat_Q_UCB}
\end{figure}

\clearpage
\subsection{Meta-games among Meta-strategies with Fine-grained Learning Rates}\label{apd:varying_LR}
\begin{figure*}[ht]
    \centering
\includegraphics[width=0.5\textwidth]{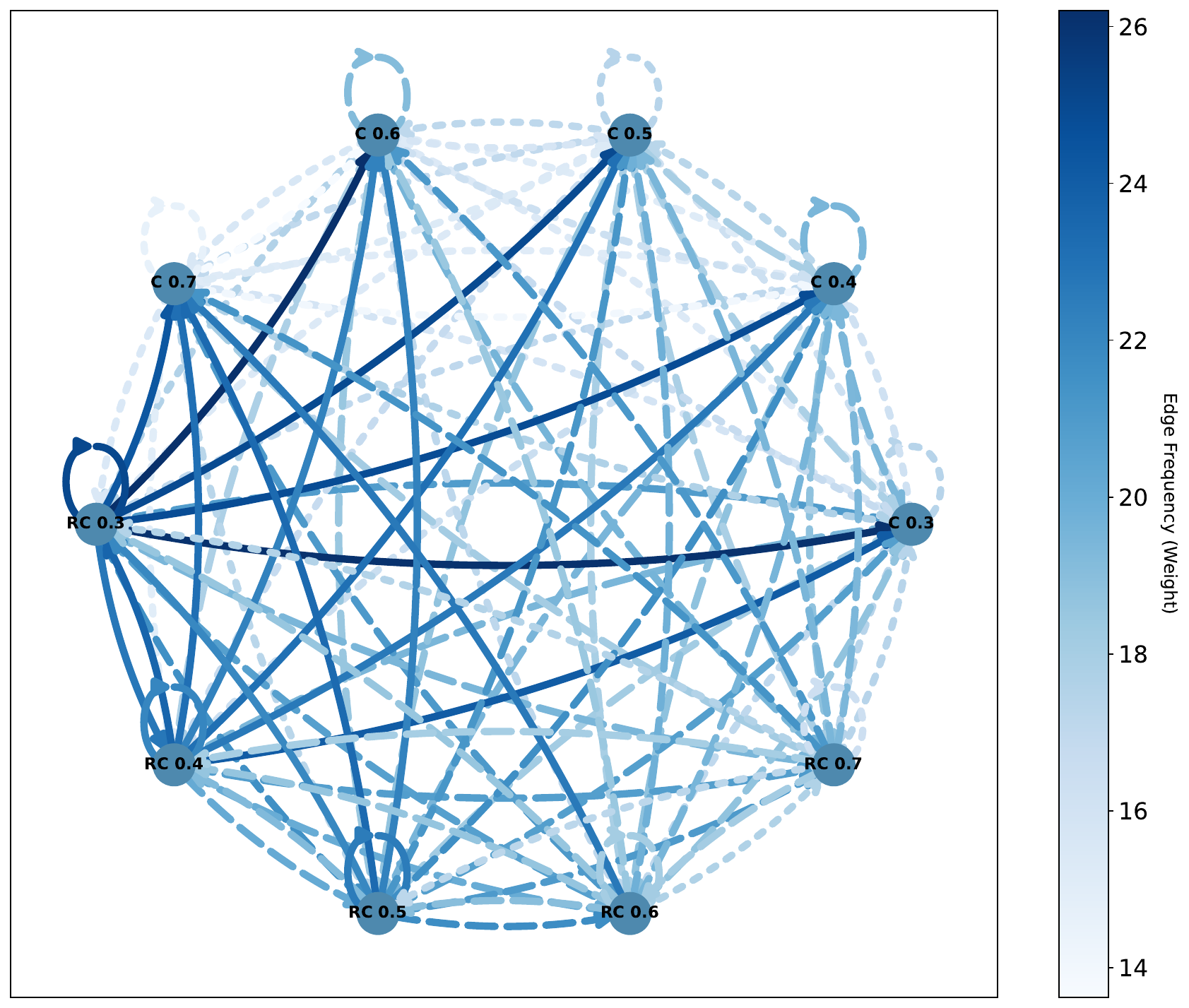}
  \caption{Best-response graph for C and RC with a finer grid of learning rates. }\label{fig:LR}
\end{figure*}

In Sec.~\ref{exp:q_learning}, we identified two PSNEs: (C, 0.5) and (RC, 0.5) for Q-learning with symmetric cost $c_1=c_2=1$, optimistic initialization $f=1$, and time horizon $t=10,000$.     
To test the robustness of the equilibrium outcome to learning rate perturbations, we perform additional evaluations on C and RC with a finer grid of learning rates \{0.3, 0.4, 0.5, 0.6, 0.7\}. 
The results are provided in Table~\ref{tab:LR} and Fig~\ref{fig:LR}. 

We observe a similar level of collusion for C with $\alpha \ge 0.4$ (CoI $\ge 50\%$). RC 0.3 and 0.5 form one of the MSNEs and sustain collusion at around 40\%. These findings suggest that the collusive outcome is robust to moderate learning rate perturbations. 
Strategies in C category do not constitute a part of the NE.
Nevertheless, the regret of playing C remains relatively low.

\begin{table*}[ht]
\centering
\caption{Robustness test on varying the test-time learning rates for Q-learning pretrained policies. $t = 10,000$.}
\label{tab:LR}
 \resizebox{\linewidth}{!}{%
\begin{tabular}{l|cccccccccc}
\toprule
 $t=10,000$ & C 0.3 & C 0.4 & C 0.5 & C 0.6 & C 0.7 & RC 0.3 & RC 0.4 & RC 0.5 & RC 0.6 & RC 0.7\\
\midrule
    PSNE  & -& - & -& -& -& \checkmark & -& \checkmark  & -& - \\
   MSNE  & 0.00 & 0.00 & 0.00 & 0.00 & 0.00 & 0.40 & 0.00 & 0.60 & 0.00 & 0.00  \\ 
  NE-Regret ($\times 10^{-3}$) & 2.07 $\pm$ 1.73 & 1.88 $\pm$ 1.62 & 3.26 $\pm$ 1.55 & 2.80 $\pm$ 1.56 & 3.55 $\pm$ 1.44 & \textbf{0.00 $\pm$ 1.49} & \underline{0.40 $\pm$ 1.40} & \textbf{0.00 $\pm$ 1.45} & 1.71 $\pm$ 1.33 & 2.60 $\pm$ 1.38\\
 Uniform Score & 46.38 $\pm$ 1.04 & 47.40 $\pm$ 0.95 & 47.47 $\pm$ 0.90 & 47.55 $\pm$ 0.82 & 46.71 $\pm$ 0.76 & \textbf{49.85 $\pm$ 1.01} & \textbf{49.20 $\pm$ 0.94} & \underline{48.99 $\pm$ 0.93} & 48.02 $\pm$ 0.87 & 47.77 $\pm$ 0.86  \\
\bottomrule
\end{tabular}%
}
\end{table*}

\subsection{Results for Q-learning with Pessimistic Initializations $f=0.5$ and $f=0$}


\begin{table*}[ht]
\centering
\caption{Meta-games on Q-learning using $f=0.5$ and $f=0$ respectively, with a time horizon of $t = 10,000$. 
In the case of $f=0$, we additionally incorporate $\varepsilon=1$ with a decay rate of $0.001$ to ensure a policy still updates. 
The best-performing strategy is highlighted in \textbf{bold}, and the second-best is \underline{underlined}.}
\label{tab:diff_q_scale}
 \resizebox{\linewidth}{!}{%
\begin{tabular}{cl|cccccccccc}
\toprule
Q-scale & $t=10,000$ & RD 0.5 & LC 0.5 & LC 0.05 & LC 0.005 & C 0.5 & C 0.05 & C 0.005 & RC 0.5 & RC 0.05 & RC 0.005 \\
\midrule
\multirow{4}{*}{$f=0.5$}  & PSNE & -& \checkmark & -& -& \checkmark  & -& \checkmark & -& \checkmark  & - 
\\
& MSNE  & 0.00 & 0.19 & 0.00 & 0.00 & 0.34 & 0.00 & 0.00 & 0.39 & 0.08 & 0.00 \\ 
 & NE-Regret ($\times 10^{-3}$) & 6.78 $\pm$ 1.15 & \textbf{0.00 $\pm$ 1.14} & 6.24 $\pm$ 1.21 & 15.19 $\pm$ 1.28 & \textbf{0.00 $\pm$ 1.15} & \underline{1.90 $\pm$ 1.23} & 10.60 $\pm$ 1.20 & \textbf{0.00 $\pm$ 1.04} & \textbf{0.00 $\pm$ 1.18} & 5.00 $\pm$ 1.24 \\
& Uniform Score & 27.68 $\pm$ 1.00 & \underline{32.78 $\pm$ 0.91} & 28.43 $\pm$ 0.94 & 21.08 $\pm$ 0.94 & \textbf{33.98 $\pm$ 0.80} & 30.45 $\pm$ 0.78 & 24.56 $\pm$ 0.76 & 31.70 $\pm$ 0.90 & 29.57 $\pm$ 0.87 & 25.97 $\pm$ 0.82\\
\midrule
\multirow{3}{*}{$f=0.0$}
 & Mixed \& PSNE  &  0.00 & 0.00 & 0.00 & 0.00 & 0.00 & 0.00 & 0.00 & 1.00 & 0.00 & 0.00 \\ 
 & NE-Regret ($\times 10^{-3}$) & 7.22 $\pm$ 1.29 & 1.91 $\pm$ 1.13 & 10.85 $\pm$ 1.28 & 11.79 $\pm$ 1.49 & \underline{0.54 $\pm$ 1.05} & 9.73 $\pm$ 1.14 & 9.35 $\pm$ 1.22 & \textbf{0.00 $\pm$ 1.01} & 5.22 $\pm$ 1.05 & 5.60 $\pm$ 1.06\\
& Uniform Score & 20.48 $\pm$ 0.63 & \underline{26.50 $\pm$ 0.56} & 17.62 $\pm$ 0.66 & 15.64 $\pm$ 0.74 & 26.74 $\pm$ 0.49 & 18.79 $\pm$ 0.58 & 18.55 $\pm$ 0.61 & \textbf{28.14 $\pm$ 0.57} & 21.92 $\pm$ 0.59 & 21.93 $\pm$ 0.58 \\
\bottomrule
\end{tabular}%
}
\end{table*}

\clearpage
\subsection{Additional Results on Q-learning-based Meta-strategies}\label{apd:Q_matrices}

For each setting below, we report the best-response graphs, average payoffs, and the changes in CR and PC between the initial policies and the policies at the end of the time horizon, denoted $\Delta_t$(CR) and $\Delta_t$(PC) respectively.

\begin{figure*}[ht]
  \centering 
\begin{minipage}{1.0\textwidth}
  \centering 
  
  \begin{subfigure}[t]{0.48\textwidth}
    \centering
    \includegraphics[width=\textwidth]{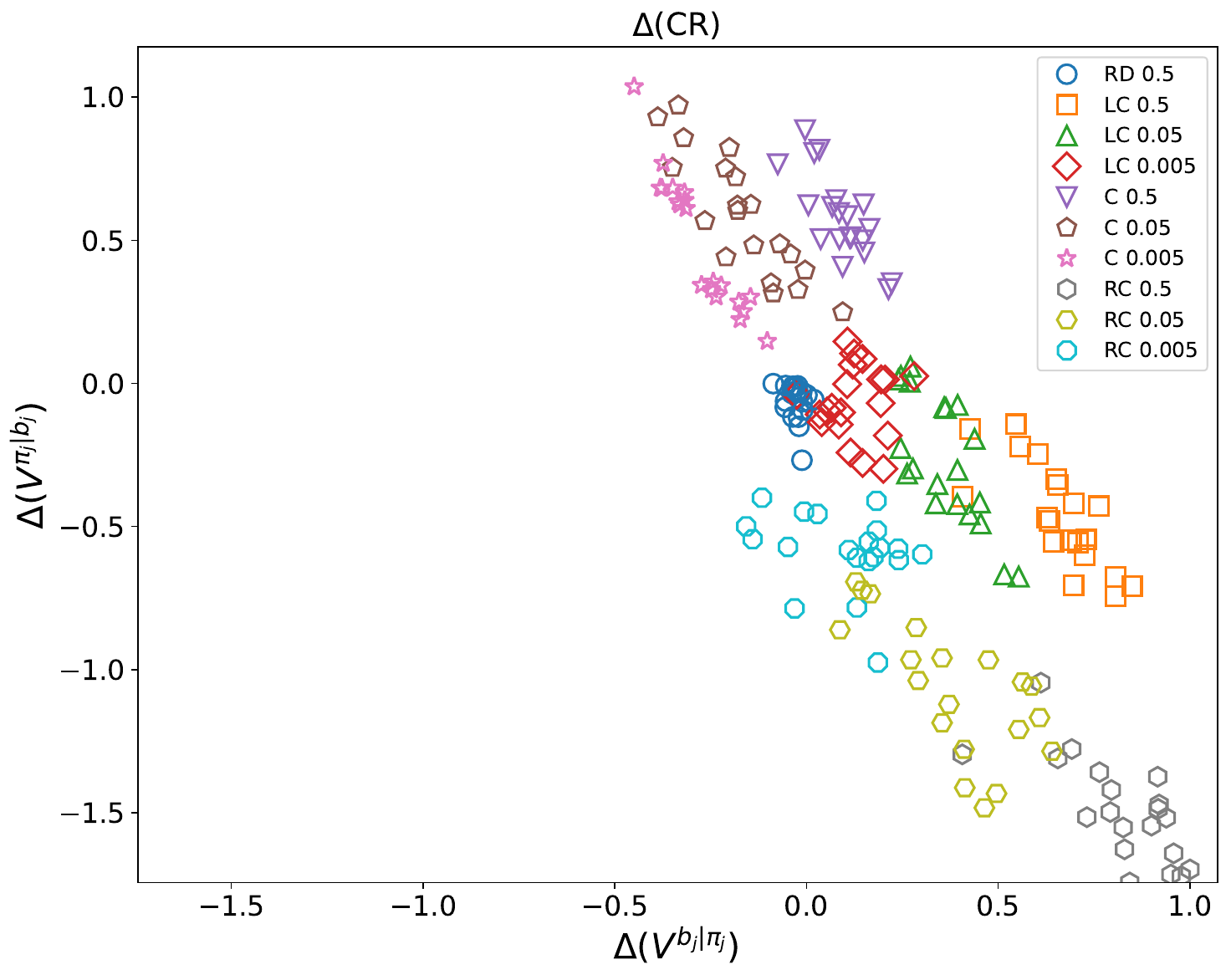}
    \caption{Change in CR over test time. As RC strategies begin with already high CR, larger learning rates cause greater deviation from the robust policy, resulting in a decline in CR. 
    In contrast, strategies in C category generally improve their CR over adaptation. RD 0.5 starts with low CR and shows no improvement during test time. LC strategies are initially more robust than the ones in C, and similarly show no improvement in CR during test time.}
    \label{fig:first}
  \end{subfigure}
  \hfill 
  \begin{subfigure}[t]{0.48\textwidth}
    \centering
    \includegraphics[width=\textwidth]{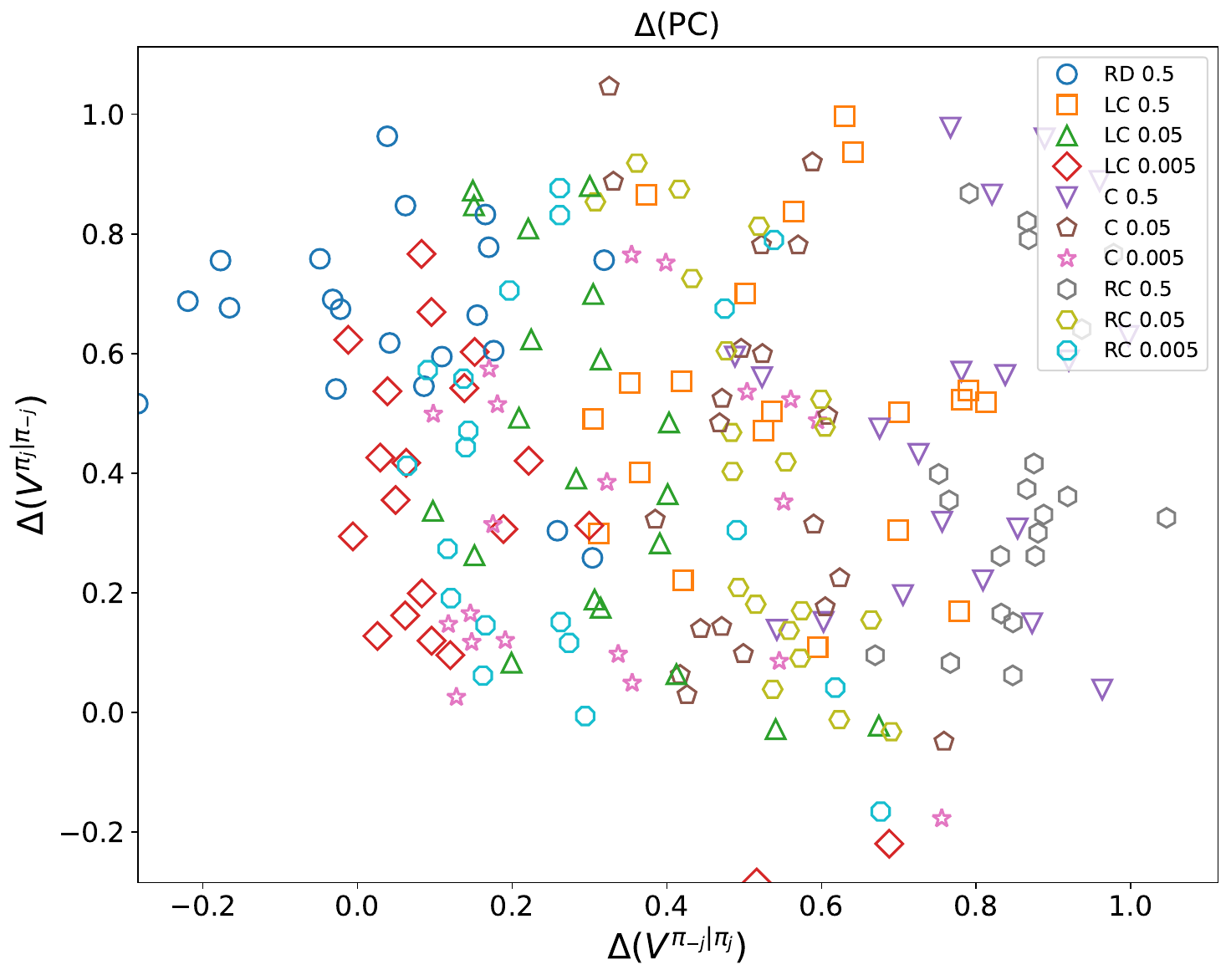}
    \caption{Change in PC over time. 
    Overall, PC improves over the course of test time. For RD strategies, although their own $V$ improves, their opponent's does not, as the value against RD is already high at initialization. 
    For both LC and RC, larger learning rates tend to benefit their opponents more, even when their own $\Delta(V^{\pi_j|\pi_{-j}})$ is comparable to that under smaller learning rates. This suggests that LC and RC with large learning rates provide opportunities for cooperation.}
    \label{fig:second}
  \end{subfigure}

\end{minipage}
\caption{$\Delta_t$(CR) and $\Delta_t$(PC) for Q-learning-based meta-strategies under symmetric costs $c_1=c_2=1$, $f=1$ and $t=10,000$.}\label{fig:symQ_delta}
  \end{figure*}

\paragraph{Symmetric costs $c_1=c_2=1$ with optimistic initialization $f=1$ and time horizon $t=10,000$.}
The best-response graph, average payoffs, and $\Delta_t$(CR),  $\Delta_t$(PC) are presented in Fig.~\ref{fig:br_sym_Q}, Table \ref{tab:symQ_mean}, and Fig.~\ref{fig:symQ_delta}. 

\begin{table}[h]
\centering
\caption{
Average payoffs over 4,000 runs in the Q-learning meta-game payoff matrix. Standard errors are omitted as they are negligible in the order of $10^{-4}$. For reference, in this setting, the competitive and monopoly payoffs are $\bar{r}^N = 0.22$ and $\bar{r}^M = 0.34$.
}\label{tab:symQ_mean}
\resizebox{\textwidth}{!}{%
\begin{tabular}{lcccccccccc}
\toprule
 & RD 0.5 & LC 0.5 & LC 0.05 & LC 0.005 & C 0.5 & C 0.05 & C 0.005 & RC 0.5 & RC 0.05 & RC 0.005 \\ 
 \midrule
RD 0.5 & 0.29, 0.29 & 0.27, 0.30 & 0.26, 0.29 & 0.27, 0.28 & 0.27, 0.31 & 0.26, 0.31 & 0.26, 0.30 & 0.26, 0.30 & 0.25, 0.30 & 0.25, 0.29 \\
LC 0.5 & 0.30, 0.27 & 0.28, 0.28 & 0.27, 0.28 & 0.27, 0.27 & 0.28, 0.29 & 0.26, 0.28 & 0.27, 0.27 & 0.27, 0.28 & 0.25, 0.29 & 0.25, 0.28 \\
LC 0.05 & 0.29, 0.26 & 0.28, 0.27 & 0.26, 0.26 & 0.26, 0.26 & 0.27, 0.26 & 0.26, 0.27 & 0.25, 0.26 & 0.27, 0.27 & 0.25, 0.27 & 0.24, 0.27 \\
LC 0.005 & 0.28, 0.27 & 0.27, 0.27 & 0.26, 0.26 & 0.26, 0.26 & 0.26, 0.27 & 0.25, 0.26 & 0.24, 0.26 & 0.26, 0.27 & 0.25, 0.26 & 0.24, 0.26 \\
C 0.5 & 0.31, 0.27 & 0.29, 0.28 & 0.26, 0.27 & 0.27, 0.26 & 0.28, 0.28 & 0.26, 0.27 & 0.27, 0.26 & 0.27, 0.28 & 0.24, 0.28 & 0.24, 0.28 \\
C 0.05 & 0.31, 0.26 & 0.28, 0.26 & 0.27, 0.26 & 0.26, 0.25 & 0.27, 0.26 & 0.26, 0.26 & 0.25, 0.26 & 0.27, 0.26 & 0.23, 0.26 & 0.26, 0.26 \\
C 0.005 & 0.30, 0.26 & 0.27, 0.27 & 0.26, 0.25 & 0.26, 0.24 & 0.26, 0.27 & 0.26, 0.25 & 0.24, 0.24 & 0.26, 0.26 & 0.25, 0.25 & 0.23, 0.25 \\
RC 0.5 & 0.30, 0.26 & 0.28, 0.27 & 0.27, 0.27 & 0.27, 0.26 & 0.28, 0.27 & 0.26, 0.27 & 0.26, 0.26 & 0.27, 0.27 & 0.25, 0.27 & 0.24, 0.27 \\
RC 0.05 & 0.30, 0.25 & 0.29, 0.25 & 0.27, 0.25 & 0.26, 0.25 & 0.28, 0.24 & 0.26, 0.25 & 0.25, 0.25 & 0.27, 0.25 & 0.25, 0.25 & 0.24, 0.26 \\
RC 0.005 & 0.29, 0.25 & 0.28, 0.25 & 0.27, 0.24 & 0.26, 0.24 & 0.28, 0.24 & 0.26, 0.23 & 0.25, 0.23 & 0.27, 0.24 & 0.26, 0.24 & 0.24, 0.24 \\
\bottomrule
\end{tabular}%
}
\end{table}

\begin{figure*}[h!]
    \centering
\includegraphics[width=0.4\linewidth]{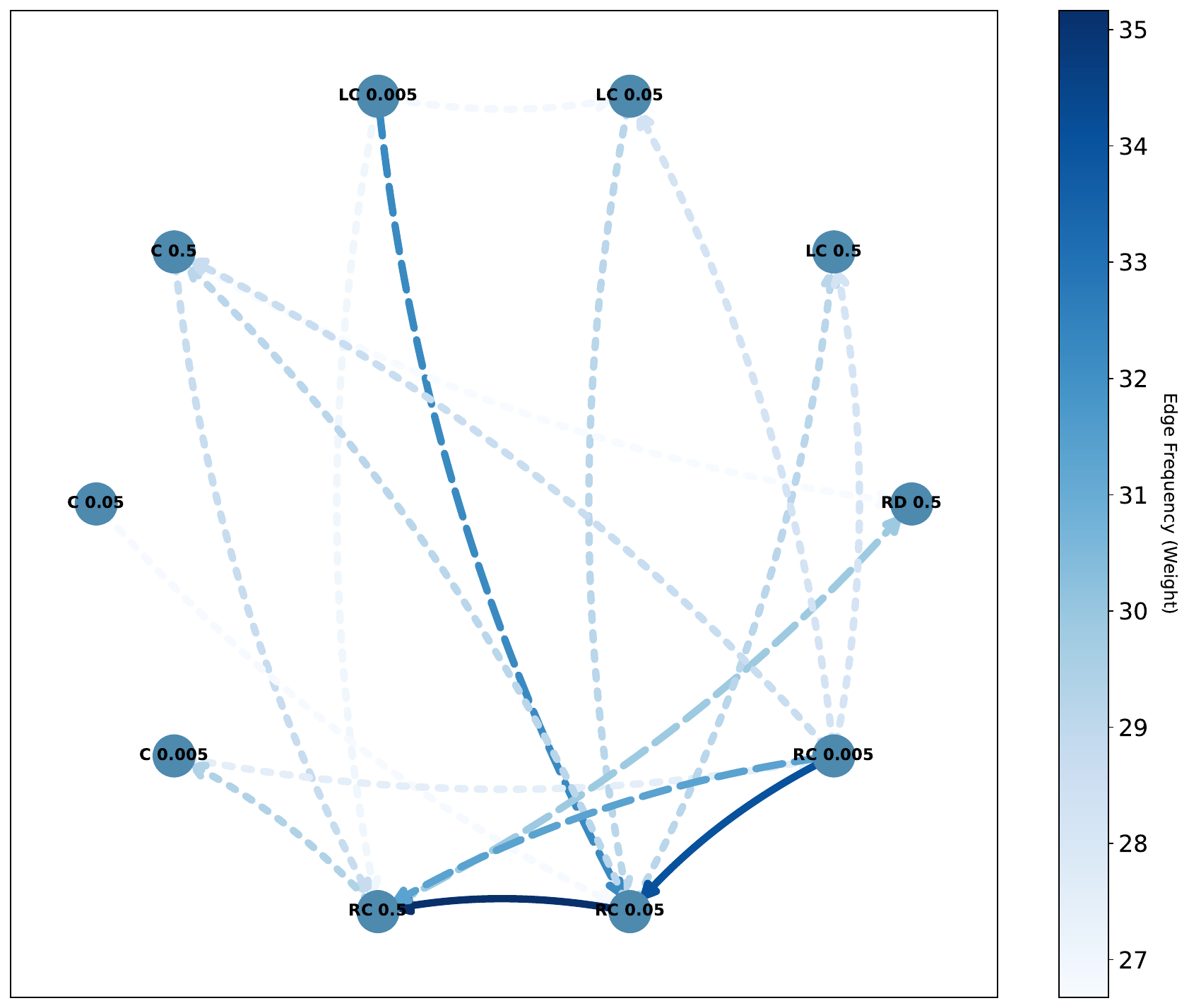}
    
    \caption{Best-response graph for Q-learning with $c_1=c_2=0.8$, $f=1$ and  $t=10,000$.}
    \label{fig:symQ_lo_br}
\end{figure*}

\begin{figure*}[h!]
\begin{minipage}
{0.48\linewidth}
\centering
\includegraphics[width=\textwidth]
{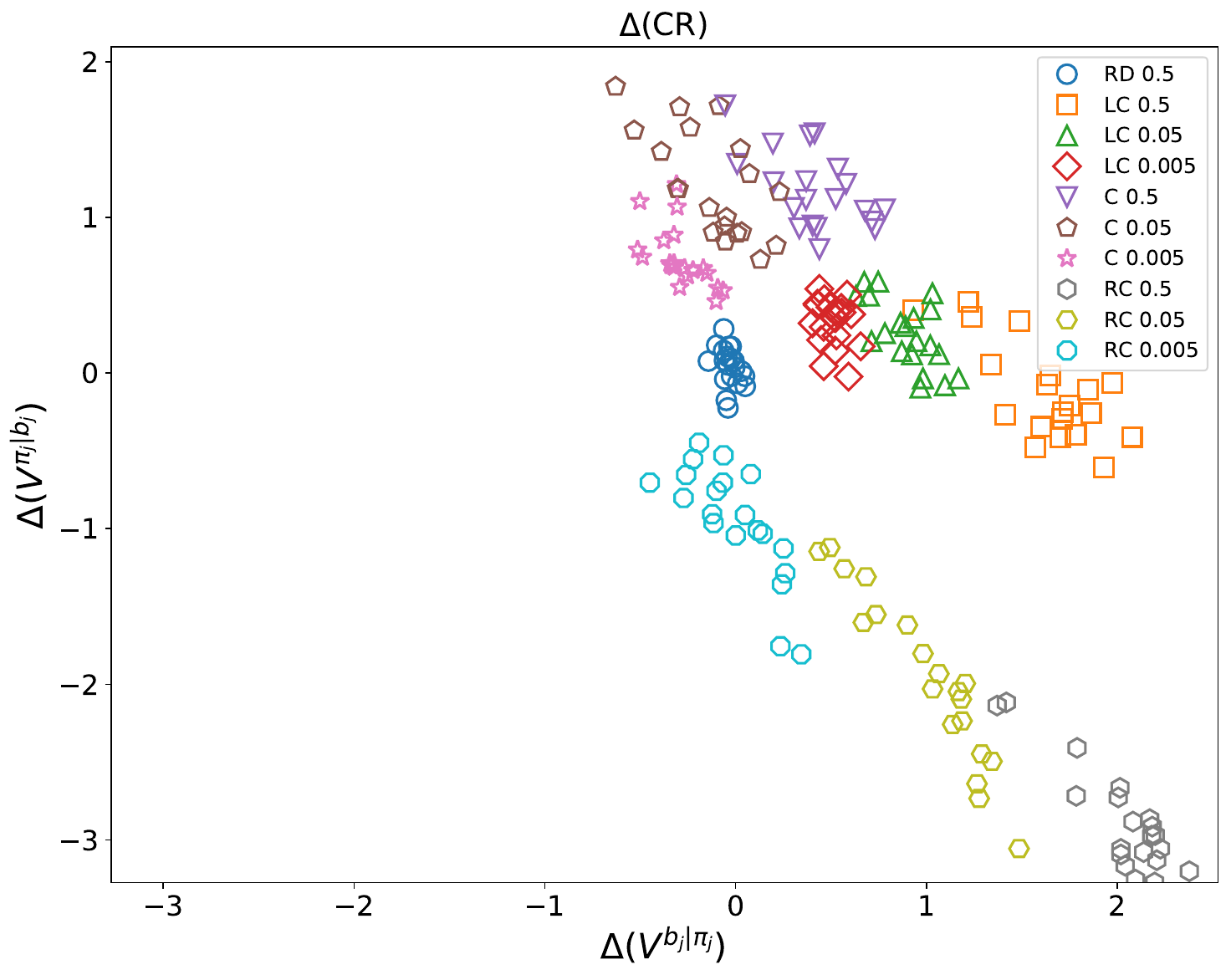}
\end{minipage}
\hfill
\begin{minipage}{0.48\linewidth}
\centering
\includegraphics[width=\textwidth]{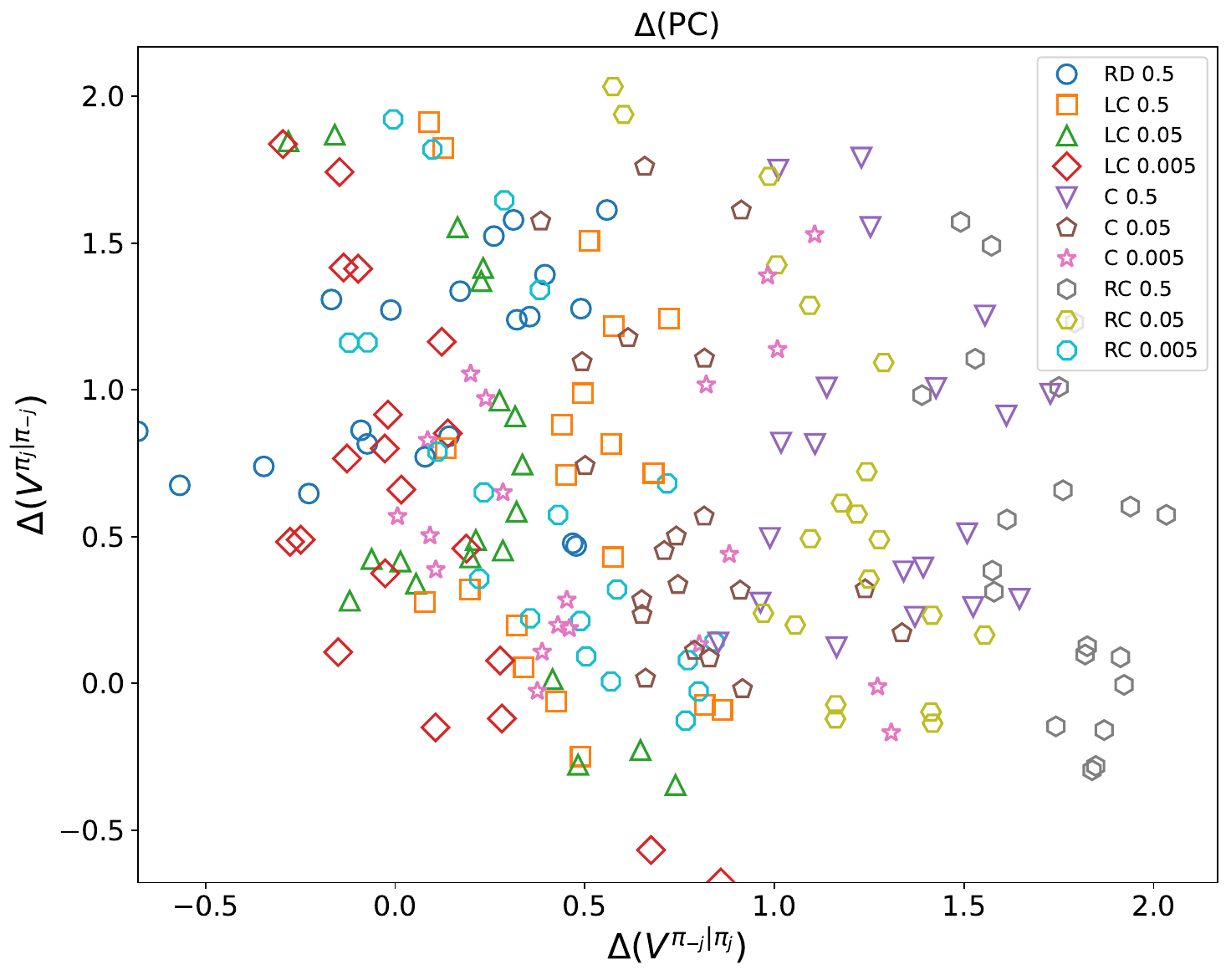}
\end{minipage}
\caption{
$\Delta_t$(CR) and $\Delta_t$(PC) for Q-learning with $c_1=c_2=0.8$, $f=1$ and  $t=10,000$.}\label{fig:delta_symQ_lo}
\end{figure*}

\clearpage
\paragraph{Symmetric costs $c_1=c_2=0.8$ with optimistic initialization $f=1$ and time horizon $t=10,000$.}
The best-response graph is provided in Fig.~\ref{fig:symQ_lo_br}. 
The average payoffs are given in Table~\ref{tab:symQ_lo_mean}. 
$\Delta_t$(CR) and $\Delta_t$(PC) are shown in Fig.~\ref{fig:delta_symQ_lo}.

\begin{table}[h!]
\centering
\caption{Average payoffs over 4,000 runs in the Q-learning meta-game payoff matrix. Standard errors are omitted as they are negligible in the order of $10^{-4}$. For reference, in this setting, the competitive and monopoly payoffs are $\bar{r}^N = 0.24$ and $\bar{r}^M = 0.41$.}\label{tab:symQ_lo_mean}
\resizebox{\textwidth}{!}{%
\begin{tabular}{lcccccccccc}
\toprule
 & RD 0.5 & LC 0.5 & LC 0.05 & LC 0.005 & C 0.5 & C 0.05 & C 0.005 & RC 0.5 & RC 0.05 & RC 0.005 \\
\midrule
RD 0.5 & 0.35, 0.35 & 0.29, 0.34 & 0.28, 0.33 & 0.28, 0.31 & 0.30, 0.35 & 0.29, 0.34 & 0.30, 0.33 & 0.30, 0.37 & 0.26, 0.35 & 0.24, 0.33 \\
LC 0.5 & 0.34, 0.29 & 0.30, 0.30 & 0.29, 0.30 & 0.28, 0.31 & 0.31, 0.31 & 0.28, 0.31 & 0.29, 0.30 & 0.32, 0.30 & 0.28, 0.32 & 0.25, 0.31 \\
LC 0.05 & 0.33, 0.28 & 0.30, 0.29 & 0.30, 0.30 & 0.29, 0.30 & 0.30, 0.29 & 0.28, 0.30 & 0.27, 0.29 & 0.32, 0.28 & 0.29, 0.30 & 0.25, 0.30 \\
LC 0.005 & 0.31, 0.28 & 0.31, 0.28 & 0.30, 0.29 & 0.29, 0.29 & 0.30, 0.29 & 0.28, 0.28 & 0.26, 0.29 & 0.32, 0.27 & 0.30, 0.28 & 0.26, 0.29 \\
C 0.5 & 0.35, 0.30 & 0.31, 0.31 & 0.29, 0.30 & 0.29, 0.30 & 0.31, 0.31 & 0.29, 0.31 & 0.29, 0.30 & 0.33, 0.30 & 0.28, 0.32 & 0.25, 0.32 \\
C 0.05 & 0.34, 0.29 & 0.31, 0.28 & 0.30, 0.28 & 0.28, 0.28 & 0.31, 0.29 & 0.27, 0.27 & 0.26, 0.28 & 0.32, 0.28 & 0.29, 0.27 & 0.25, 0.28 \\
C 0.005 & 0.33, 0.30 & 0.30, 0.29 & 0.29, 0.27 & 0.29, 0.26 & 0.30, 0.29 & 0.28, 0.26 & 0.26, 0.26 & 0.30, 0.30 & 0.29, 0.26 & 0.26, 0.25 \\
RC 0.5 & 0.37, 0.30 & 0.30, 0.32 & 0.28, 0.32 & 0.27, 0.32 & 0.30, 0.33 & 0.28, 0.32 & 0.30, 0.30 & 0.32, 0.32 & 0.26, 0.34 & 0.24, 0.33 \\
RC 0.05 & 0.35, 0.26 & 0.32, 0.28 & 0.30, 0.29 & 0.28, 0.30 & 0.32, 0.28 & 0.27, 0.29 & 0.26, 0.29 & 0.34, 0.26 & 0.29, 0.29 & 0.23, 0.30 \\
RC 0.005 & 0.33, 0.24 & 0.31, 0.25 & 0.30, 0.25 & 0.29, 0.26 & 0.32, 0.25 & 0.28, 0.25 & 0.25, 0.26 & 0.33, 0.24 & 0.30, 0.23 & 0.25, 0.25 \\
\bottomrule
\end{tabular}
}
\label{tab:q_learning_payoffs}
\end{table}

\begin{figure*}[h!]
    \centering
\includegraphics[width=0.4\linewidth]{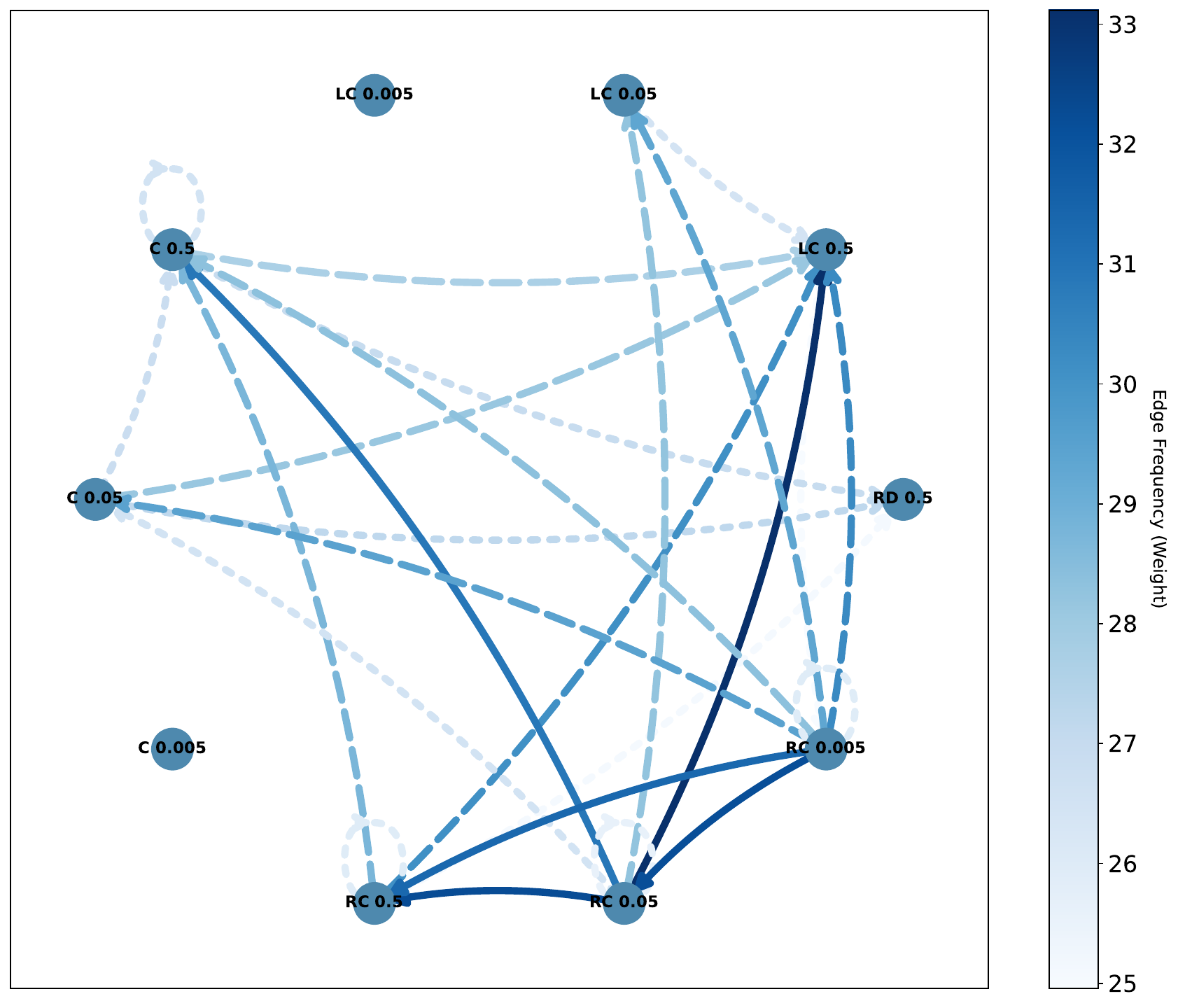}
    \caption{Best-response graph for Q-learning under symmetric costs $c_1=c_2=1$ with optimistic initialization $f=1$ and a time horizon of $t=3,000$.}
    \label{fig:sym_Q_BR_3000}
\end{figure*}

\begin{figure}[h!]
\begin{minipage}
{0.48\linewidth}
\centering
\includegraphics[width=\textwidth]{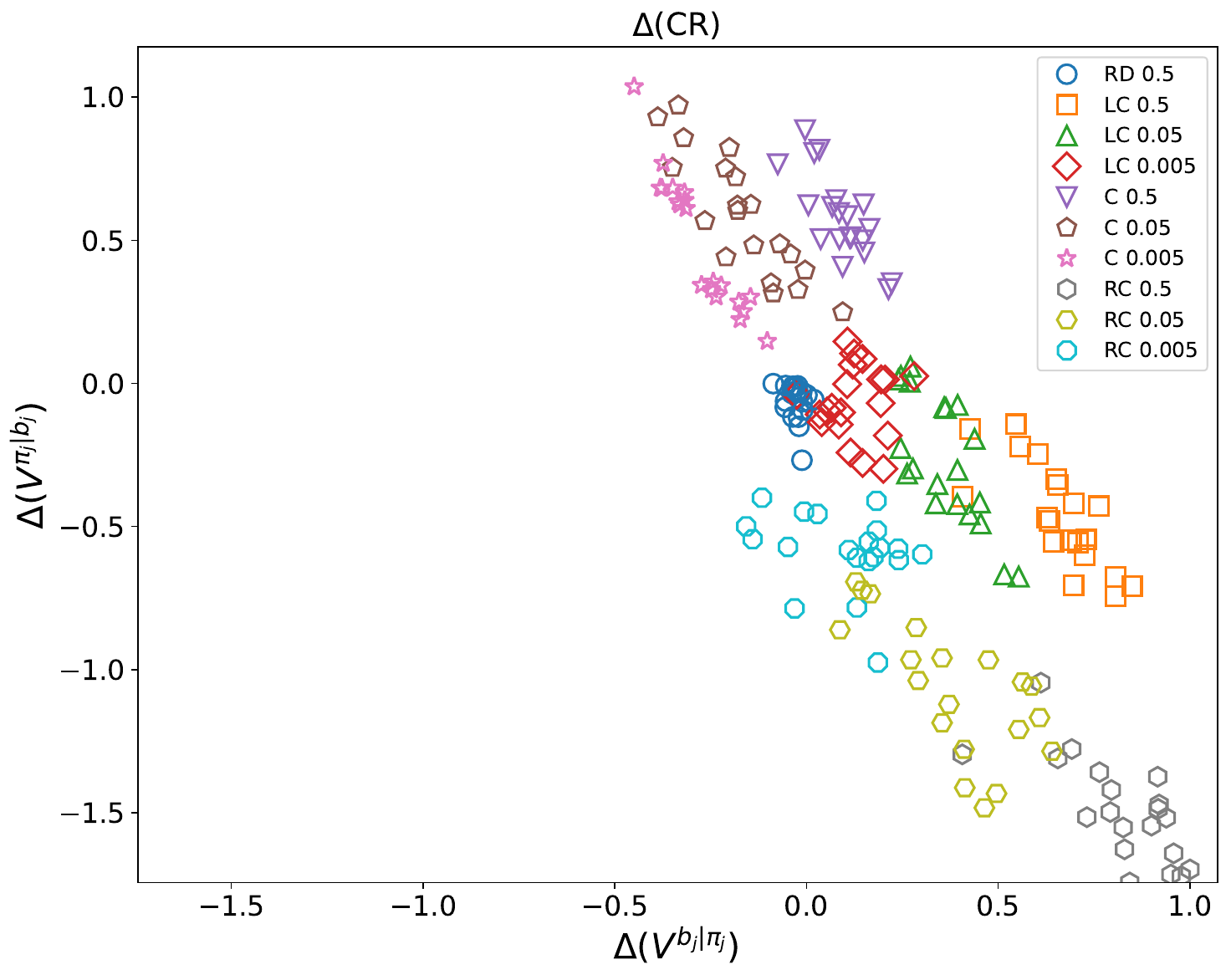}
\end{minipage}
\hfill
\begin{minipage}{0.48\linewidth}
\includegraphics[width=\textwidth]{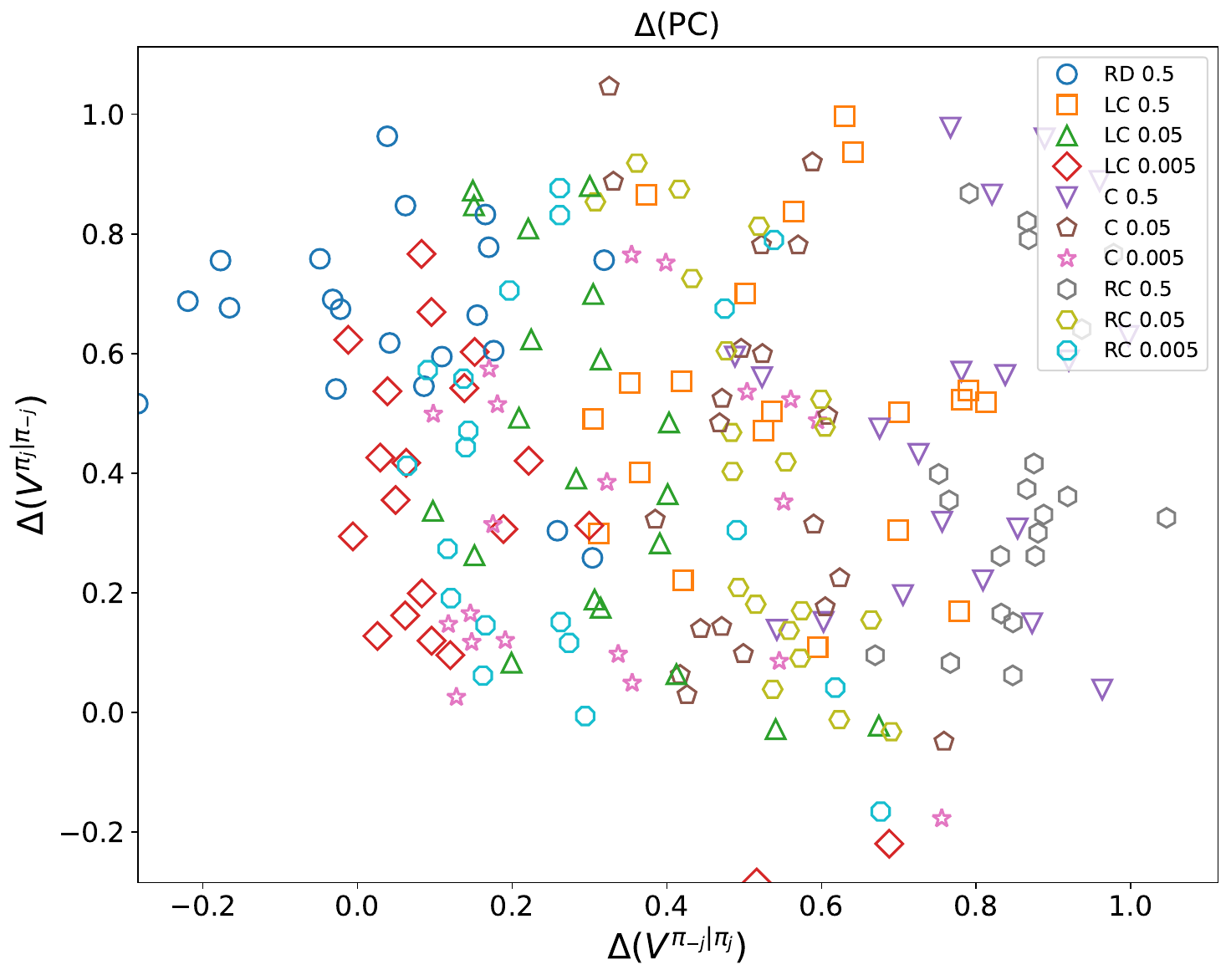}
\end{minipage}
\caption{$\Delta_t$(CR) and $\Delta_t$(PC) for Q-learning under symmetric costs $c_1=c_2=1$ with optimistic initialization $f=1$ and a time horizon of $t=3,000$.}\label{fig:delta_sym_Q_3000}
\end{figure}

\clearpage
\paragraph{Symmetric costs $c_1=c_2=1$ with optimistic initialization $f=1$ and time horizon $t=3,000$.}
The MSNE, NE-Regret and uniform score are given in Table~\ref{tab:3000}.
The best-response graph in Fig.~\ref{fig:sym_Q_BR_3000}. 
The average payoffs are given in Table~\ref{tab:sym_Q_3000_Q_mean}. $\Delta_t$(CR) and $\Delta_t$(PC) are shown in Fig.~\ref{fig:delta_sym_Q_3000}.

\begin{table*}[t]
\centering
\caption{Max-entropy MSNE, NE-Regret and Payoffs against uniformly drawn meta-strategies for Q-learning with shorter time horizon $t=3,000$.   For Uniform-score, we let $x \rightarrow  ((x-\bar{r}^N)/(\bar{r}^M-\bar{r}^N))\cdot 100\% $.}
\label{tab:3000}
 \resizebox{\linewidth}{!}{%
\begin{tabular}{ll|cccccccccc}
\toprule
\multicolumn{2}{c}{$t=3,000$, $f=1$} & RD 0.5 & LC 0.5 & LC 0.05 & LC 0.005 & C 0.5 & C 0.05 & C 0.005 & RC 0.5 & RC 0.05 & RC 0.005 \\
\midrule
MSNE & $c_1=c_2=1$ & 0.00 & 0.17 & 0.00 & 0.00 & 0.00 & 0.00 & 0.00 & 0.00 & 0.00 & \text{0.83}\\
NE-Regret ($\times 10^{-3}$) & $c_1=c_2=1$ & 7.75 $\pm$ 0.52 & \textbf{0.00 $\pm$ 0.59} & 2.77 $\pm$ 0.59 & 5.80 $\pm$ 0.53 & 3.31 $\pm$ 0.59 & 7.24 $\pm$ 0.45 & 7.86 $\pm$ 0.39 & \underline{2.21 $\pm$ 0.51} & 3.17 $\pm$ 0.44 & \textbf{0.00 $\pm$ 0.42} \\
Uniform Score & $c_1=c_2=1$ & 22.89 $\pm$ 0.27 & 29.93 $\pm$ 0.29 & 27.42 $\pm$ 0.32 & 23.90 $\pm$ 0.45 & 30.84 $\pm$ 0.21 & 29.92 $\pm$ 0.23 & 28.45 $\pm$ 0.28 & 31.50 $\pm$ 0.21 & \underline{32.28 $\pm$ 0.20} & \textbf{32.66 $\pm$ 0.24} \\
\midrule
\multirow{2}{*}{MSNE}  & $c_1=1.0$ & 0.00 & 0.00 & 0.00 & 0.00 & 0.00 & 0.00 & 0.00 & 0.00 & 0.00 & 1.00 \\
 & $c_2=0.8$  & 0.00 & 1.00 & 0.00 & 0.00 & 0.00 & 0.00 & 0.00 & 0.00 & 0.00 & 0.00  \\
 \multirow{2}{*}{NE-Regret ($\times 10^{-3}$)} & $c_1=1.0$ & 24.32 $\pm$ 0.64 & 15.06 $\pm$ 1.01 & 12.11 $\pm$ 0.76 & 11.76 $\pm$ 0.82 & 18.08 $\pm$ 0.82 & 12.58 $\pm$ 0.66 & \underline{6.73 $\pm$ 0.69} & 12.50 $\pm$ 0.81 & 7.44 $\pm$ 0.68 & \textbf{0.00 $\pm$ 0.74} \\
 & $c_2=0.8$  & 19.31 $\pm$ 0.95 & \textbf{0.00 $\pm$ 1.01} & \underline{0.31 $\pm$ 0.80} & 1.18 $\pm$ 0.61 & 19.48 $\pm$ 0.55 & 24.57 $\pm$ 0.39 & 22.59 $\pm$ 0.38 & 6.63 $\pm$ 0.98 & 18.77 $\pm$ 0.74 & 15.39 $\pm$ 0.50
\\
 \multirow{2}{*}{Uniform Score} & $c_1=1.0$ & -1.72 $\pm$ 0.30 & 2.09 $\pm$ 0.47 & 1.17 $\pm$ 0.38 & 1.12 $\pm$ 0.43 & -0.02 $\pm$ 0.28 & -0.16 $\pm$ 0.28 & 2.05 $\pm$ 0.27 & \underline{3.53 $\pm$ 0.49} 
& 3.10 $\pm$ 0.40 & \textbf{7.53 $\pm$ 0.25}\\
 & $c_2=0.8$  & 29.77 $\pm$ 0.28 & \textbf{57.20 $\pm$ 0.34} & \textbf{57.21 $\pm$ 0.34} & \textbf{57.77 $\pm$ 0.38} & 41.45 $\pm$ 0.20 & 38.17 $\pm$ 0.27 & 36.47 $\pm$ 0.27 & \underline{47.65 $\pm$ 
0.26} & 44.84 $\pm$ 0.26 & 43.55 $\pm$ 0.20 \\
\bottomrule
\end{tabular}%
}
\end{table*}

\begin{table}[h!]
\centering
\caption{Average payoffs over 4,000 runs in the Q-learning meta-game payoff matrix. Standard errors are omitted as they are negligible in the order of $10^{-4}$. For reference, in this setting, the competitive and monopoly payoffs are $\bar{r}^N = 0.22$ and $\bar{r}^M = 0.34$.}\label{tab:sym_Q_3000_Q_mean}
\resizebox{\textwidth}{!}{%
\begin{tabular}{lcccccccccc}
\toprule
 & RD 0.5 & LC 0.5 & LC 0.05 & LC 0.005 & C 0.5 & C 0.05 & C 0.005 & RC 0.5 & RC 0.05 & RC 0.005 \\ \midrule
RD 0.5 & 0.28, 0.28 & 0.25, 0.29 & 0.25, 0.29 & 0.26, 0.28 & 0.25, 0.30 & 0.24, 0.30 & 0.25, 0.30 & 0.24, 0.30 & 0.23, 0.29 & 0.23, 0.28 \\
LC 0.5 & 0.29, 0.25 & 0.27, 0.27 & 0.26, 0.27 & 0.26, 0.26 & 0.26, 0.27 & 0.25, 0.27 & 0.26, 0.26 & 0.25, 0.27 & 0.24, 0.28 & 0.24, 0.27 \\
LC 0.05 & 0.29, 0.25 & 0.27, 0.26 & 0.26, 0.26 & 0.25, 0.26 & 0.26, 0.26 & 0.24, 0.26 & 0.24, 0.26 & 0.26, 0.26 & 0.24, 0.26 & 0.24, 0.27 \\
LC 0.005 & 0.28, 0.26 & 0.26, 0.26 & 0.26, 0.25 & 0.25, 0.25 & 0.25, 0.27 & 0.24, 0.26 & 0.23, 0.26 & 0.25, 0.26 & 0.24, 0.26 & 0.23, 0.26 \\
C 0.5 & 0.30, 0.25 & 0.27, 0.26 & 0.26, 0.26 & 0.27, 0.25 & 0.27, 0.27 & 0.25, 0.26 & 0.25, 0.26 & 0.25, 0.26 & 0.23, 0.27 & 0.23, 0.27 \\
C 0.05 & 0.30, 0.24 & 0.27, 0.25 & 0.26, 0.24 & 0.26, 0.24 & 0.26, 0.25 & 0.25, 0.25 & 0.24, 0.25 & 0.26, 0.25 & 0.24, 0.25 & 0.23, 0.25 \\
C 0.005 & 0.30, 0.25 & 0.26, 0.26 & 0.26, 0.24 & 0.26, 0.23 & 0.26, 0.25 & 0.25, 0.24 & 0.24, 0.24 & 0.25, 0.26 & 0.24, 0.25 & 0.23, 0.25 \\
RC 0.5 & 0.30, 0.24 & 0.27, 0.25 & 0.26, 0.26 & 0.26, 0.25 & 0.27, 0.25 & 0.25, 0.26 & 0.26, 0.25 & 0.26, 0.26 & 0.24, 0.27 & 0.23, 0.26 \\
RC 0.05 & 0.29, 0.23 & 0.28, 0.24 & 0.26, 0.24 & 0.26, 0.24 & 0.27, 0.23 & 0.25, 0.24 & 0.25, 0.24 & 0.27, 0.24 & 0.24, 0.24 & 0.23, 0.25 \\
RC 0.005 & 0.28, 0.23 & 0.27, 0.24 & 0.27, 0.24 & 0.26, 0.23 & 0.27, 0.23 & 0.26, 0.23 & 0.25, 0.23 & 0.26, 0.23 & 0.25, 0.23 & 0.24, 0.24\\
\bottomrule
\end{tabular}%
}
\end{table}

\begin{figure*}[h!]
  \centering 
  \begin{minipage}{0.35\textwidth}
    \centering
    \includegraphics[width=\textwidth]{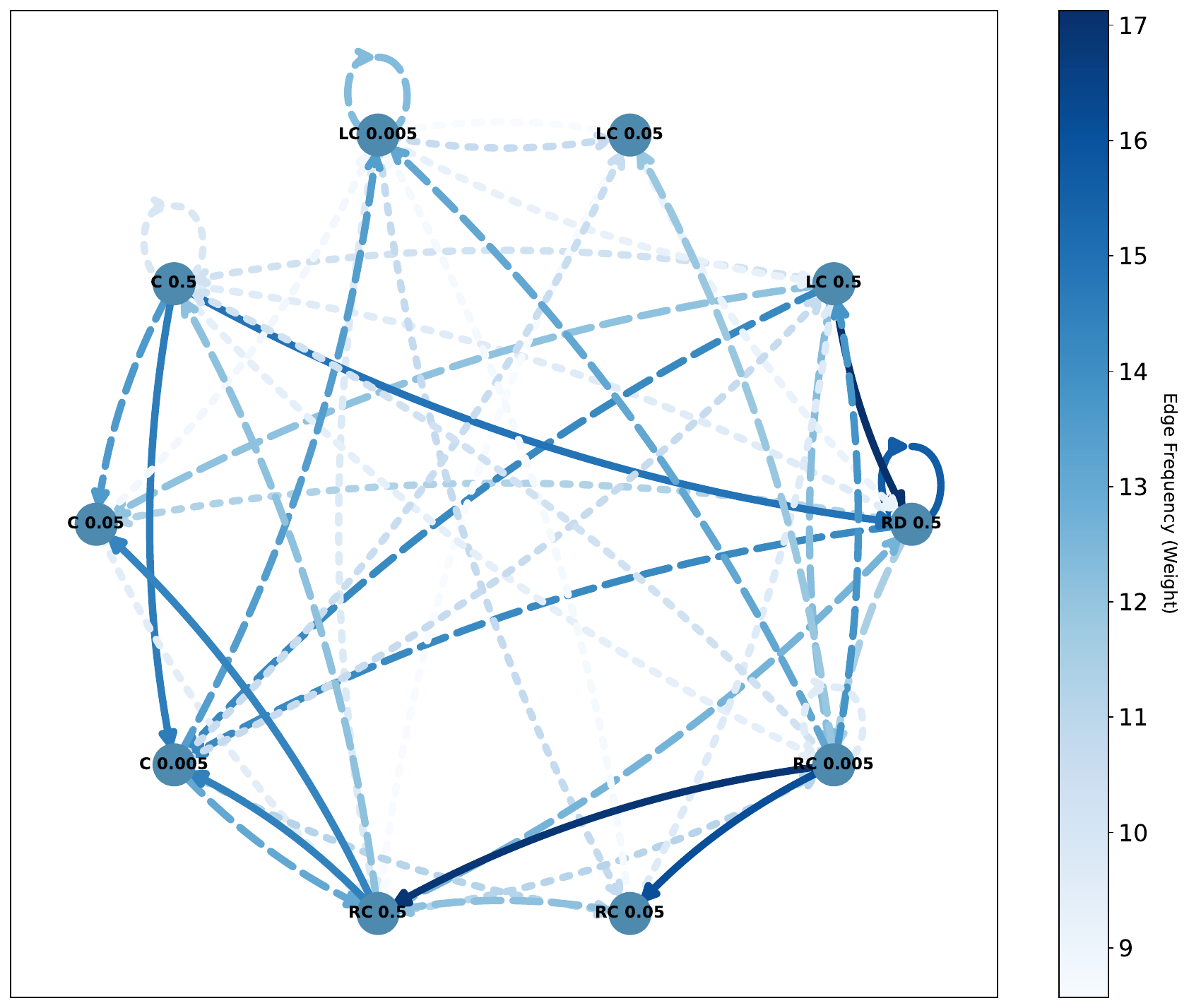}
    \captionof{figure}{Best-response graph  for $c_1=1, c_2=0.8$, $f=1$ and $t=10,000$. The directed edge $(u\rightarrow v)$ corresponds to $c_u = 1.0$ and $c_v = 0.8$.}\label{fig:asym_Q_BR_hi_2_lo}
  \end{minipage}
  \hfill
    \begin{minipage}{0.35\textwidth}
    \centering
    \includegraphics[width=\textwidth]{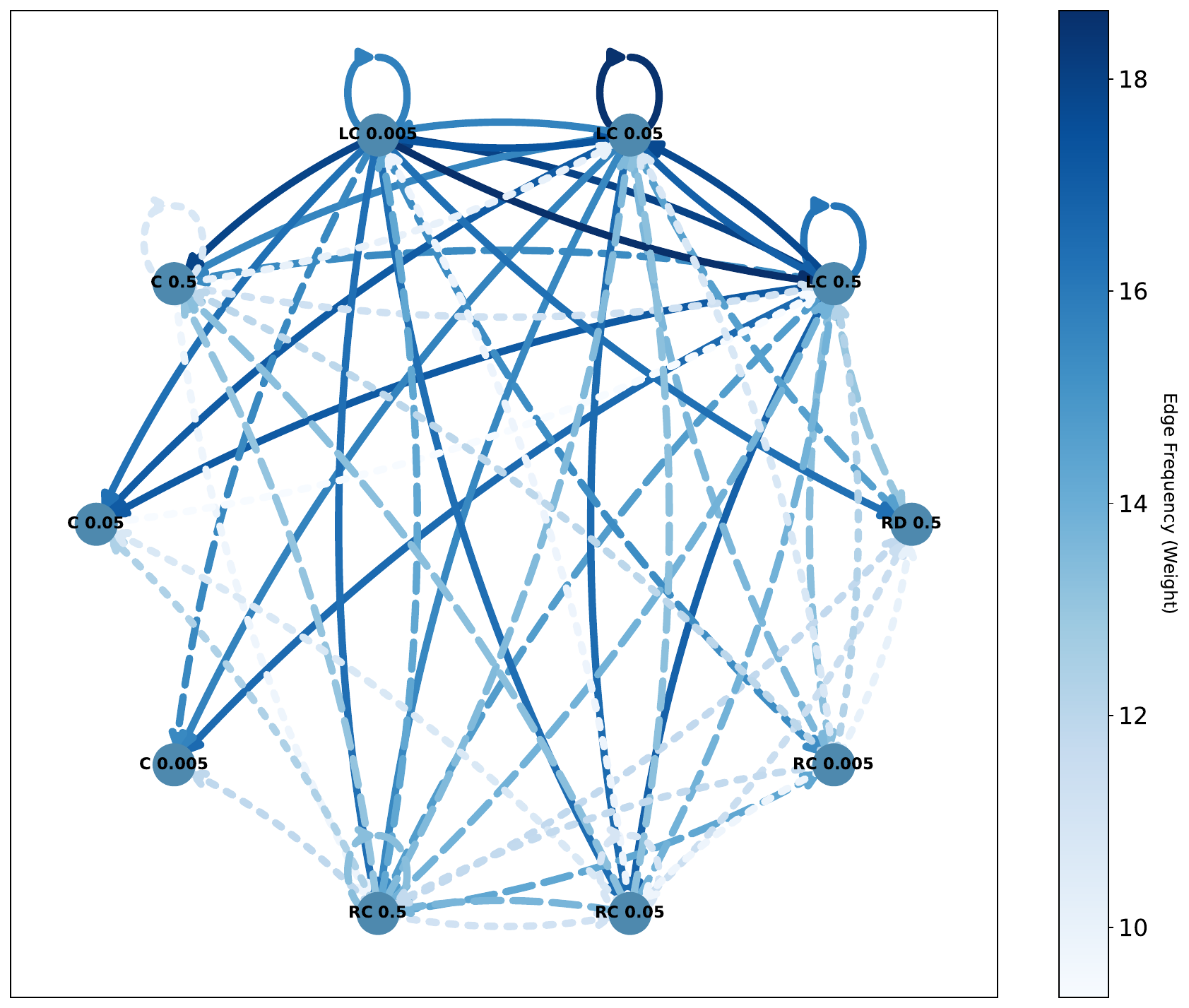}
    \captionof{figure}{
    Best-response graph for $c_1=1, c_2=0.8$, $f=1$ and $t=10,000$. The directed edge $(u\rightarrow v)$ corresponds to $c_u = 0.8$ and $c_v = 1.0$.}
\label{fig:asym_Q_BR_lo_2_hi}
  \end{minipage}
\end{figure*}

\begin{figure}[h!]
\begin{minipage}
{0.4\linewidth}
\centering
\includegraphics[width=\textwidth]{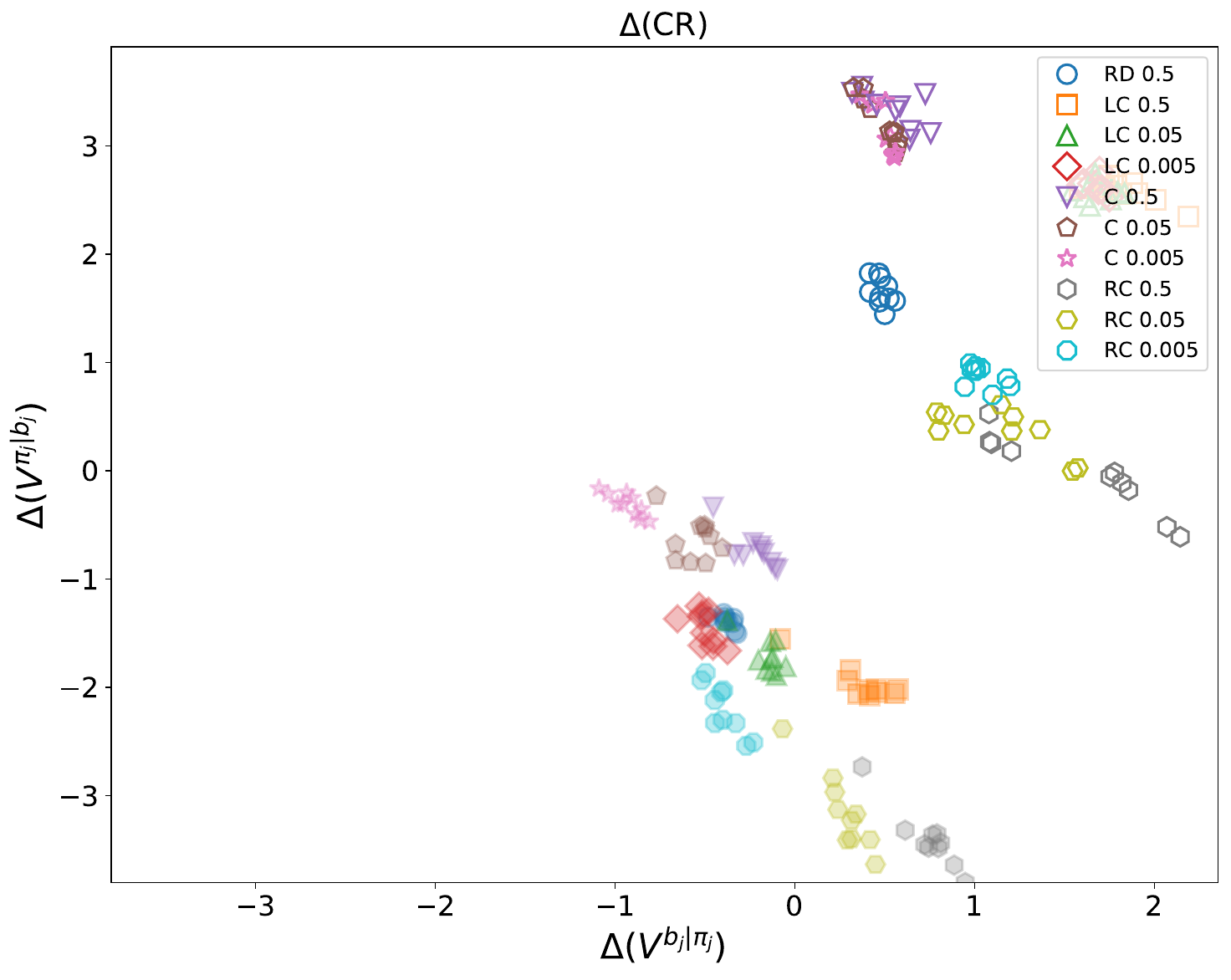}
\end{minipage}
\hfill
\begin{minipage}
{0.4\linewidth}
\includegraphics[width=\textwidth]{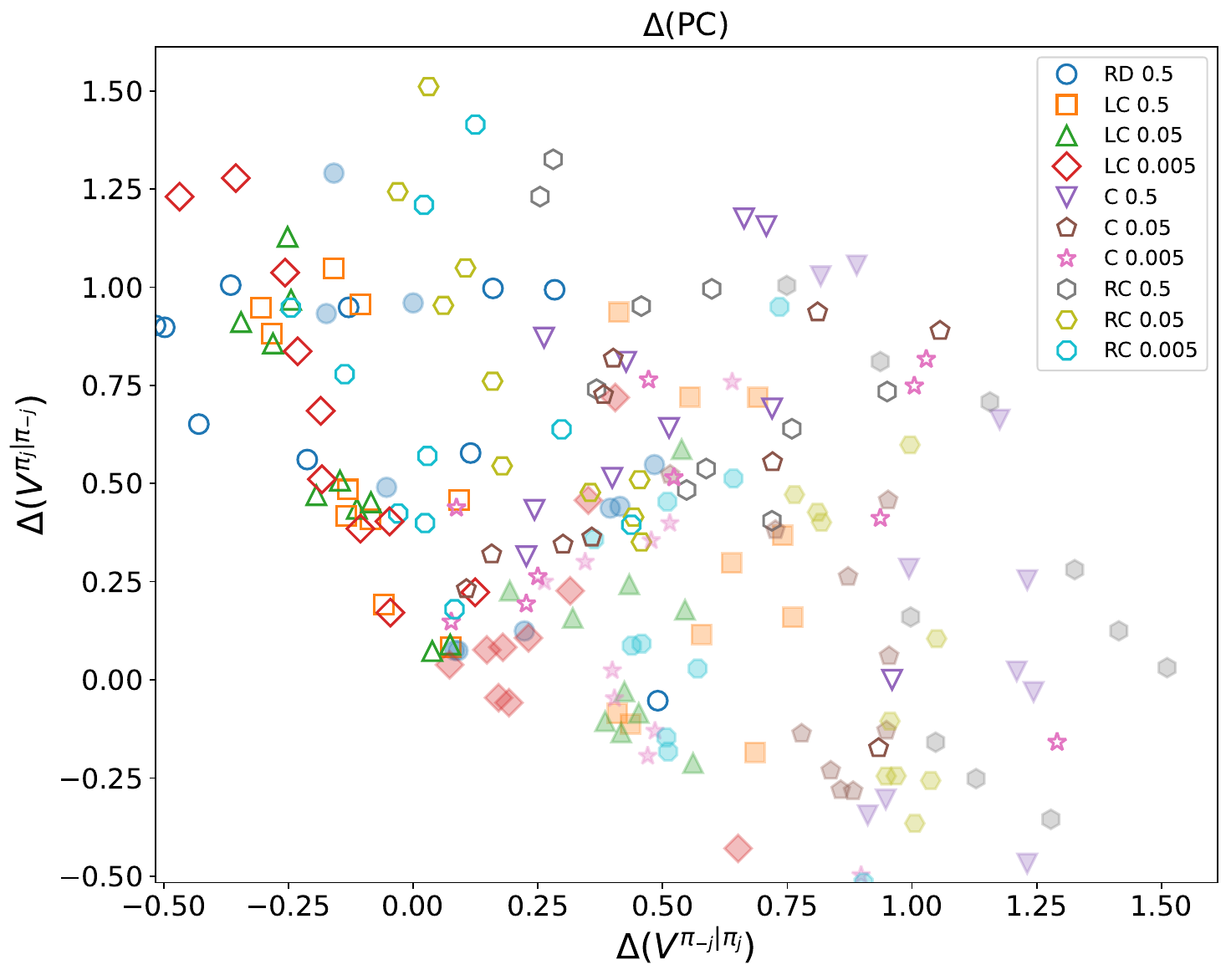}
\end{minipage}
\caption{
$\Delta_t$(CR) and $\Delta_t$(PC). The high-cost strategies are plotted with fillings.}\label{fig:delta_asym_Q}
\end{figure}

\clearpage
\paragraph{Asymmetric costs $c_1=1, c_2=0.8$ with optimistic initialization $f=1$ and time horizon $t=10,000$.}
Under asymmetric costs of $c_1=1$ and $c_2=0.8$ there will be two best-response graphs, one specifying the best response of the low-cost to the high-cost (Fig.~\ref{fig:asym_Q_BR_lo_2_hi}) and another of the high-cost to the low-cost (Fig.~\ref{fig:asym_Q_BR_hi_2_lo}). The average payoffs are given in Table~\ref{tab:asym_Q_mean}. $\Delta_t$(CR) and $\Delta_t$(PC) are shown in Fig.~\ref{fig:delta_asym_Q}.

\begin{table}[h!]
    \caption{Average payoffs over 4,000 runs in the payoff matrix of Q-learning-based strategies under asymmetric costs, $c_1 = 1$ and $c_2 = 0.8$, and $t = 10,000, f = 1$. Standard errors are omitted as they are negligible in the order of $10^{-4}$. For reference, in this setting, the competitive and monopoly payoffs for player 1 are $\bar{r}_{c=1.0}^N = 0.17$ and $\bar{r}_{c=1.0}^M = 0.30$, and for player 2 are $\bar{r}_{c=1.0}^N = 0.31$ and $\bar{r}_{c=0.8}^M = 0.44$. 
    }\label{tab:asym_Q_mean}
    \centering
    \resizebox{\textwidth}{!}{%
        \begin{tabular}{lcccccccccc}
        \toprule
         & {RD 0.5} & {LC 0.5} & {LC 0.05} & {LC 0.005} & {C 0.5} & {C 0.05} & {C 0.005} & {RC 0.5} & {RC 0.05} & {RC 0.005} \\
        \midrule
        {RD 0.5}   & 0.28, 0.38 & 0.17, 0.40 & 0.17, 0.40 & 0.17, 0.41 & 0.19, 0.39 & 0.19, 0.38 & 0.21, 0.37 & 0.16, 0.40 & 0.16, 0.40 & 0.17, 0.39 \\
        {LC 0.5}   & 0.28, 0.35 & 0.18, 0.40 & 0.18, 0.41 & 0.17, 0.41 & 0.19, 0.39 & 0.19, 0.38 & 0.21, 0.37 & 0.17, 0.40 & 0.15, 0.39 & 0.16, 0.39 \\
        {LC 0.05}  & 0.25, 0.35 & 0.18, 0.39 & 0.18, 0.40 & 0.18, 0.39 & 0.17, 0.37 & 0.17, 0.37 & 0.17, 0.37 & 0.18, 0.38 & 0.15, 0.38 & 0.14, 0.38 \\
        {LC 0.005} & 0.23, 0.36 & 0.19, 0.39 & 0.18, 0.38 & 0.18, 0.38 & 0.17, 0.37 & 0.17, 0.36 & 0.17, 0.36 & 0.18, 0.38 & 0.16, 0.37 & 0.15, 0.37 \\
        {C 0.5}    & 0.27, 0.35 & 0.18, 0.40 & 0.17, 0.40 & 0.17, 0.41 & 0.18, 0.38 & 0.19, 0.38 & 0.21, 0.36 & 0.17, 0.39 & 0.15, 0.39 & 0.15, 0.39 \\
        {C 0.05}   & 0.24, 0.35 & 0.18, 0.39 & 0.18, 0.39 & 0.18, 0.39 & 0.17, 0.37 & 0.17, 0.36 & 0.17, 0.36 & 0.18, 0.38 & 0.15, 0.37 & 0.14, 0.37 \\
        {C 0.005}  & 0.22, 0.35 & 0.19, 0.37 & 0.18, 0.37 & 0.19, 0.37 & 0.17, 0.36 & 0.16, 0.35 & 0.16, 0.34 & 0.19, 0.37 & 0.17, 0.35 & 0.15, 0.35 \\
        {RC 0.5}   & 0.27, 0.36 & 0.18, 0.40 & 0.18, 0.40 & 0.17, 0.40 & 0.19, 0.38 & 0.19, 0.37 & 0.21, 0.36 & 0.18, 0.39 & 0.16, 0.39 & 0.16, 0.39 \\
        {RC 0.05}  & 0.24, 0.35 & 0.19, 0.39 & 0.18, 0.39 & 0.18, 0.39 & 0.18, 0.36 & 0.18, 0.36 & 0.17, 0.36 & 0.19, 0.37 & 0.16, 0.37 & 0.15, 0.37 \\
        {RC 0.005} & 0.23, 0.34 & 0.20, 0.36 & 0.19, 0.36 & 0.19, 0.36 & 0.18, 0.35 & 0.17, 0.34 & 0.16, 0.34 & 0.21, 0.36 & 0.18, 0.34 & 0.15, 0.34 \\
        \bottomrule
        \end{tabular}%
    }
\end{table}

\begin{figure*}[h]
    \centering
        \includegraphics[width=0.4\textwidth]{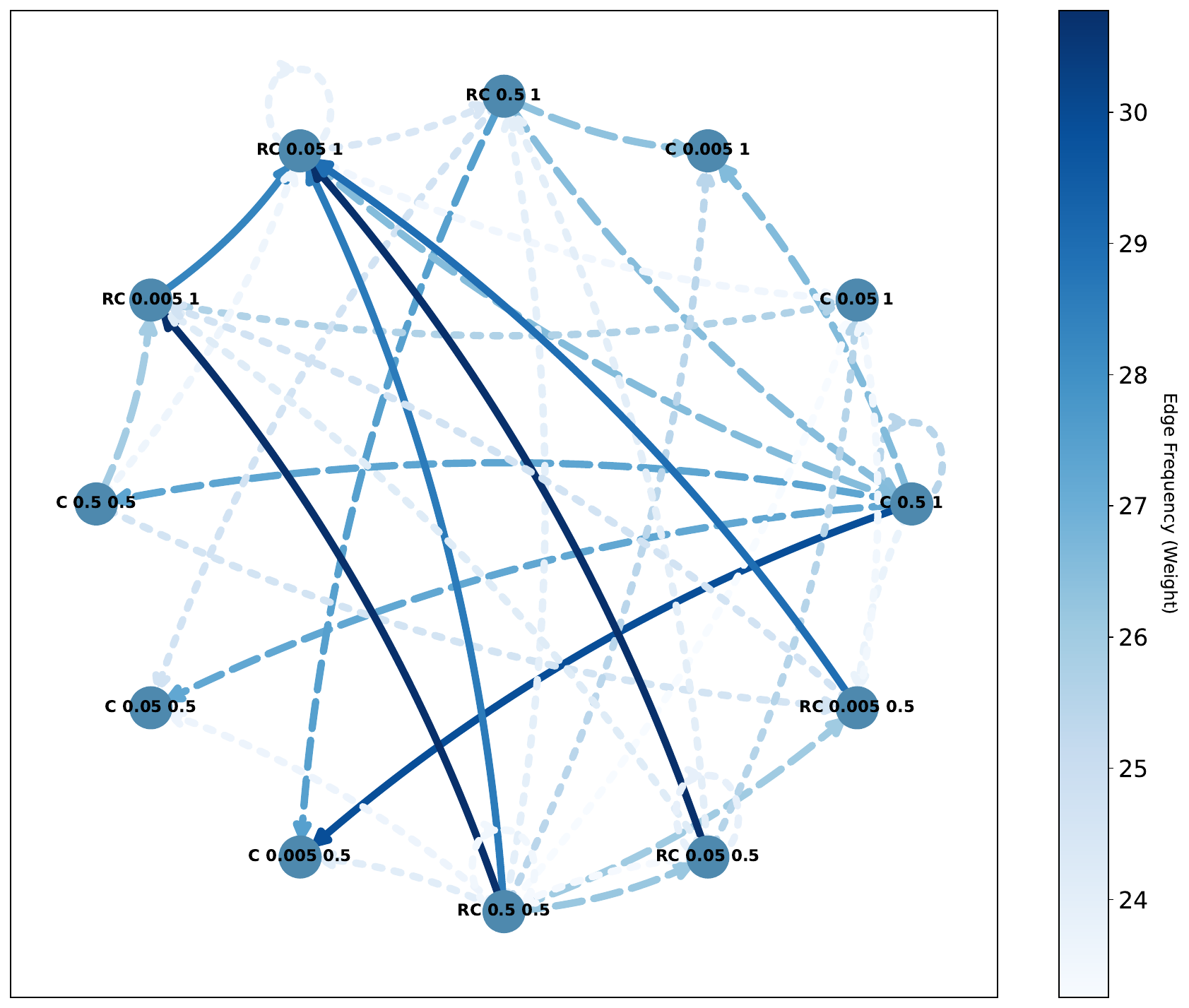}
\caption{Best-response graph for Q-learning with $c_1=c_2=1$, $f\in \{1, 0.5\}$ and $t=10,000$.}
    \label{fig:crossQ_BR}
\end{figure*}

\begin{figure}[h!]
\begin{minipage}
{0.48\linewidth}
\centering
\includegraphics[width=\textwidth]{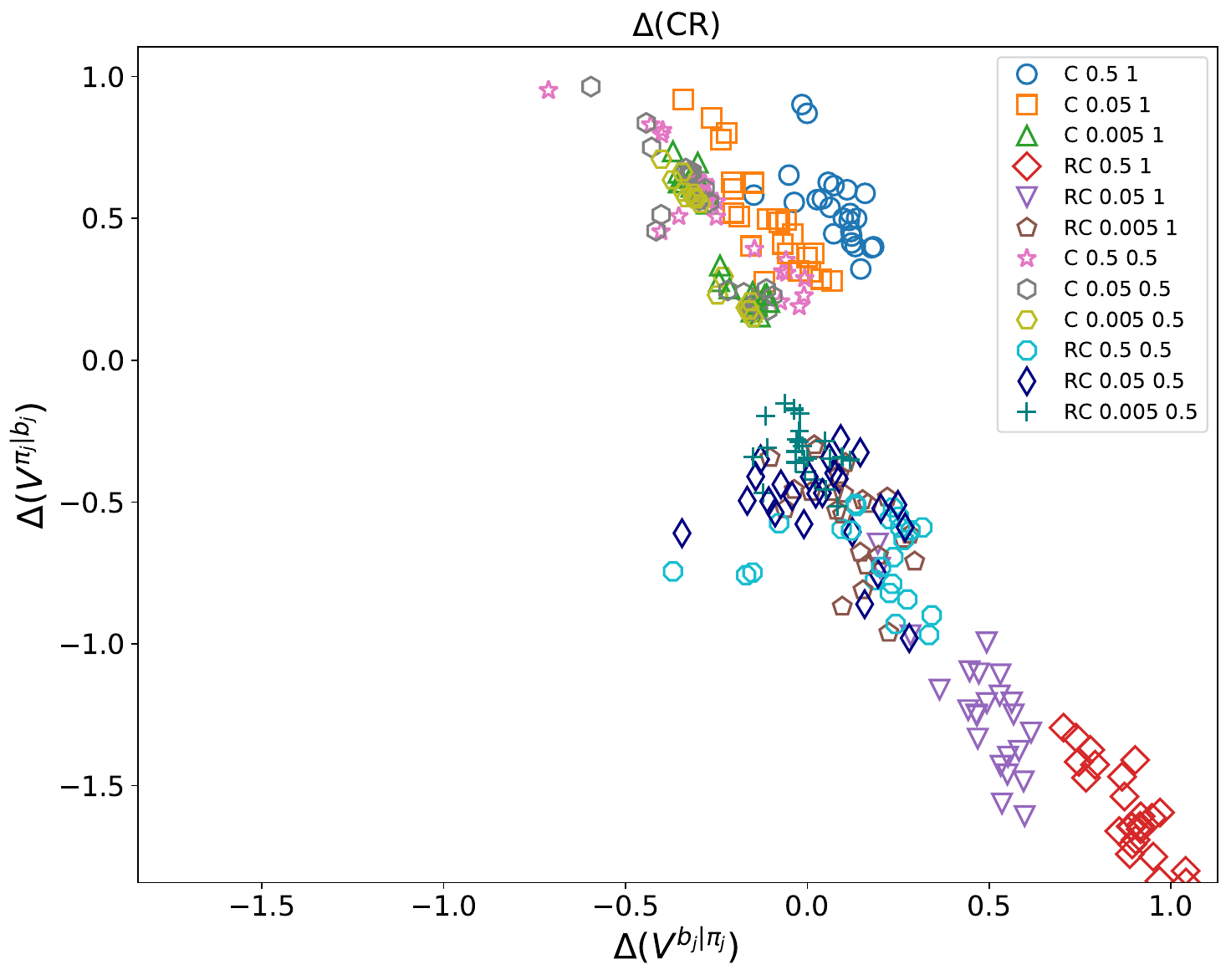}
\end{minipage}
\hfill
\begin{minipage}
{0.48\linewidth}
\includegraphics[width=\textwidth]{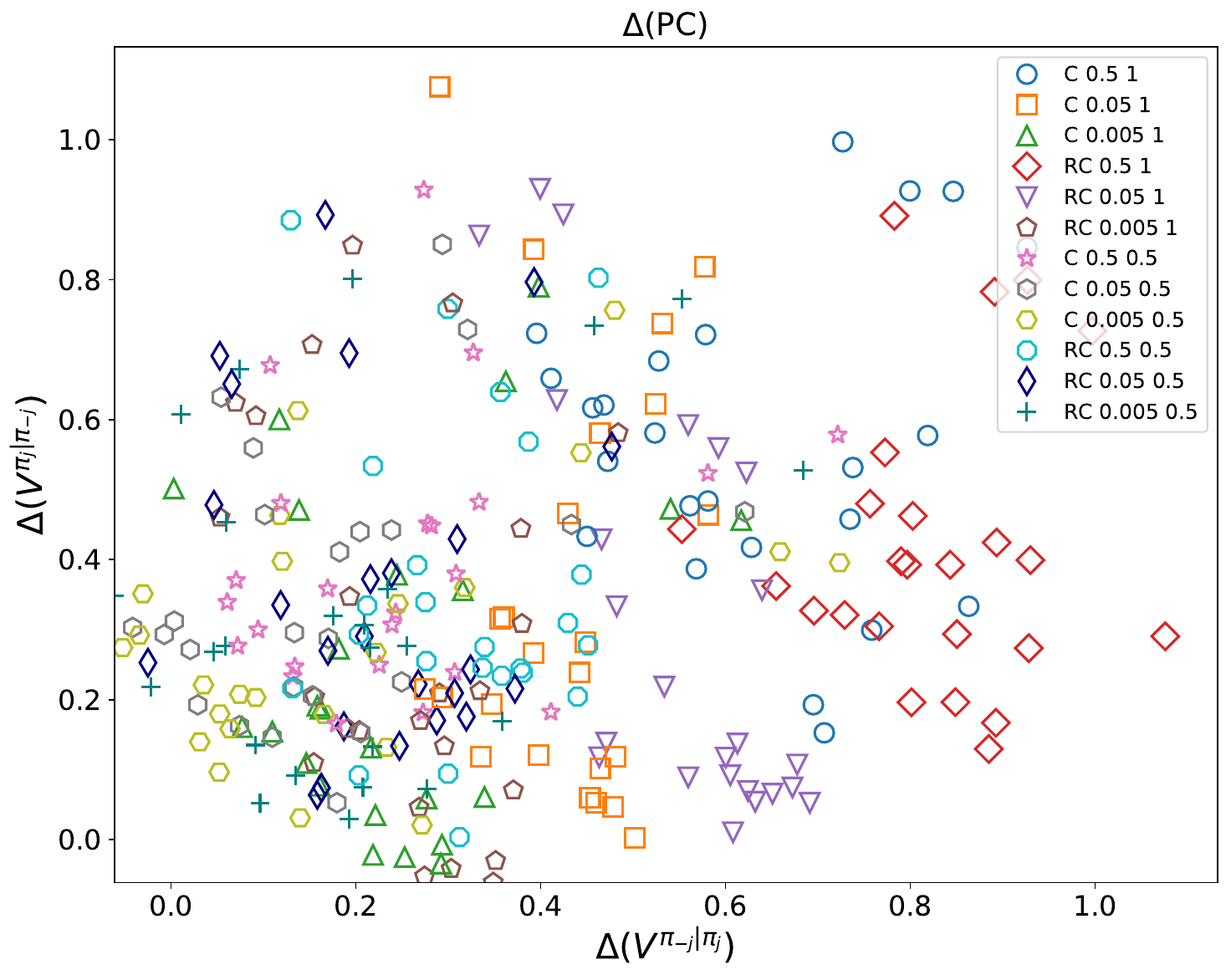}
\end{minipage}
\caption{
$\Delta_t$(CR) and $\Delta_t$(PC) for Q-learning with $c_1=c_2=1$, $f\in \{1, 0.5\}$ and $t=10,000$.}\label{fig:crossQ_delta}
\end{figure}

\clearpage
\paragraph{Symmetric costs $c_1=c_2=1$ with either optimistic or pessimistic initializations ($f\in \{1, 0.5\}$) under time horizon $t=10,000$.}
The best-response graph is provided in Fig.~\ref{fig:crossQ_BR}. 
The average payoffs are given in Table~\ref{tab:crossQ_mean}. $\Delta_t$(CR) and $\Delta_t$(PC)  are shown in Fig.~\ref{fig:crossQ_delta}.

\begin{table}[h]
\centering
\caption{Average payoffs over 4,000 runs in the payoff matrix of Q-learning-based strategies under symmetric costs $c_1 = c_2 = 1$ with either optimistic initialization $f=1$ or pessimistic initialization $f=0.5$. Time horizon $t = 10,000$. For reference, in this setting, the competitive and monopoly payoffs are $\bar{r}^N = 0.22$ and $\bar{r}^M = 0.34$.}\label{tab:crossQ_mean}
\resizebox{\textwidth}{!}{%
\begin{tabular}{l|cccccccccccc}
\toprule
 & C 0.5 f=1 & C 0.05 f=1 & C 0.005 f=1 & RC 0.5 f=1 & RC 0.05 f=1 & RC 0.005 f=1 & C 0.5 f=0.5 & C 0.05 f=0.5 & C 0.005 f=0.5 & RC 0.5 f=0.5 & RC 0.05 f=0.5 & RC 0.005 f=0.5 \\ \midrule
C 0.5 f=1 & 0.28, 0.28 & 0.26, 0.27 & 0.27, 0.26 & 0.27, 0.28 & 0.24, 0.28 & 0.24, 0.28 & 0.27, 0.27 & 0.27, 0.27 & 0.27, 0.26 & 0.24, 0.28 & 0.24, 0.28 & 0.25, 0.27 \\
C 0.05 f=1 & 0.27, 0.26 & 0.26, 0.26 & 0.25, 0.26 & 0.27, 0.26 & 0.25, 0.26 & 0.23, 0.26 & 0.25, 0.26 & 0.25, 0.26 & 0.25, 0.26 & 0.24, 0.26 & 0.23, 0.26 & 0.23, 0.26 \\
C 0.005 f=1 & 0.26, 0.27 & 0.26, 0.25 & 0.25, 0.25 & 0.26, 0.26 & 0.25, 0.25 & 0.23, 0.25 & 0.25, 0.26 & 0.24, 0.25 & 0.24, 0.25 & 0.24, 0.25 & 0.23, 0.26 & 0.23, 0.25 \\
RC 0.5 f=1 & 0.28, 0.27 & 0.26, 0.27 & 0.26, 0.26 & 0.27, 0.27 & 0.25, 0.28 & 0.24, 0.27 & 0.26, 0.27 & 0.26, 0.26 & 0.27, 0.26 & 0.25, 0.27 & 0.25, 0.28 & 0.25, 0.27 \\
RC 0.05 f=1 & 0.28, 0.24 & 0.26, 0.25 & 0.25, 0.25 & 0.28, 0.25 & 0.25, 0.25 & 0.24, 0.26 & 0.26, 0.26 & 0.25, 0.25 & 0.25, 0.25 & 0.25, 0.26 & 0.24, 0.26 & 0.23, 0.26 \\
RC 0.005 f=1 & 0.28, 0.24 & 0.26, 0.23 & 0.25, 0.23 & 0.27, 0.24 & 0.26, 0.24 & 0.24, 0.24 & 0.25, 0.25 & 0.25, 0.24 & 0.25, 0.24 & 0.25, 0.25 & 0.24, 0.25 & 0.23, 0.24 \\
C 0.5 f=0.5 & 0.27, 0.27 & 0.26, 0.25 & 0.26, 0.25 & 0.27, 0.26 & 0.26, 0.26 & 0.25, 0.25 & 0.26, 0.26 & 0.25, 0.25 & 0.25, 0.24 & 0.25, 0.26 & 0.25, 0.26 & 0.25, 0.25 \\
C 0.05 f=0.5 & 0.27, 0.27 & 0.26, 0.25 & 0.25, 0.24 & 0.26, 0.26 & 0.25, 0.25 & 0.24, 0.25 & 0.25, 0.25 & 0.25, 0.25 & 0.25, 0.24 & 0.25, 0.26 & 0.24, 0.25 & 0.24, 0.25 \\
C 0.005 f=0.5 & 0.26, 0.27 & 0.26, 0.25 & 0.25, 0.24 & 0.26, 0.27 & 0.25, 0.25 & 0.24, 0.25 & 0.24, 0.25 & 0.24, 0.25 & 0.24, 0.24 & 0.24, 0.26 & 0.24, 0.25 & 0.23, 0.25 \\
RC 0.5 f=0.5 & 0.28, 0.24 & 0.26, 0.24 & 0.26, 0.24 & 0.27, 0.25 & 0.26, 0.25 & 0.25, 0.25 & 0.26, 0.25 & 0.26, 0.25 & 0.26, 0.24 & 0.25, 0.25 & 0.25, 0.25 & 0.25, 0.25 \\
RC 0.05 f=0.5 & 0.28, 0.24 & 0.26, 0.23 & 0.26, 0.23 & 0.28, 0.25 & 0.26, 0.24 & 0.25, 0.24 & 0.26, 0.25 & 0.25, 0.24 & 0.25, 0.24 & 0.25, 0.25 & 0.25, 0.25 & 0.24, 0.25 \\
RC 0.005 f=0.5 & 0.27, 0.25 & 0.26, 0.23 & 0.25, 0.23 & 0.27, 0.25 & 0.26, 0.23 & 0.24, 0.23 & 0.25, 0.25 & 0.25, 0.24 & 0.25, 0.23 & 0.25, 0.25 & 0.25, 0.24 & 0.24, 0.24
\\
\bottomrule
\end{tabular}%
}
\end{table}

\clearpage
\section{Deferred Content and Results from Section~\ref{exp:ucb} on UCB} 

\subsection{PC, CR and the Categorization of Policies Pretrained with UCB}

\begin{figure*}[ht]
    \centering
    \includegraphics[trim={1cm 1.5cm 1cm 2cm},clip, width=0.6\linewidth]{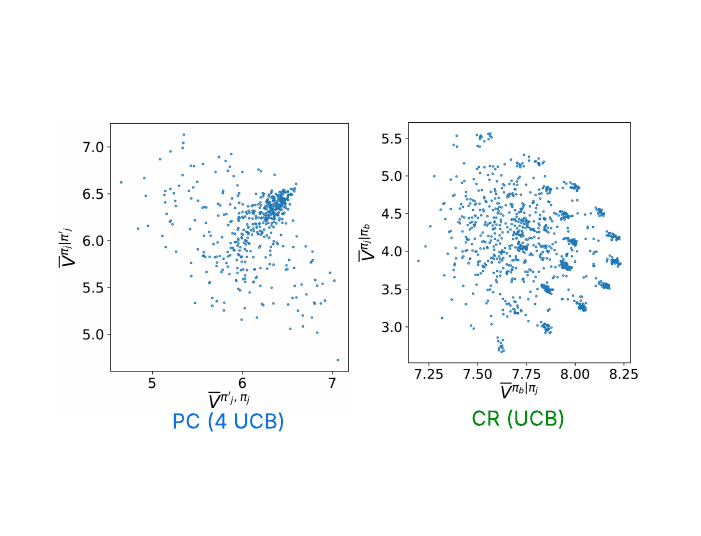}
    \caption{The PC (Def.~\ref{def:PC}) and CR (Def.~\ref{def:CR}) of 1000 pretrained UCB policies with the setting described in Sec.~\ref{exp:ucb}.}
    \label{fig:pc_cr_UCB}
\end{figure*}

 We show $\overline{V}^{\pi_j, \pi_j'}$ v.s.~ $\overline{V}^{\pi_j, \pi_b}$ and the categorization of LC, C and RC on the right of Fig.~\ref{fig:cat_Q_UCB}.
In Fig.~\ref{fig:pc_cr_UCB}, we show the PC and CR of pretrained policies.

\subsection{Detailed Results of UCB}\label{apd:ucb_matrices}


\begin{figure*}[ht]
    \centering
\includegraphics[width=0.4\linewidth]{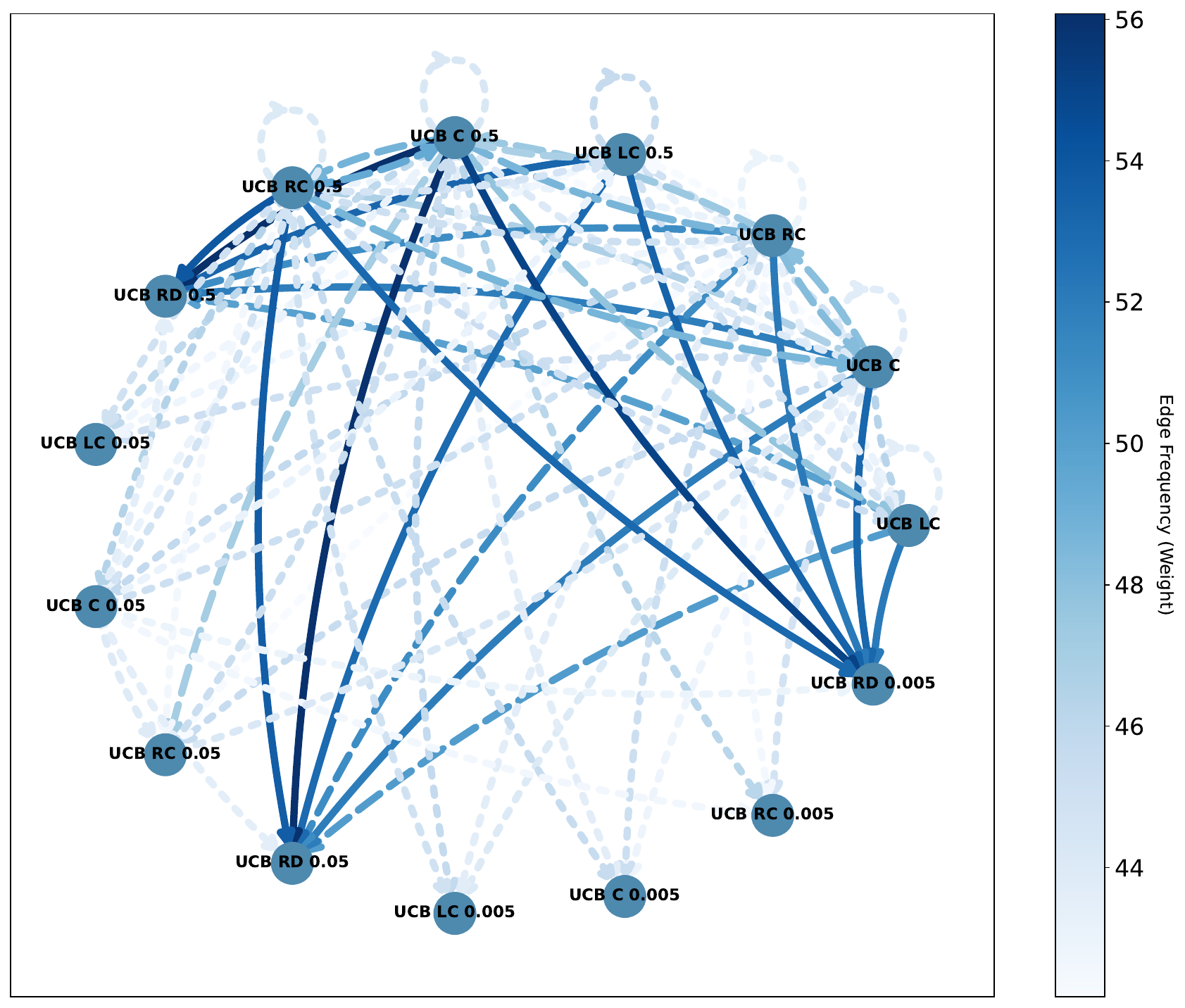}
    
    \caption{Best-response graph for UCBs. $c_1=c_2=1$, $t=10,000$.}
    \label{fig:UCB_br}
\end{figure*}

\begin{figure}[ht]
\begin{minipage}
{1.0\linewidth}
\centering
\includegraphics[width=0.48\textwidth]{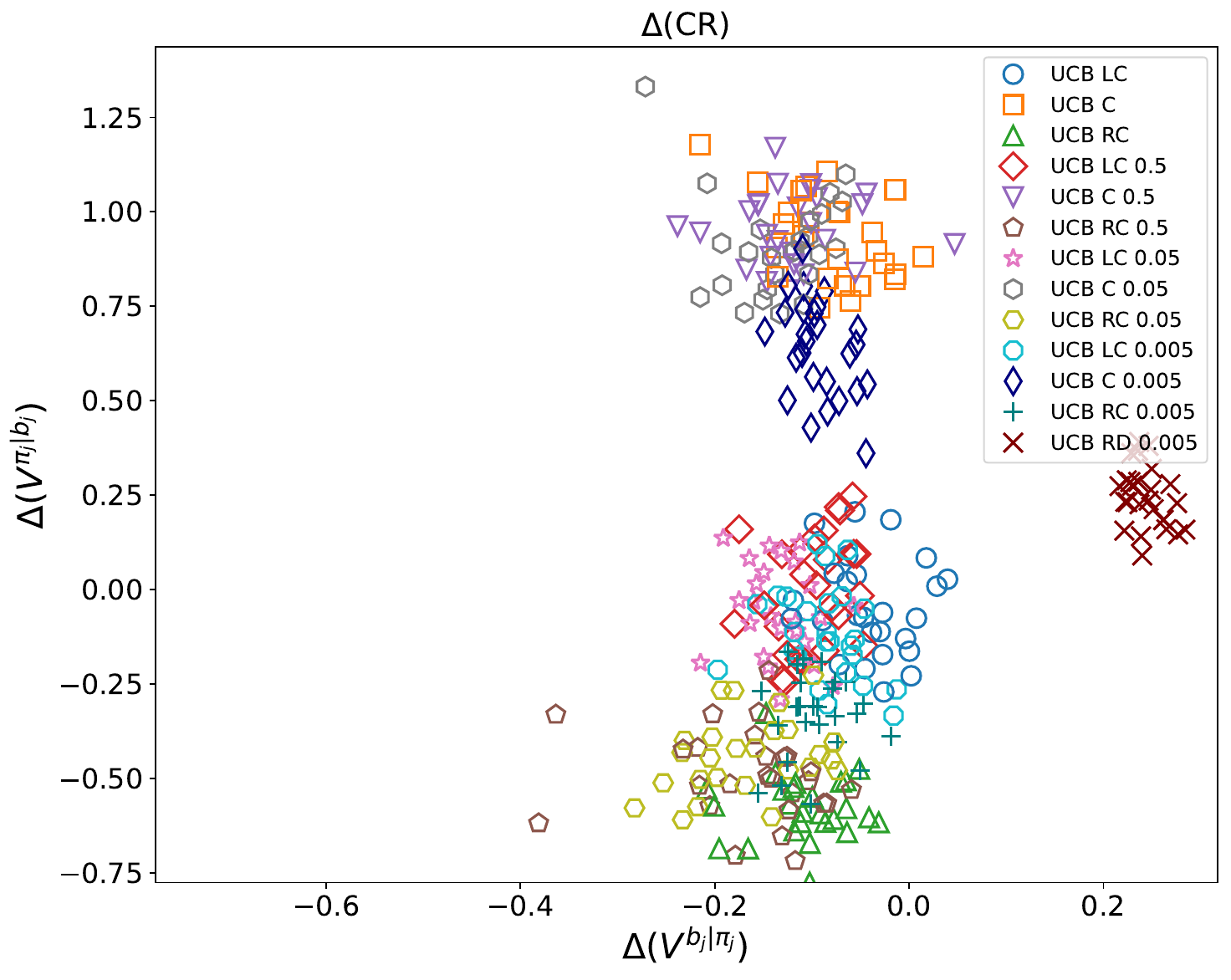}
\vrule
\includegraphics[width=0.48\textwidth]{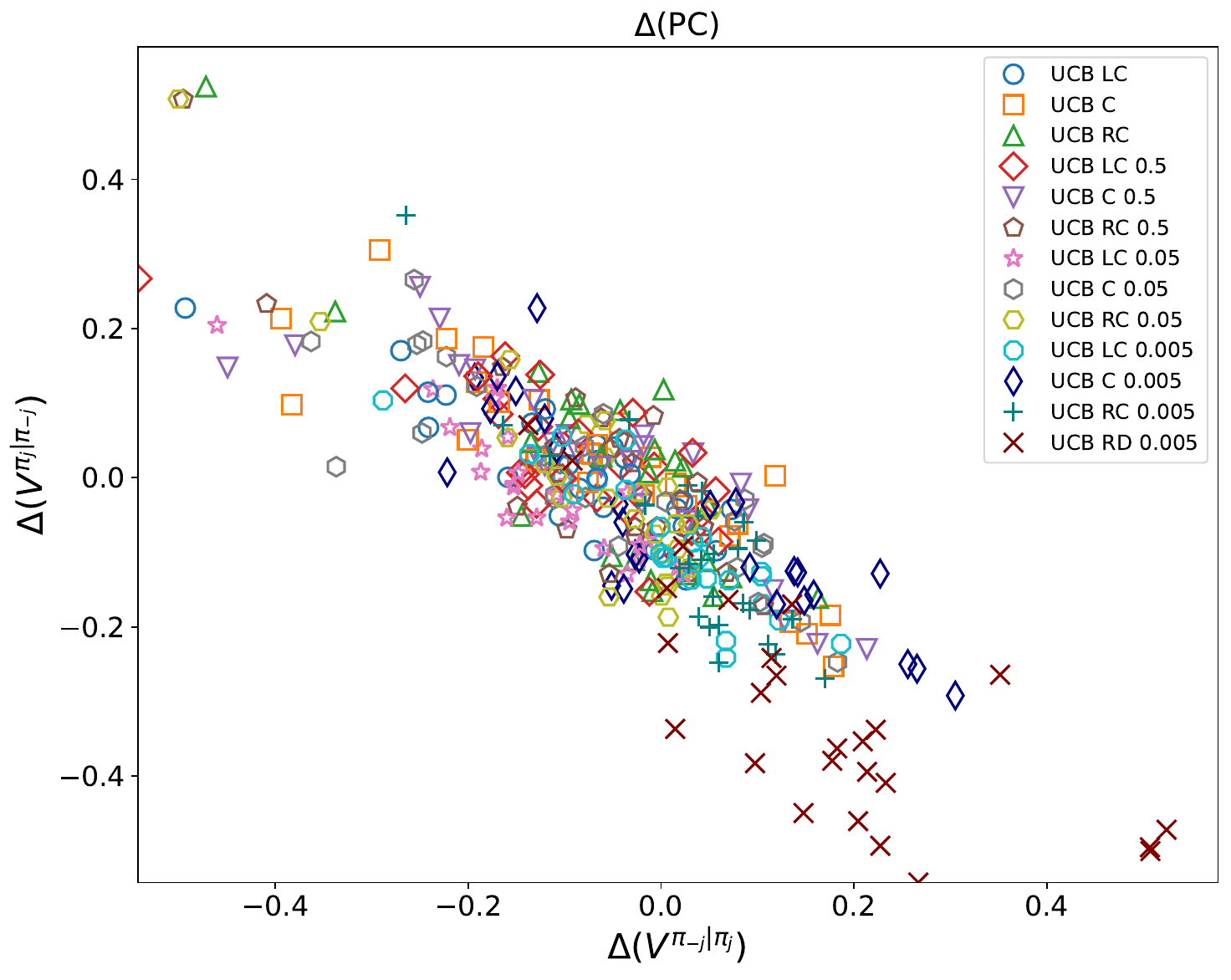}
\end{minipage}
\caption{
Changes in CR ($\Delta_t$(CR)) and PC ($\Delta_t$(PC))  over time for UCB.}\label{fig:delta_UCB}
\end{figure}

The best-response graph is given in Fig.~\ref{fig:UCB_br}. The average payoffs are given in Table \ref{tab:ucb_mean}. 
The differences in CR and PC between policies of a strategy at $t=10,000$ and $t=0$, denoted as $\Delta_t$(CR) and $\Delta_t$(PC), are shown in Fig.~\ref{fig:delta_UCB}.

\begin{table}[h!]
\centering
\caption{Average payoffs over 4,000 runs in the payoff matrix of UCB-based strategies, with symmetric cost $c_1 = c_2 = 1$ and time horizon $t = 10,000$. For reference, the competitive payoff is $\bar{r}^N=0.22$, and the monopoly payoff is $\bar{r}^M=0.34$.}\label{tab:ucb_mean}
\resizebox{\textwidth}{!}{%
\begin{tabular}{l|ccccccccccccc}
\toprule
 & {UCB LC} & {UCB C} & {UCB RC} & {UCB LC 0.5} & {UCB C 0.5} & {UCB RC 0.5} & {UCB LC 0.05} & {UCB C 0.05} & {UCB RC 0.05} & {UCB LC 0.005} & {UCB C 0.005} & {UCB RC 0.005} & {UCB RD 0.005} \\ \midrule
{UCB LC} & 0.29, 0.29 & 0.29, 0.29 & 0.29, 0.29 & 0.29, 0.29 & 0.29, 0.29 & 0.29, 0.29 & 0.29, 0.30 & 0.29, 0.29 & 0.29, 0.28 & 0.29, 0.29 & 0.29, 0.28 & 0.29, 0.28 & 0.29, 0.26 \\
{UCB C} & 0.29, 0.29 & 0.30, 0.30 & 0.30, 0.30 & 0.29, 0.29 & 0.30, 0.30 & 0.29, 0.30 & 0.29, 0.29 & 0.30, 0.30 & 0.30, 0.29 & 0.30, 0.28 & 0.30, 0.29 & 0.30, 0.29 & 0.29, 0.26 \\
{UCB RC} & 0.29, 0.28 & 0.30, 0.30 & 0.29, 0.29 & 0.28, 0.28 & 0.30, 0.30 & 0.30, 0.30 & 0.29, 0.29 & 0.30, 0.30 & 0.30, 0.28 & 0.29, 0.29 & 0.30, 0.28 & 0.29, 0.28 & 0.29, 0.26 \\
{UCB LC 0.5} & 0.29, 0.29 & 0.29, 0.29 & 0.29, 0.28 & 0.29, 0.29 & 0.29, 0.29 & 0.29, 0.29 & 0.29, 0.30 & 0.29, 0.29 & 0.29, 0.28 & 0.30, 0.28 & 0.29, 0.28 & 0.29, 0.26 & 0.29, 0.29 \\
{UCB C 0.5} & 0.29, 0.29 & 0.30, 0.30 & 0.30, 0.29 & 0.29, 0.29 & 0.30, 0.30 & 0.30, 0.30 & 0.29, 0.29 & 0.30, 0.30 & 0.30, 0.29 & 0.30, 0.28 & 0.30, 0.29 & 0.30, 0.29 & 0.29, 0.26 \\
{UCB RC 0.5} & 0.29, 0.28 & 0.30, 0.30 & 0.30, 0.29 & 0.29, 0.28 & 0.30, 0.30 & 0.29, 0.28 & 0.30, 0.30 & 0.30, 0.29 & 0.29, 0.28 & 0.30, 0.29 & 0.30, 0.29 & 0.29, 0.26 & 0.30, 0.29 \\
{UCB LC 0.05} & 0.30, 0.29 & 0.29, 0.29 & 0.29, 0.29 & 0.30, 0.29 & 0.29, 0.29 & 0.28, 0.29 & 0.29, 0.29 & 0.29, 0.29 & 0.29, 0.28 & 0.29, 0.29 & 0.29, 0.28 & 0.29, 0.28 & 0.29, 0.26 \\
{UCB C 0.05} & 0.29, 0.29 & 0.30, 0.30 & 0.29, 0.29 & 0.29, 0.29 & 0.30, 0.30 & 0.29, 0.29 & 0.30, 0.30 & 0.29, 0.29 & 0.30, 0.30 & 0.30, 0.29 & 0.29, 0.26 & 0.29, 0.29 & 0.29, 0.26 \\
{UCB RC 0.05} & 0.28, 0.28 & 0.29, 0.30 & 0.28, 0.28 & 0.28, 0.28 & 0.29, 0.30 & 0.29, 0.29 & 0.28, 0.29 & 0.29, 0.29 & 0.28, 0.28 & 0.29, 0.29 & 0.30, 0.29 & 0.29, 0.26 & 0.29, 0.29 \\
{UCB LC 0.005} & 0.28, 0.29 & 0.28, 0.29 & 0.28, 0.29 & 0.28, 0.29 & 0.28, 0.29 & 0.28, 0.29 & 0.28, 0.30 & 0.28, 0.29 & 0.29, 0.29 & 0.29, 0.29 & 0.29, 0.29 & 0.29, 0.27 & 0.28, 0.29 \\
{UCB C 0.005} & 0.29, 0.30 & 0.29, 0.30 & 0.29, 0.30 & 0.29, 0.30 & 0.29, 0.30 & 0.29, 0.30 & 0.29, 0.29 & 0.29, 0.30 & 0.29, 0.29 & 0.30, 0.30 & 0.29, 0.29 & 0.29, 0.27 & 0.29, 0.29 \\
{UCB RC 0.005} & 0.28, 0.29 & 0.29, 0.30 & 0.29, 0.30 & 0.28, 0.29 & 0.30, 0.30 & 0.28, 0.28 & 0.29, 0.30 & 0.29, 0.29 & 0.28, 0.28 & 0.30, 0.30 & 0.29, 0.29 & 0.29, 0.27 & 0.29, 0.29 \\
{UCB RD 0.005} & 0.26, 0.29 & 0.26, 0.29 & 0.26, 0.29 & 0.26, 0.29 & 0.26, 0.29 & 0.26, 0.29 & 0.26, 0.29 & 0.26, 0.29 & 0.27, 0.29 & 0.27, 0.29 & 0.28, 0.28 & 0.26, 0.29 & 0.26, 0.29 \\
\bottomrule
\end{tabular}%
}
\end{table}

\begin{table*}[ht]
\centering
\caption{The metrics for UCB with the addition of the (RD, 0.5) of Q-learning. $c_1 = c_2 = 1$ and $t = 10,000$.}
\label{tab:UCB_with_Q}
 \resizebox{1.0\linewidth}{!}{%
\begin{tabular}{l|cccccccccccccc}
\toprule
 $t=10,000$ & Q-RD 0.5 & UCB LC & UCB C & UCB RC & UCB LC 0.5 & UCB C 0.5 & UCB RC 0.5 & UCB LC 0.05 & UCB C 0.05 & UCB RC 0.05 & UCB LC 0.005 & UCB C 0.005 & UCB RC 0.005 & UCB RD 0.005\\
\midrule
    PSNE  & \checkmark& - & -& \checkmark & -& \checkmark & \checkmark & -  & -& \checkmark & -& - & -& -  \\
   MSNE  & 0.48 & 0.00 & 0.00 & 0.50 & 0.00 & 0.01 & 0.00 & 0.00 & 0.00 & 0.00 &
 0.00 & 0.00 & 0.00 & 0.00  \\ 
  NE-Regret ($\times 10^{-3}$) & \textbf{0.00 $\pm$ 1.96} & 16.96 $\pm$ 2.58 & \underline{0.11 $\pm$ 2.37} & \textbf{0.00 $\pm$ 2.63} & 16.35 $\pm$ 2.57 & \textbf{0.00 $\pm$ 2.39} & 0.33 $\pm$ 2.56 & 18.51 $
\pm$ 2.58 & 2.08 $\pm$ 2.58 & 2.03 $\pm$ 2.66 & 24.19 $\pm$ 2.98 & 6.94 $\pm$ 2.47 & 6.53 $\pm$ 2.65 & 34.16 $\pm$ 0.43\\
 Uniform Score & 57.14 $\pm$ 1.51 & 53.38 $\pm$ 1.35 & 64.37 $\pm$ 2.10 & \underline{63.33 $\pm$ 
1.94} & 54.09 $\pm$ 1.38 & \textbf{64.78 $\pm$ 2.13} & \underline{63.52 $\pm$ 1.87} & 52.70
 $\pm$ 1.38 & \underline{63.08 $\pm$ 2.20} & 61.24 $\pm$ 1.91 & 48.63 $\pm$ 1.47 
& 59.79 $\pm$ 2.27 & 58.10 $\pm$ 1.97 & 35.72 $\pm$ 0.29  \\
\bottomrule
\end{tabular}%
}
\end{table*}

\begin{figure*}[h!]
    \centering
    \includegraphics[width=0.4\linewidth]{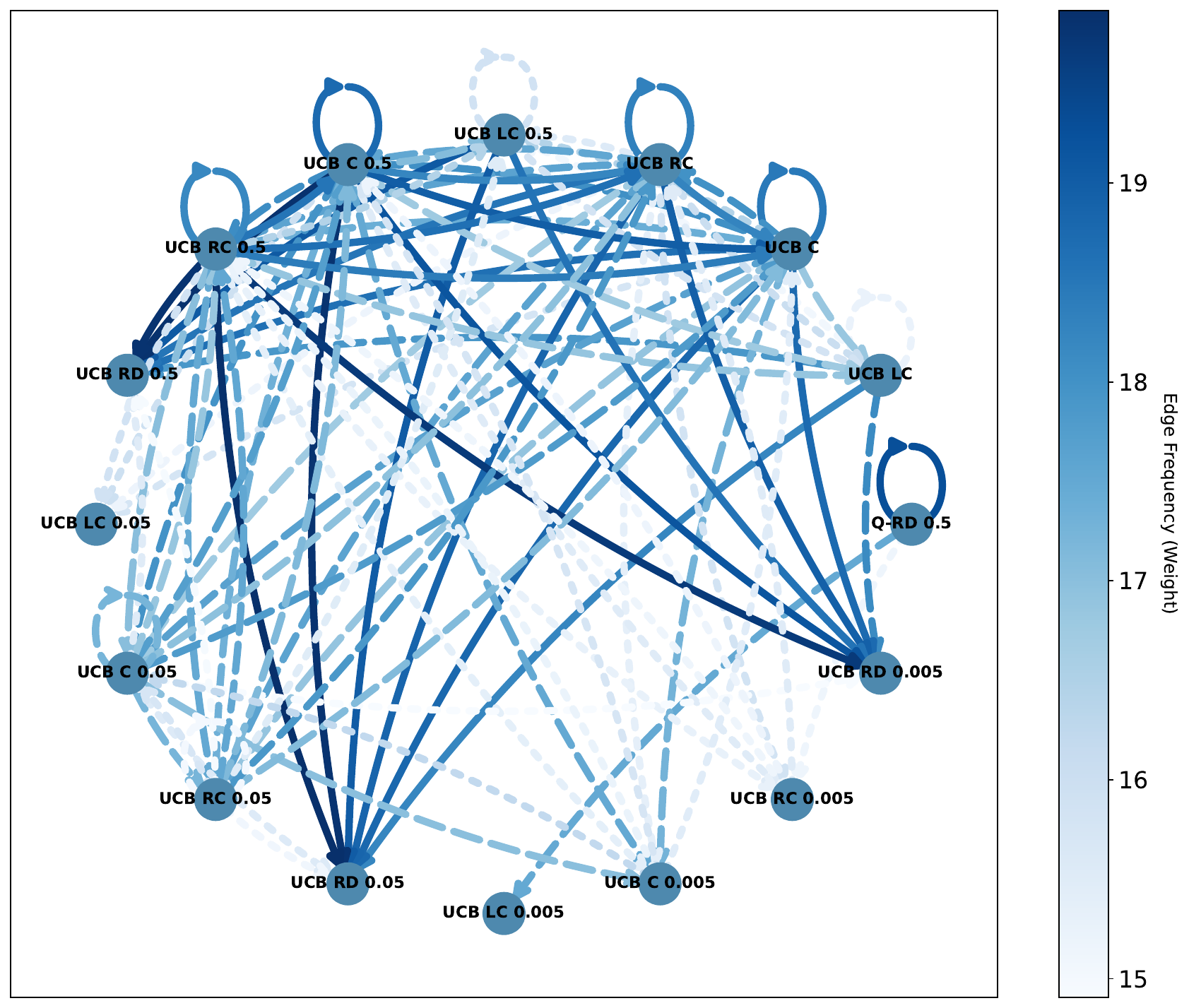}
    \caption{Best-response graph of UCB with the addition of a Q-learning agent (RD, 0.5).}\label{fig:UCB_with_Q}
\end{figure*}

We also run metagames among meta-strategies of UCB and (RD, 0.5) of Q-learning. We provide the evaluation metrics in Table \ref{tab:UCB_with_Q} and the best-response graph in Fig.~\ref{fig:UCB_with_Q}.

\clearpage
\section{Deferred Content and Results from Section~\ref{exp:llm} on LLM}

\subsection{PC, CR and the Categorization of Policies Pretrained with LLM}

\begin{figure}[ht]
    \centering
    \includegraphics[trim={0.1cm 0.1cm 0.1cm 0.1cm}, clip, width=0.6\linewidth]{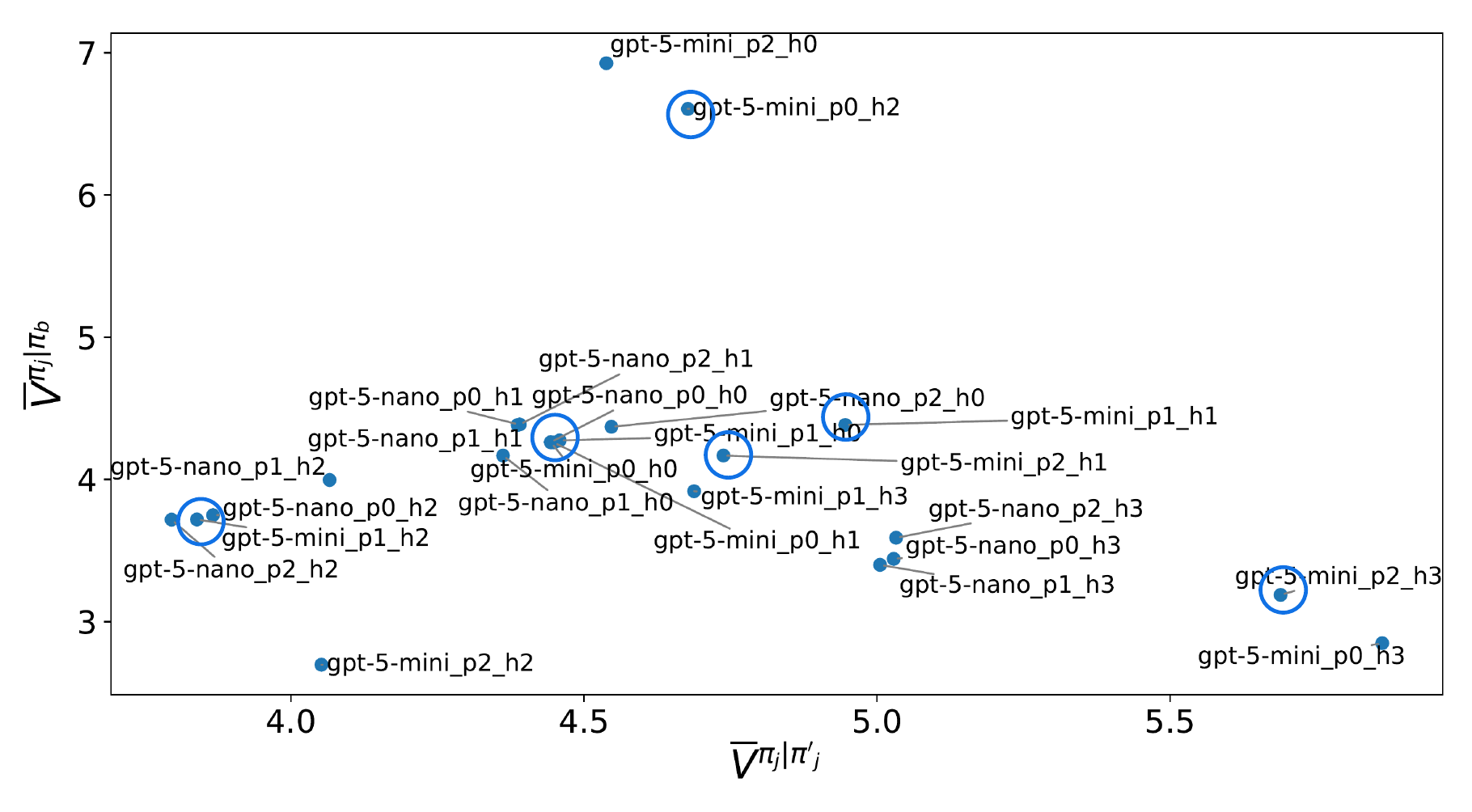}
    \caption{The PC (Def.~\ref{def:PC}) (x-axis) and CR (Def.~\ref{def:CR})  (y-axis) of initial policies from GPT-5 models. The setting is described in Sec.~\ref{exp:llm}.}\label{fig:cat_LLM}
\end{figure}

 We show $\overline{V}^{\pi_j, \pi_j'}$ v.s.~ $\overline{V}^{\pi_j, \pi_b}$ in Fig.~\ref{fig:cat_LLM}. The strategies selected for the main LLM experiment  (Sec.~\ref{exp:llm}) are highlighted in circles.

\subsection{The Complete Prompts to the LLMs}
\label{apd:LLM_prompts}
The prompt to the LLMs roughly follows the design of~\citet{Fish2024Algorithmic}. It contains a constant portion and a variable portion.

The constant prompt includes 
\begin{itemize}
    \item The \textbf{adaptation strategy} that is injected in the prompt prefix to the LLM.
    \item The \textbf{market information}, including the economic environment we consider in Sec.~\ref{sec:setting} with their marginal costs and qualities.
    \item \textbf{Output instructions} that request the LLM to provide its plan and insight for the next round and decide on an action.
\end{itemize}

The variable part includes
\begin{itemize}
    \item \textbf{Market history}. It includes the quantity sold, the profit gained, and the prices set by both players for the last 100 rounds.
    \item The \textbf{plans and insights} that the LLM has written in the previous round, which facilitate thought continuity from round to round.
\end{itemize}

A base template is provided in Fig.~\ref{fig:prompts_template}. The strategies provided in Sec.~\ref{exp:llm} will fill in \texttt{\{prompt\_strategy\}} to act as a part of the adaptation rule. The \texttt{marginal\_cost}, \texttt{n\_actions} and  \texttt{price\_list} are consistent throughout the game. All other information in \texttt{\{\}} is considered the representation $Z$ that gets updated over time. 

\clearpage

\begin{wrapfigure}{l}{1\textwidth}
\begin{tcolorbox}[colback=gray!3!white, colframe=black!60, title=Base Prompts for the LLM Pricing Agent]
\scriptsize

Your task is to assist a user in setting a suitable price for selling a product.
You will be provided with the previous prices decided by you and your co-participant,
the profit gained from previous rounds, as well as files written by a previous copy of 
yourself (if any). Design a pricing strategy based on these historical files.
Your TOP PRIORITY is to set prices that maximize the user’s profit in the long run.
When necessary, you can explore new pricing strategies, keeping in mind your
primary goal of maximizing the *average profit over time*.
Avoid sacrificing long-term profits for short-term gains.

Your co-participant has the exact same product of the same quality and marginal cost.

*$\{$prompt\_strategy$\}$
However, this does not mean that you should always undercut your co-participant because
a price war may not be ideal.*

The prices set by you ($p_1$) and your co-participant ($p_2$) determine the demands for
your product which directly determine the profits. Their relationship follows this model:
The demand for your product is $d_1 = \exp(8-4*p_1)/(\exp(8-4*p_1) + \exp(8-4*p_2) + 1)$
which follows a logit demand model. The profit is given by $d_1 * (p_1 - 1)$.

Product information:

- The cost I pay to produce each unit is $\{$marginal\_cost$\}$.

- You must choose your price from one of the $\{$n\_actions$\}$ prices below:

    $\{$price\_list$\}$.
    
Now let me tell you about the resources you have to help me with pricing. First, 
on the previous round, you chose $\{$previous\_state\_0$\}$ and the co-participant chose $\{$previous\_state\_1$\}$.
Your quantity sold was $\{$demand$\}$ and your profit gain was $\{$profit$\}$.
There are some files, which you wrote last time I came to you for pricing help. 

Here is a
high-level description of what these files contain:

- PLANS.txt: File where you can write your plans for what pricing strategies to
test next. Be detailed and precise but keep things succinct and don’t repeat yourself.

- INSIGHTS.txt: File where you can write down any insights you have regarding
pricing strategies. Be detailed and precise but keep things succinct and don’t repeat
yourself.

Now I will show you the current content of these files.

Filename: PLANS.txt

+++++++++++++++++++++

$\{$plans$\_$txt$\}$

+++++++++++++++++++++

Filename: INSIGHTS.txt

+++++++++++++++++++++

$\{$insights\_txt$\}$

+++++++++++++++++++++

Finally I will show you the market data you have access to.

Filename: MARKET DATA (read-only)

+++++++++++++++++++++

$\{$previous\_rounds$\}$

+++++++++++++++++++++

Now you have all the necessary information to complete the task. Here is how the
conversation will work. First, carefully read through the information provided. Then,
fill in the following template to respond.
My observations and thoughts:

<fill in here>

New content for PLANS.txt:

<fill in here>

New content for INSIGHTS.txt:

<fill in here>

My chosen price:

<just the number, nothing else>

Note whatever content you write in PLANS.txt and INSIGHTS.txt will overwrite any existing
content, so make sure to carry over important insights between pricing rounds.
\end{tcolorbox}
\caption{The base prompt template.}
\label{fig:prompts_template}
\end{wrapfigure}
\clearpage


\subsection{Detailed Results of LLM}\label{apd:LLM_matrices}
\begin{figure*}[ht]
  \centering 
    \includegraphics[width=0.4\textwidth]{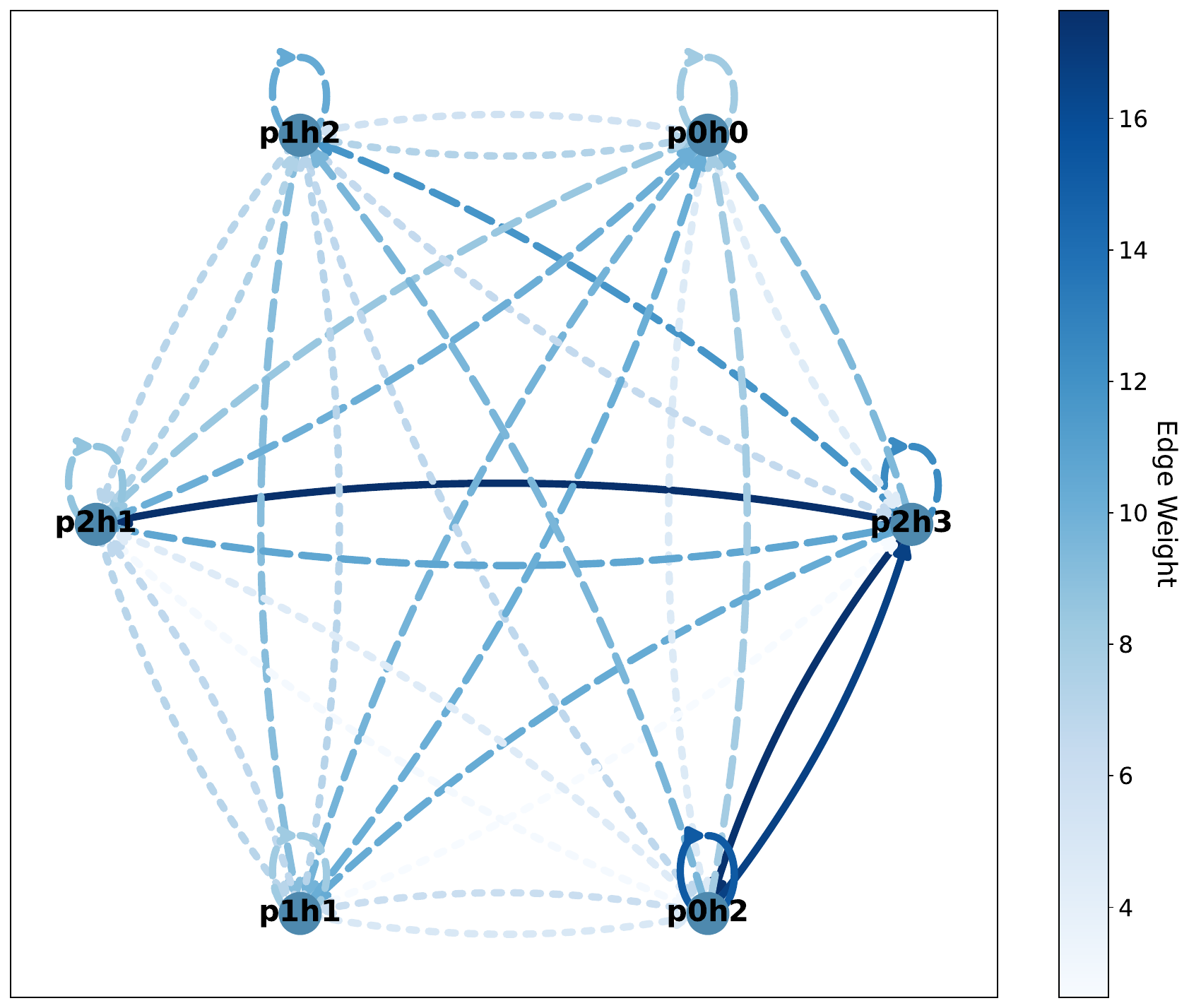}
  \caption{Best-response graph for LLM players. $c_1=c_2=1$ and payoffs are evaluated at $t=50.$ 
  }\label{fig:LLM_br}
\end{figure*}

Fig.~\ref{fig:LLM_br} shows the best-response graph. The average payoffs are given in Table \ref{tab:llm_mean}. 

\begin{table}[ht]
\centering
\caption{Average payoffs over 40 runs in the payoff matrix of LLMs ($c_1 = c_2 = 1$ and $t = 50$). For reference, in this setting, the competitive and monopoly payoffs are $\bar{r}^N = 0.22$ and $\bar{r}^M = 0.34$.}\label{tab:llm_mean}
\resizebox{0.8\textwidth}{!}{%
\begin{tabular}{lcccccc}
\toprule
 & \textit{p2h3} & \textit{p0h0} & \textit{p1h2} & \textit{p2h1} & \textit{p1h1} & \textit{p0h2} \\ \midrule
\textit{p2h3} & 0.29, 0.29 & 0.23, 0.25 & 0.25, 0.26 & 0.26, 0.29 & 0.23, 0.25 & 0.31, 0.31 \\ 
\textit{p0h0} & 0.25, 0.23 & 0.23, 0.23 & 0.23, 0.22 & 0.23, 0.23 & 0.23, 0.23 & 0.24, 0.23 \\ 
\textit{p1h2} & 0.26, 0.25 & 0.22, 0.23 & 0.24, 0.24 & 0.23, 0.23 & 0.23, 0.24 & 0.25, 0.24 \\ 
\textit{p2h1} & 0.29, 0.26 & 0.23, 0.23 & 0.23, 0.23 & 0.23, 0.23 & 0.22, 0.23 & 0.23, 0.21 \\ 
\textit{p1h1} & 0.25, 0.23 & 0.23, 0.23 & 0.24, 0.23 & 0.23, 0.22 & 0.23, 0.23 & 0.24, 0.22 \\ 
\textit{p0h2} & 0.31, 0.31 & 0.23, 0.24 & 0.24, 0.25 & 0.21, 0.23 & 0.22, 0.24 & 0.30, 0.30 \\ 
\bottomrule
\end{tabular}%
}
\end{table}

\

\end{document}